\documentclass[aps,prd,preprint,tightenlines,superscriptaddress,showpacs,byrevtex]{revtex4}
\usepackage{graphicx,color} 
\usepackage{dcolumn}  
\usepackage{bm} 

\graphicspath{{ps}}

\renewcommand{\arraystretch}{1.1}

\def\ra{\!\rightarrow\!}
\def\Taupipi0{\tau^{-}\!\ra\pi^{-}\pi^{0}\,\nu_{\tau}}
\def\Tauhpi0{\tau^{-}\!\ra h^{-}\pi^{0}\,\nu_{\tau}}
\def\Pipi0{\pi^{-}\pi^{0}}
\def\2piamu{a_{\mu}^{{\rm had},2\pi}}

\def\MassSQ{M^{2}_{\pi\pi^{0}}}

\def\MassSQgen{M^{2}_{\rm gen}}

\def\GeVCC{{\rm GeV}/c^{2}}
\def\GeVcc2{({\rm GeV}/c^{2})^{2}}
\def\GeVinvcc2{{\rm GeV}^{-2}c^{4}}

\definecolor{mycolor}{gray}{0.8}

\begin{document}

\vspace*{-3\baselineskip}
\resizebox{!}{3cm}{\includegraphics{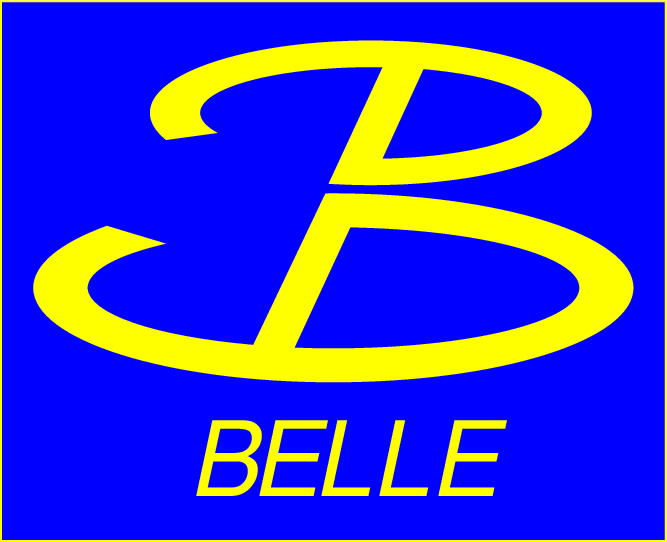}}

\preprint{\vbox{ \hbox{   }
    \hbox{}
    \hbox{}
    \hbox{}
    \hbox{}
    \hbox{}
    \hbox{}
    \hbox{BELLE Preprint 2008-16}
    \hbox{KEK Preprint 2008-10}
}}


\title{ \quad\\[0.5cm] 
 High-Statistics Study of the  $\tau^{-}\!\rightarrow \pi^{-}\pi^{0}\,\nu_{\tau}$  Decay  }

\begin{abstract}
We report a high-statistics measurement of the branching fraction 
for $\tau^{-}\!\ra\pi^{-}\pi^{0}\,\nu_{\tau}$ and the invariant mass spectrum 
of the produced $\pi^{-}\pi^{0}$ system 
using $72.2~{\rm fb}^{-1}$ of data recorded 
with the Belle detector at the KEKB  asymmetric-energy  $e^+e^-$ collider.
The branching fraction obtained is $ (25.24 \pm 0.01 \pm 0.39)\% $,
where the first error is statistical and the second is systematic.
The unfolded $\pi^{-}\pi^{0}$ mass spectrum
is used to determine resonance parameters for the 
$\rho(770)$, $\rho'(1450)$, and $\rho''(1700)$ mesons.
We also use this spectrum to estimate the hadronic ($2\pi$) contribution 
to the anomalous magnetic moment of the muon ($a_{\mu}^{\pi\pi}$).
Our result for $a_{\mu}^{\pi\pi}$ integrated over the mass range
$\sqrt{s}=2m_{\pi}-1.8~$ ${\rm GeV}/c^{2}$  is
$
a_{\mu}^{\pi\pi} = (523.5 \pm 1.5\,{\rm (exp)} 
 \pm 2.6\, {\rm (Br)} 
 \pm 2.5 \, 
{\rm (isospin)}  )
\times 10^{-10},
$
where the first error is due to the experimental uncertainties,  
the second is due to the uncertainties in the branching fractions and 
 the  third is due to the uncertainties in the isospin-violating 
corrections.

\end{abstract}

\affiliation{Budker Institute of Nuclear Physics, Novosibirsk}
\affiliation{University of Cincinnati, Cincinnati, Ohio 45221}
\affiliation{Justus-Liebig-Universit\"at Gie\ss{}en, Gie\ss{}en}
\affiliation{The Graduate University for Advanced Studies, Hayama}
\affiliation{Hanyang University, Seoul}
\affiliation{University of Hawaii, Honolulu, Hawaii 96822}
\affiliation{High Energy Accelerator Research Organization (KEK), Tsukuba}
\affiliation{Hiroshima Institute of Technology, Hiroshima}
\affiliation{University of Illinois at Urbana-Champaign, Urbana, Illinois 61801}
\affiliation{Institute of High Energy Physics, Chinese Academy of Sciences, Beijing}
\affiliation{Institute of High Energy Physics, Vienna}
\affiliation{Institute of High Energy Physics, Protvino}
\affiliation{Institute for Theoretical and Experimental Physics, Moscow}
\affiliation{J. Stefan Institute, Ljubljana}
\affiliation{Kanagawa University, Yokohama}
\affiliation{Korea University, Seoul}
\affiliation{Kyungpook National University, Taegu}
\affiliation{\'Ecole Polytechnique F\'ed\'erale de Lausanne (EPFL), Lausanne}
\affiliation{Faculty of Mathematics and Physics, University of Ljubljana, Ljubljana}
\affiliation{University of Maribor, Maribor}
\affiliation{University of Melbourne, School of Physics, Victoria 3010}
\affiliation{Nagoya University, Nagoya}
\affiliation{Nara Women's University, Nara}
\affiliation{National Central University, Chung-li}
\affiliation{National United University, Miao Li}
\affiliation{Department of Physics, National Taiwan University, Taipei}
\affiliation{H. Niewodniczanski Institute of Nuclear Physics, Krakow}
\affiliation{Nippon Dental University, Niigata}
\affiliation{Niigata University, Niigata}
\affiliation{University of Nova Gorica, Nova Gorica}
\affiliation{Osaka City University, Osaka}
\affiliation{Osaka University, Osaka}
\affiliation{Panjab University, Chandigarh}
\affiliation{RIKEN BNL Research Center, Upton, New York 11973}
\affiliation{Saga University, Saga}
\affiliation{University of Science and Technology of China, Hefei}
\affiliation{Seoul National University, Seoul}
\affiliation{Shinshu University, Nagano}
\affiliation{Sungkyunkwan University, Suwon}
\affiliation{University of Sydney, Sydney, New South Wales}
\affiliation{Tohoku Gakuin University, Tagajo}
\affiliation{Department of Physics, University of Tokyo, Tokyo}
\affiliation{Tokyo Institute of Technology, Tokyo}
\affiliation{Tokyo Metropolitan University, Tokyo}
\affiliation{Tokyo University of Agriculture and Technology, Tokyo}
\affiliation{Virginia Polytechnic Institute and State University, Blacksburg, Virginia 24061}
\affiliation{Yonsei University, Seoul}
 \author{M.~Fujikawa}\affiliation{Nara Women's University, Nara} 
  \author{H.~Hayashii}\affiliation{Nara Women's University, Nara} 
 \author{S.~Eidelman}\affiliation{Budker Institute of Nuclear Physics, Novosibirsk} 
  \author{I.~Adachi}\affiliation{High Energy Accelerator Research Organization (KEK), Tsukuba} 
  \author{H.~Aihara}\affiliation{Department of Physics, University of Tokyo, Tokyo} 
  \author{K.~Arinstein}\affiliation{Budker Institute of Nuclear Physics, Novosibirsk} 
  \author{V.~Aulchenko}\affiliation{Budker Institute of Nuclear Physics, Novosibirsk} 
  \author{T.~Aushev}\affiliation{\'Ecole Polytechnique F\'ed\'erale de Lausanne (EPFL), Lausanne}\affiliation{Institute for Theoretical and Experimental Physics, Moscow} 
  \author{A.~M.~Bakich}\affiliation{University of Sydney, Sydney, New South Wales} 
 \author{V.~Balagura}\affiliation{Institute for Theoretical and Experimental Physics, Moscow} 
% ...
\collaboration{The Belle Collaboration}

%
\maketitle

{\renewcommand{\thefootnote}{\fnsymbol{footnote}}}
\setcounter{footnote}{0}

\section{Introduction}

Hadronic decays of the $\tau$ lepton provide a clean environment for  
studying the dynamics of hadronic states with various quantum numbers.
Among the decay channels of the $\tau$ lepton, $\Taupipi0$
has the largest branching fraction~\cite{CCG}. 
The decay is dominated by intermediate resonances and thus
can be used to extract  
information on the properties of the $\rho(770)$, $\rho^{\prime}(1450)$, and 
$\rho^{\prime\prime}(1700)$ mesons and their mutual interference.

From the conservation of vector current~(CVC) theorem,
the  $\pi^{-}\pi^{0}$ mass spectrum in this range
can be related to the cross section for the process $e^+e^- \to \pi^+\pi^-$
and thus used to improve the theoretical error 
on the anomalous magnetic moment of the muon 
$a_{\mu}=(g_{\mu}-2)/2$.  Recent reviews of  
calculations of $a_{\mu}$ are given in Refs.~\cite{DM2004,PA2005,JE2007}.   
It is known that the theoretical error on $a_{\mu}$ is dominated by 
the contribution from the leading-order hadronic vacuum polarization 
$a_{\mu}^{\rm had,LO}$.
This contribution cannot be derived within the framework of
perturbative QCD and is usually evaluated using dispersion relations
and the experimental cross section for $e^+e^-$ annihilation 
to hadrons~\cite{DEHZ, DEHZ2, HMNT,DHZ2006}.
 Alternatively, 
CVC 
relates the properties of 
the $\pi^{+}\pi^{-}$ system produced 
in $e^+e^-\rightarrow\pi^+\pi^{-}$ to those of the 
$\pi^{-}\pi^{0}$ system produced in $\Taupipi0$ decay; 
thus, using CVC and correcting for 
isospin-violating effects, 
$\tau$ data have also been used to obtain a
 more precise
 prediction for $a^{\rm had,LO}_{\mu}$~\cite{DH98,DEHZ,DEHZ2,DHZ2006}.

Recently, new precise data on $e^+e^-\!\rightarrow\!\pi^+\pi^-$  
have become available from the CMD-2, KLOE, 
and SND experiments~\cite{CMD2002,CMD2004,CMD2005,CMD2006,CMD2007,
KLOE2005,SND2005,SND2006}.   
ALEPH~\cite{ALEPH97,ALEPH05}, CLEO~\cite{CLEO2000,CLEO94}, 
and OPAL~\cite{OPAL1999,OPAL98M} measured both the $2\pi$ spectral function and the 
branching fraction for the
$\Taupipi0$ decay; the latter was also determined by L3~\cite{L395} and DELPHI~\cite{DELPHI06}.  
Recent evaluations 
of the hadronic contribution to $a_{\mu}$ using
$e^+e^-$ data   result in
$ a^{\rm exp}_{\mu} - a^{\rm th}_{\mu} = (27.5\pm 8.4)\times 10^{-10}$~\cite{EID2006,DAV2007},
while that using the  $\tau$ lepton data 
where applicable gives
$ a^{\rm exp}_{\mu} - a^{\rm th}_{\mu} = (9.4\pm 10.5  )\times 10^{-10}$~\cite{DEHZ},
where the experimental value $a_\mu^{\rm exp}$
is dominated by the BNL E821 measurement~\cite{BNL2004}
$( 11\ 659\ 208.0 \pm 6.3)\times 10^{-10}$. 
These differences correspond to 3.3 and 0.9 standard 
deviations, respectively. 
For the evaluation based on the $e^+e^-$ data, a deviation of similar size that corresponds
 to a $3.4\, \sigma$ discrepancy is claimed 
in Ref. ~\cite{Hagiwara2007}.
To clarify these differences  between the $e^+e^-$-based and 
$\tau$-based predictions,
more data on $e^+e^-\!\rightarrow\!\pi^-\pi^+$ and 
$\tau^-\!\rightarrow\!\pi^-\pi^0\nu_{\tau}$
 decays are needed. 
In this paper we present a high-statistics measurement 
of  the $\pi^{-}\pi^{0}$ mass spectrum produced in
$\Taupipi0$ decays using 
data collected with the Belle experiment at the 
KEKB asymmetric-energy $e^+e^-$ collider operating at a 
center-of-mass (CM) energy of~10.6~GeV. 
The data sample is about 50 times larger 
than those of previous experiments.

\section{Basic formulas}

The differential decay rate for
$\Taupipi0$  can be expressed as~\cite{ISB2001}
\begin{eqnarray}
\frac{{\rm d}\Gamma(\Taupipi0)}{{\rm d}s}
 = \Gamma_{e}^{0} \cdot
\frac { 6\pi |V_{ud}|^{2} S_{\rm EW}^{\pi\pi} } {m_{\tau}^{2}} 
     \left( 1 - \frac{s}{m_{\tau}^{2}} \right)^{2}
     \left( 1 + \frac{2s}{m_{\tau}^{2}} \right)
\,v_{-}(s),
\label{eq:tauspec} 
\end{eqnarray}
with
\begin{eqnarray}
 \Gamma_{e}^{0} = \frac{G_{F}^{2} m_{\tau}^{5}}{192\pi^{3}}.
\end{eqnarray}
Here $s$ is the invariant mass squared of the $\pi^{-}\pi^{0}$ system,
$v_{-}(s)$ is the weak spectral function
characterizing the $\pi^{-}\pi^{0}$ system,   $G_{F}$ is the Fermi coupling constant,  
$|V_{ud}|=0.97377\pm 0.00027$~\cite{PDG2006} is  the Cabibbo-Kobayashi-Maskawa (CKM) 
 matrix element~\cite{CKM}, 
$m_{\tau}=1776.99 ^{+0.29}_{-0.26}~{\rm MeV/}c^{2}$~\cite{PDG2006} is
 the $\tau$ lepton mass and  
$S_{\rm EW}^{\pi\pi}$ accounts for short-distance electroweak radiative corrections for the $\pi^{-}\pi^{0}$ system.
The measured electron decay rate of the $\tau$ lepton is related to $\Gamma_{e}^{0}$ by

\begin{eqnarray}
 \Gamma(\tau^{-} \rightarrow e^- \bar{\nu_{e}}\nu_{\tau})
\equiv   \Gamma_{e}^{0} S_{\rm EW}^{e}
  = \Gamma_{e}^{0} 
\left\{ 1 + \frac{\alpha(m_{\tau})}{2\pi} \left( \frac{25}{4} -\pi^{2} \right) \right\},
\label{eq:edecay}
\end{eqnarray} 
where $S_{\rm EW}^{e}$ is the electroweak radiative correction for the decay $\tau^{-} \rightarrow e^- \bar{\nu_{e}}\nu_{\tau}$.

The corresponding $\pi^{+}\pi^{-}$spectral function $v_{0}(s)$ can
be obtained from the $e^+e^-\rightarrow \pi^+\pi^-$ cross section
\begin{eqnarray}
\sigma(e^+e^-\rightarrow\pi^+\pi^-)  = 
\frac{4\pi^{2}\alpha_{0}^{2}}{s}\,v_{0}(s),
\label{eq:eepipi}
\end{eqnarray}
where $s$ is the $e^+e^-$ CM energy squared and
$\alpha_{0}$ is the fine-structure constant  at $s=0$.
Up to isospin-violating effects, CVC allows one to 
relate the spectral function from $\tau$ decays
to the isovector part of the $e^+e^-$ spectral function~\cite{Weakf}:
\begin{eqnarray}
v_{-}(s) & = & v_{0}^{I=1}(s)\,.
\label{eq:cvc}
\end{eqnarray}

Alternatively, the mass spectrum of the two-pion system can be expressed 
in terms of pion form factors, which are useful for comparing  
resonance shapes in the charged and neutral two-pion systems. 
The spectral function $v_{j}(s)\ (j=-,0)$ is related to the 
form factor $F^{j}_{\pi}(s)$ via
\begin{equation}
v_{j}(s) = \frac{\beta_{j}^{3}(s)}{12\pi}|F_{\pi}^{j}(s)|^{2},
\label{eq:pionform}
\end{equation} 
where $\beta_{-}(s)\, (\beta_{0}(s))$ 
is the pion velocity in the $\pi^-\pi^{0}$\, ($\pi^{+}\pi^{-}$) rest system.
The velocities $\beta_{j}(s)$ are explicitly given by\
$\beta_{-}(s) =\lambda^{1/2}(1, m_{\pi^{-}}^{2}/s, m_{\pi^{0}}^{2}/s)$~\cite{KINE}\, 
and\, $\beta_{0}(s) = \lambda^{1/2}(1, m_{\pi^{-}}^{2}/s, m_{\pi^{-}}^{2}/s)= 
\left[ 1 - 4m_{\pi^{-}}^2 /s\right]$,
with 
$\lambda(x,y,z)=[x -(\sqrt{y} + \sqrt{z})^{2}] [x -(\sqrt{y} - \sqrt{z})^{2}]$.

The hadronic physics is contained within $v_{j}(s)$ or, equivalently, in
$F_{\pi}^{j}(s)$.
One goal of this analysis is to provide a high-precision determination of
 the weak form factor $|F_{\pi}^{-}(s)|$
using  $\Taupipi0$ data, so that
a comparison with $|F^{0}_{\pi }(s) |$ from the $e^+e^-$ data
can be used to test CVC.
From   Eqs.~(\ref{eq:tauspec}), (\ref{eq:edecay}) and (\ref{eq:pionform}), one can obtain the 
basic formula that expresses the form factor $F_{\pi}^{-}(s)$ in terms of the observables: 
\begin{eqnarray}
|F^{-}_{\pi}(s)|^{2}=\frac{2m_{\tau}^{2}}
           { |V_{ud}|^{2} \left( 1- \frac{s}{m_{\tau}^{2}}\right)^{2}
           \left( 1 + \frac{2s}{m_{\tau}^{2}} \right) S_{\rm EW} }
            \frac{1}{\beta_{-}^{3}}
          \left(  \frac{\mathcal B_{\pi\pi}}{\mathcal B_{e}}  \right)
           \left(  \frac{1}{N_{\pi\pi}}\frac{{\rm d}N_{\pi\pi}}{{\rm d}s} \right),
\label{eq:fpiweak}
\end{eqnarray}
where   $\mathcal{B}_{\pi\pi}$ is the branching fraction,  
 $(1/N_{\pi\pi} )({\rm d}N_{\pi\pi}/{\rm d}s)$ is  the normalized invariant mass-squared
 distribution  for the $\Taupipi0$ decay, 
$\mathcal{B}_{e}$ is the branching fraction for 
$\tau^{-}\rightarrow e^{-}\nu_{\tau}\bar{\nu}_{e}$ and 
$S_{\rm EW}= S_{\rm EW}^{\pi\pi}/S_{\rm EW}^{e}$.

In this paper, we report new measurements for both the branching fraction  $\mathcal{B}_{\pi\pi}$
 and the normalized mass spectrum $(1/N_{\pi\pi})({\rm d}N_{\pi\pi}/{\rm d}s)$. 
 These results are used to
provide a new evaluation of  the hadronic contribution to the muon
anomalous magnetic moment from the $2\pi$ channel. 

\section{Data Sample and Selection Criteria }

The data sample used was collected with the Belle detector at 
the KEKB asymmetric-energy  $e^{+}e^{-}$ collider~\cite{KEKB}. 
It is based on an integrated luminosity of $72.2~{\rm fb}^{-1}$ 
recorded at a CM energy of 10.58~GeV.
The Belle detector is a large-solid-angle magnetic spectrometer consisting
of several detector components. 
 Charged track coordinates near the collision point are measured by a 
three-layer silicon-vertex detector (SVD) that surrounds a 2~cm radius
 beryllium beam pipe. Track trajectory coordinates are reconstructed 
in a 50-layer central drift chamber (CDC), and momentum measurements 
are made together with the SVD. An array of 1188 silica-aerogel 
Cherenkov counters (ACC), a barrel-like arrangement of time-of-flight 
scintillation counters (TOF), and specific ionization measurements
 ({\it dE/dx}) in the CDC provide a capability for the identification 
of charged particles. Photon detection and energy measurement 
of the photons and electrons are provided by an electromagnetic 
calorimeter (ECL) consisting of an array of 8736 CsI(Tl) crystals 
all pointing toward the interaction point. These detector components 
are located in a magnetic field of 1.5 T provided by a
 superconducting solenoid. An iron flux-return located outside
 the coil is instrumented to identify muons and to detect $ K_L^{0}$ 
mesons (KLM). A comprehensive description of the detector is given 
in Ref.~\cite{Belle}.

To study backgrounds and determine selection criteria,
we perform Monte Carlo~(MC) simulation studies 
for various processes. 
Signal and background $\tau^+\tau^-$-pair events are 
simulated using the KKMC generator~\cite{KKMC}.
The $\tau$ decays are modeled with the TAUOLA 
program~\cite{TAUOLA,TAUOLA2004} in which
the values of the branching fractions
are updated to more recent values~\cite{PDG2004}.
The cross section for $e^+e^-\rightarrow \tau^+\tau^-(\gamma)$ 
is also updated to
the recent measurement reported in Ref.~\cite{BA2008}.
The radiative corrections to the $\tau$-hadronic decays are simulated by the PHOTOS
program~\cite{PHOTOS2003}. 
The QQ generator~\cite{QQ} is used for ${\bar B B}$ and ${\bar q q}$ continuum
processes, the BHLUMI~\cite{BHLUMI} program for radiative Bhabha events,
the KKMC~\cite{KKMC} program for radiative $\mu^+\mu^-$-pair
 events, and the
AAFH~\cite{AAFHB} program for two-photon processes.
The BHLUMI and KKMC programs include higher-order radiative 
corrections and are among the most accurate programs available. 
The detector response is simulated by a GEANT3-based program~\cite{GEANT}. 
In order to simulate beam-induced background realistically, 
detector hits taken from randomly triggered data are added 
to wire hits in the CDC and to energy deposits in the ECL.
Uncertainties due to imperfections in the Monte
Carlo generators and detector simulation are discussed in
the later sections.

\subsection{$\tau^+\tau^-$ pair selection}

The event selection consists of two steps. 
Initially, a sample of generic
$e^+e^-\rightarrow \tau^+\tau^-(\gamma)$ events is
selected with relatively loose criteria. 
From this sample
$\Taupipi0$ decays are identified. 
The number of  generic $\tau^+\tau^-$ events is used to determine
the $\Taupipi0$ branching fraction.

\begin{figure}[t]
\includegraphics[width=0.40\textwidth,clip]{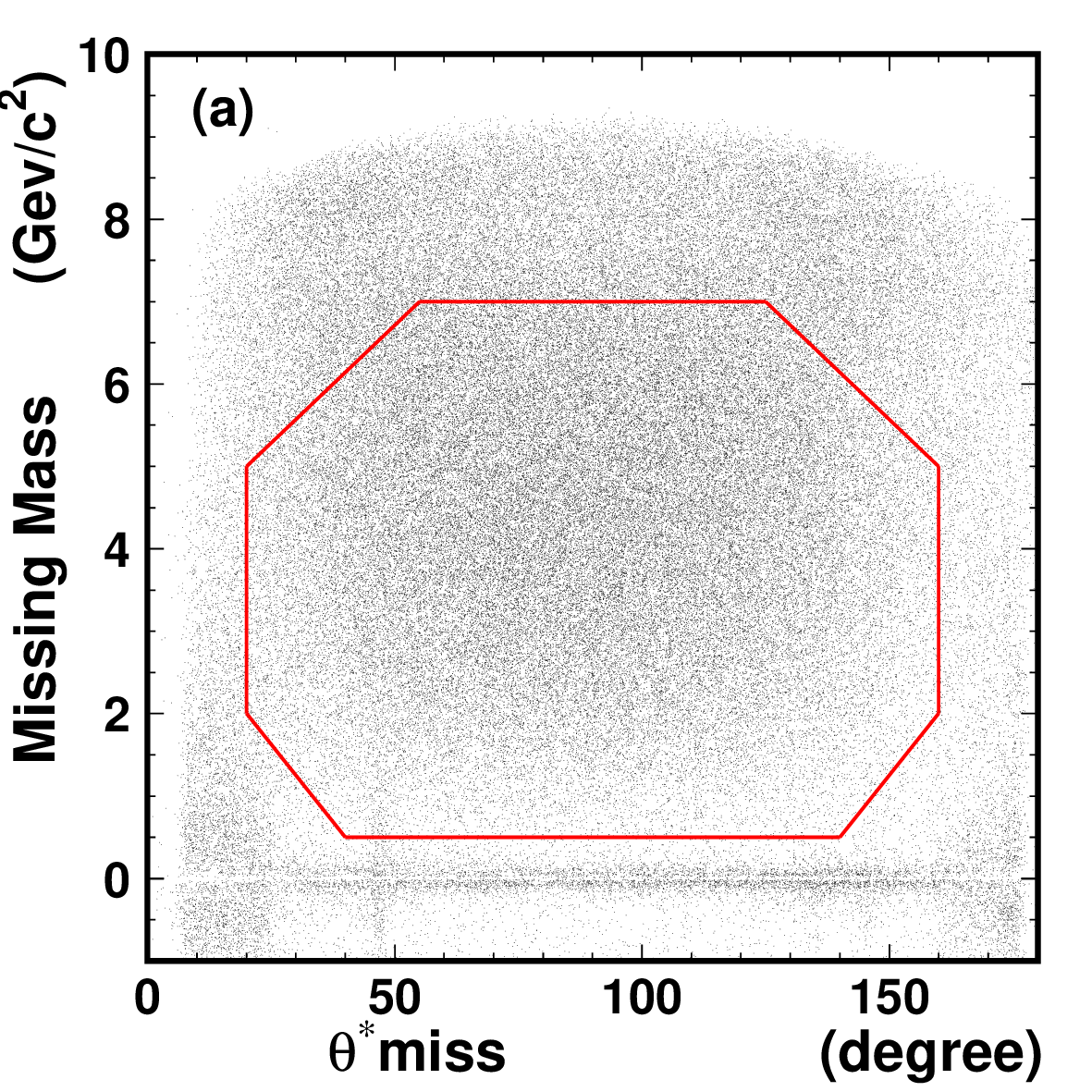}
\includegraphics[width=0.40\textwidth,clip]{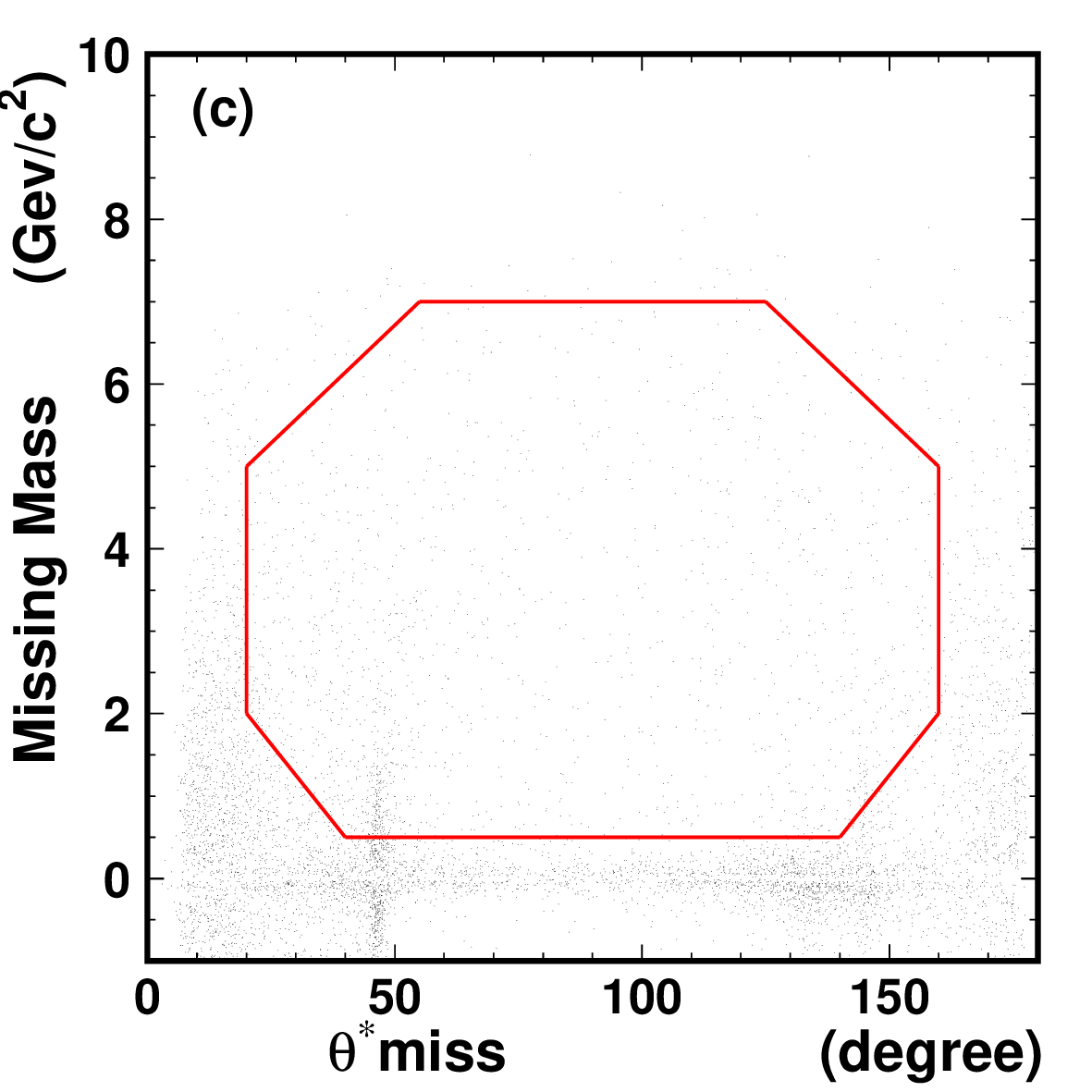}
\includegraphics[width=0.40\textwidth,clip]{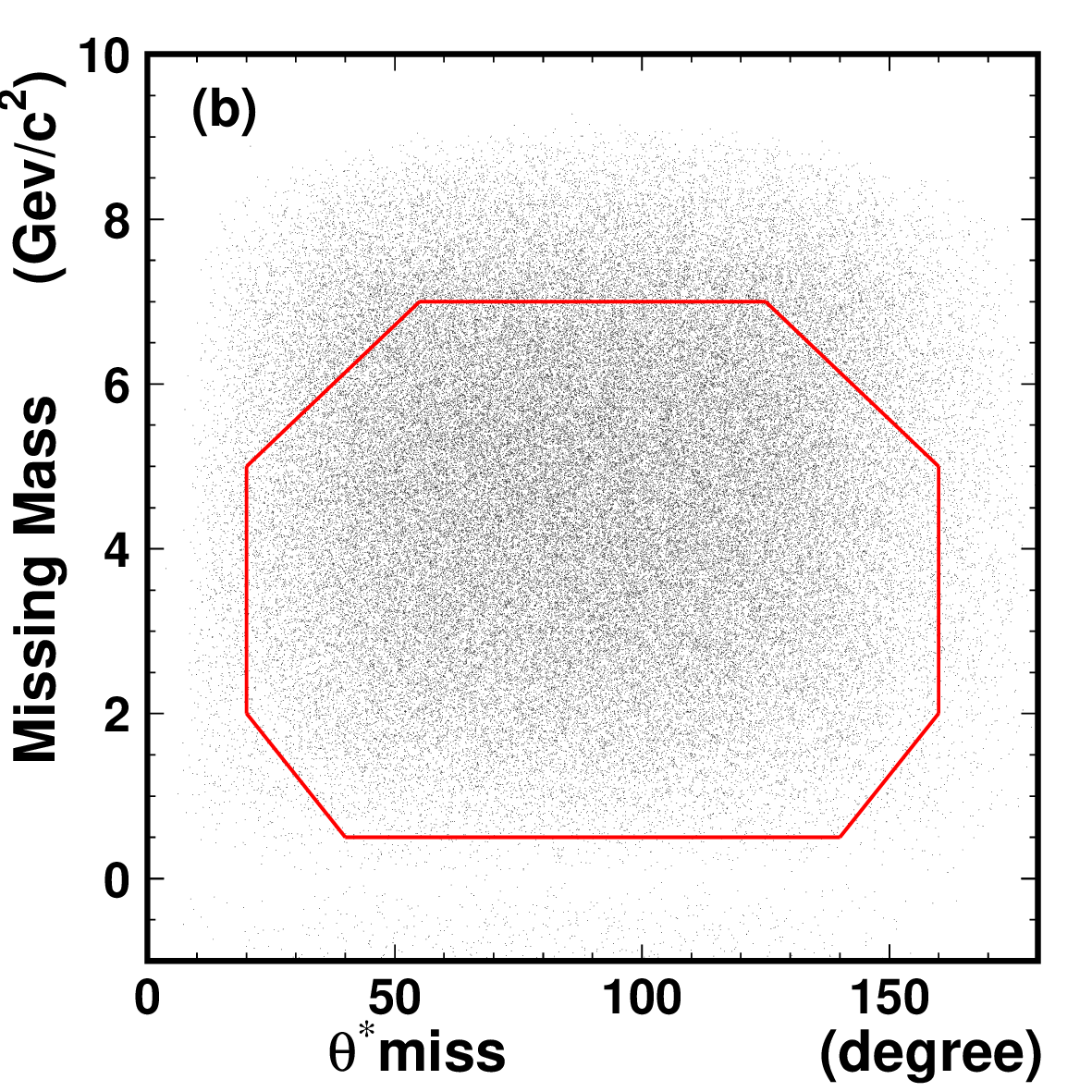}
\includegraphics[width=0.40\textwidth,clip]{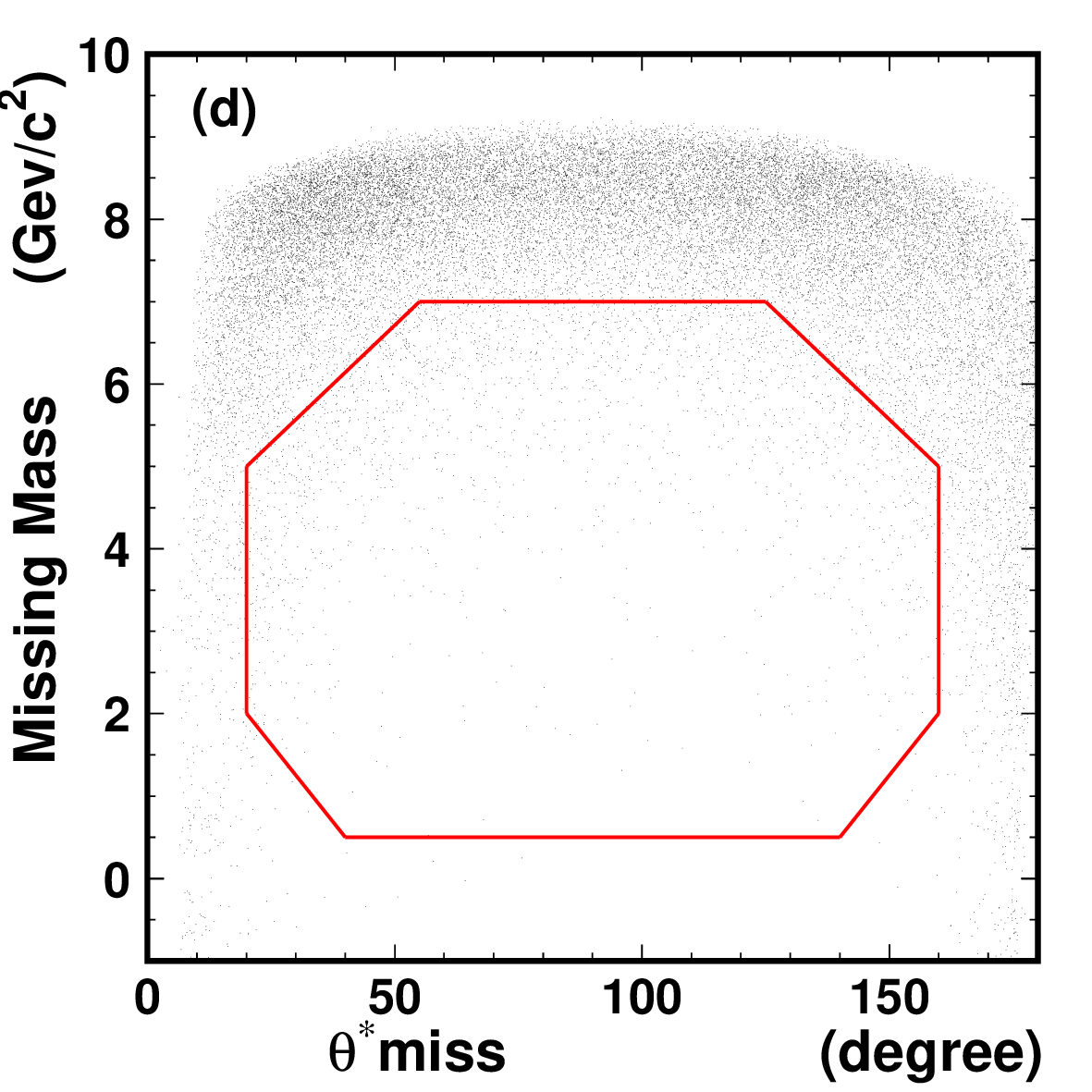}
\caption
{ 
Missing mass ($M_{\rm miss}$) versus  the polar 
angle direction 
of the missing momentum ($\theta^{*}_{\rm miss}$) for
(a) the data, (b) MC 
$e^+e^-\rightarrow \tau^+\tau^-$ events, (c) MC Bhabha  
and $\mu^+\mu^-(\gamma)$
 events and
 (d) two-photon processes. 
Events inside the octagonal region are selected 
as $\tau^+\tau^-$-pair candidates.}
\label{data-mm}
\end{figure}

Generic $\tau^+\tau^-$ events are selected by requiring that
the number of charged tracks in an event  
be two or four with zero net charge; 
that each track have a momentum
transverse to the beam axis ($p_T$)
of greater than 0.1~GeV/$c$ 
to avoid  tracks reentering  the CDC; and that
each track extrapolate to the interaction point (IP) 
within ${\pm1}$ cm transversely 
and within ${\pm5}$ cm along the beam direction to suppress tracks that originate from 
beam-particle interactions with the residual gas in the vacuum chamber.
To suppress background from Bhabha and $\mu^+\mu^-$ events, 
the reconstructed CM energies and
the sum of the momenta of
 the first and the second highest momentum tracks are required to be less 
than 9.0 GeV/$c$. The maximum $p_T$ among the tracks is 
required to be greater than 0.5~GeV/$c$.
Beam-related background is rejected by requiring that the position 
of the reconstructed event vertex be less than 0.5~cm from the IP
in the transverse direction and less than 2.5~cm from the IP
along the beam direction.
The polar angle of the leading particle with respect to the
beam axis ($\theta^*$) in the CM frame is required
to be in the fiducial region of the detector:
${35^{\circ} < \theta^* < 145^{\circ} }$.

To reduce the remaining background from Bhabha, $\mu^+\mu^-(\gamma$),
and two-photon events,
a requirement is imposed in the plane of the
missing mass $M_{\rm miss}$ and the direction of missing momentum 
in CM ${\theta^{*}_{\rm miss}}$, where
$M_{\rm miss}$ is evaluated from the four-momenta
of the measured tracks and  photons:
${(M_{\rm miss})^{2} = (p_{\rm in} - p_{\rm tr} 
- p_{\gamma})^{2} }$. 
In this expression $p_{\rm in}$
is the four-momentum of the initial $e^+e^-$ system, while $p_{\rm tr}$ and $p_{\rm \gamma}$ are
 the sum of the momenta of measured tracks and photons, respectively.
 A pion mass is assumed for the charged tracks if they are not identified as electrons or muons.
Each photon (reconstructed from clusters in the calorimeter) must 
be separated 
from the nearest track projection by at least 20~cm
and have an energy greater than 0.05~GeV in the 
barrel region  
($-0.63\le\cos\theta< 0.85$),\, and greater than 0.1~GeV in the endcap 
region ($-0.90\le\cos\theta< -0.62$ and $0.85\le\cos\theta<0.95$).
Photons near the 
edge of the detector fiducial volume are rejected.  Scatter plots of 
$M_{\rm miss}$ versus \ $\theta_{\rm miss}$ 
for data, the $\tau$-signal MC, the Bhabha and the two-photon MC are shown in Figs.~\ref{data-mm}-(a),
  (b), (c) and (d), respectively.
The band of events in data at $M_{\rm miss}\approx 0$ is due to backgrounds from Bhabha 
and $\mu^{+}\mu^{-}(\gamma)$ processes.
Small vertical bands at $\theta^{*}_{\rm miss} = 45^{\circ}$ and $ = 150^{\circ}$ 
are Bhabha events where the energy of one of the final state electron/positron is
 poorly measured
because it has scattered in the material at the boundary of the barrel and endcap calorimeters.
 The events in the high-$M_{\rm miss}$ region ($\ge7 \GeVCC$) are from the two-photon 
processes.

Events within the octagonal region
are selected as $\tau^+\tau^-$candidates to avoid the tail from background processes.

\begin{figure}[!ht]
\begin{center}
\includegraphics[width=0.45\textwidth,clip]
{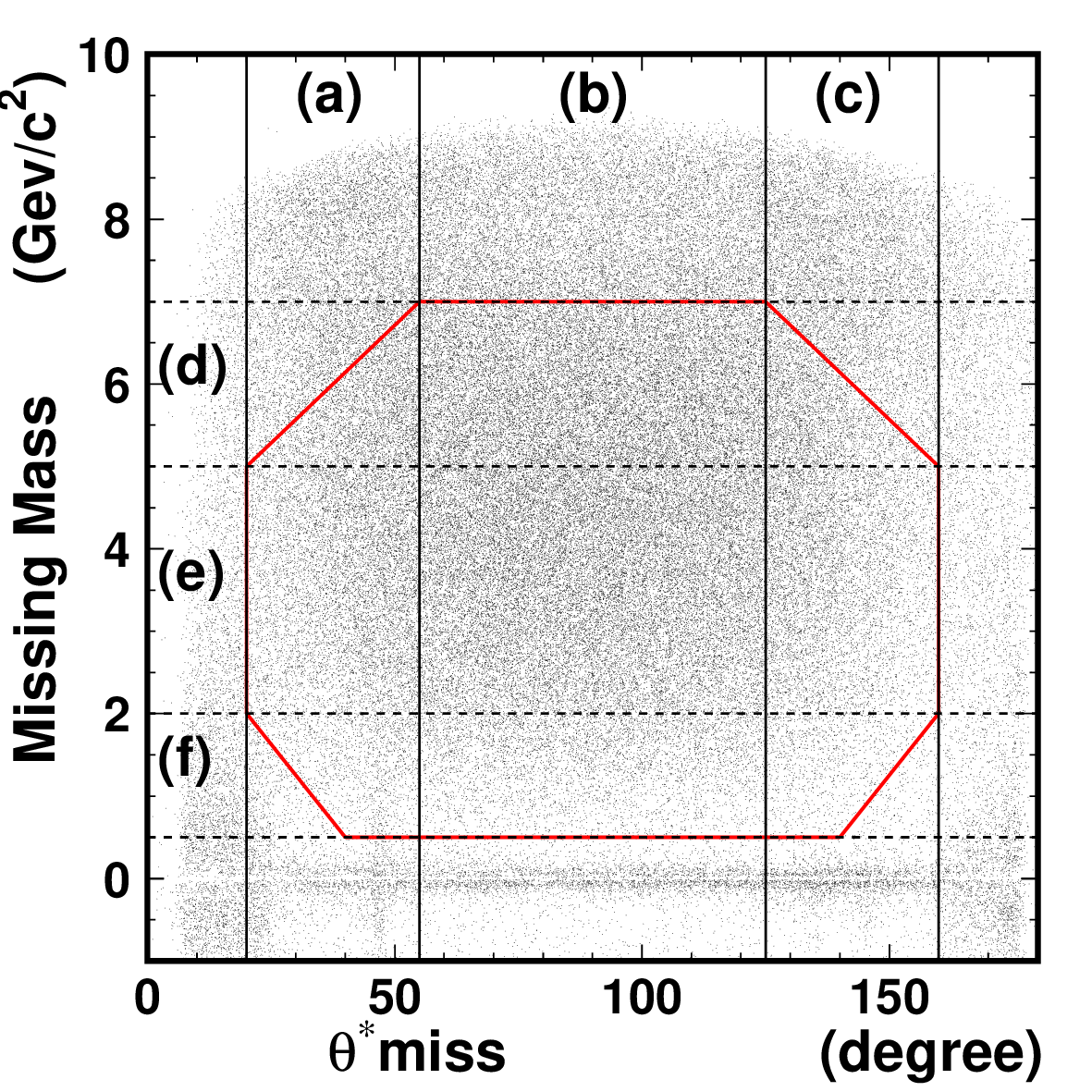}
\caption{Missing mass ($M_{\rm miss}$) versus  the polar 
angle direction 
of the missing momentum ($\theta^{*}_{\rm miss}$) for data.
The solid (dashed) lines show  three vertical (horizontal) slices
that are used to present the projections in Fig.~\ref{mm-sub}.
The coordinates of the vertical (horizontal) lines are $\theta^{*}_{\rm miss}=20^{\circ},\, 
55^{\circ},\,
125^{\circ},\, 160^{\circ}$
($M_{\rm miss}= 0.5~\GeVCC,\, 2.0~\GeVCC,\, 5.0~\GeVCC,\, 7.0~\GeVCC$).}
\label{mm-region}
\end{center}
\end{figure}

\begin{figure}[!htp]
\includegraphics[width=0.39\textwidth,clip]{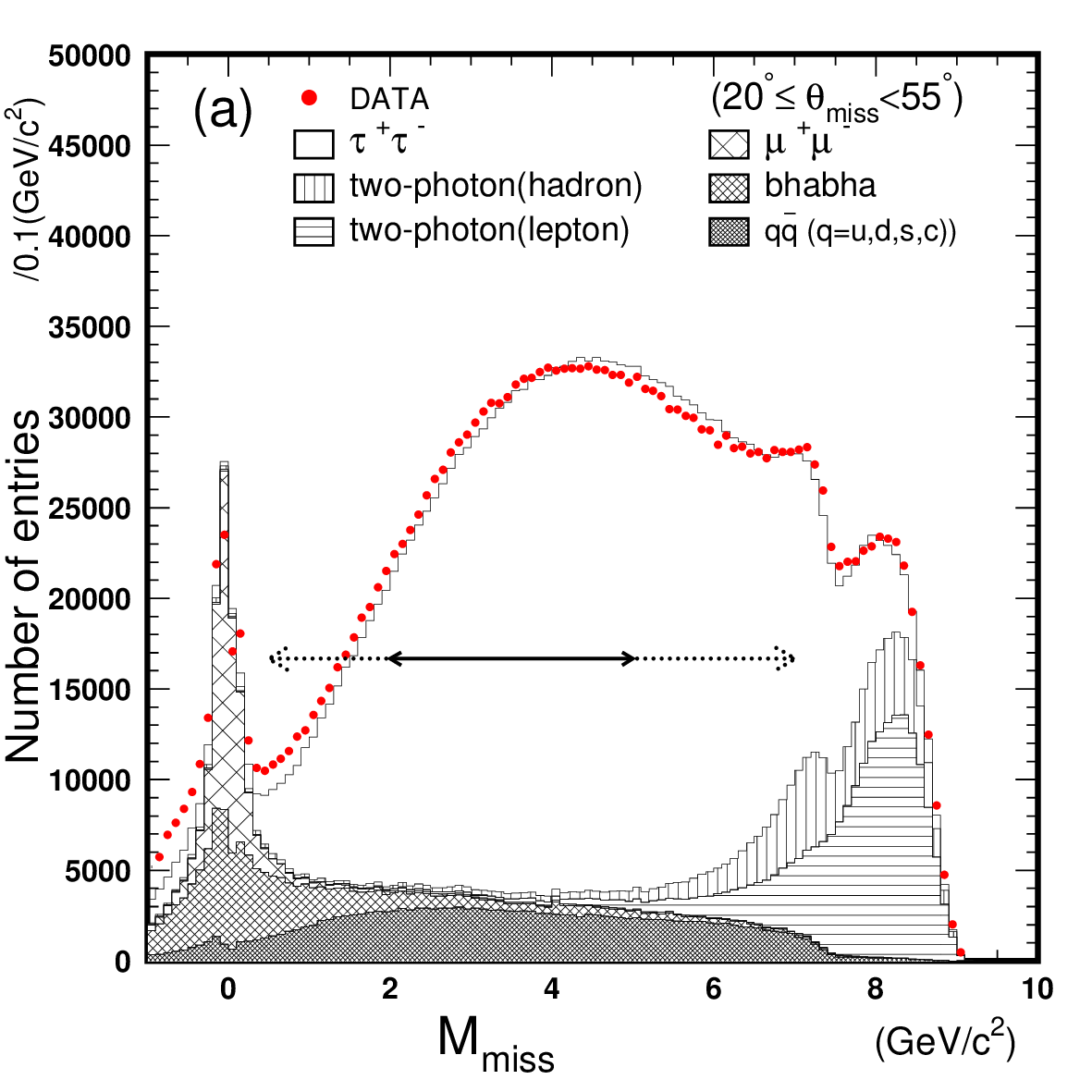}
\includegraphics[width=0.39\textwidth,clip]{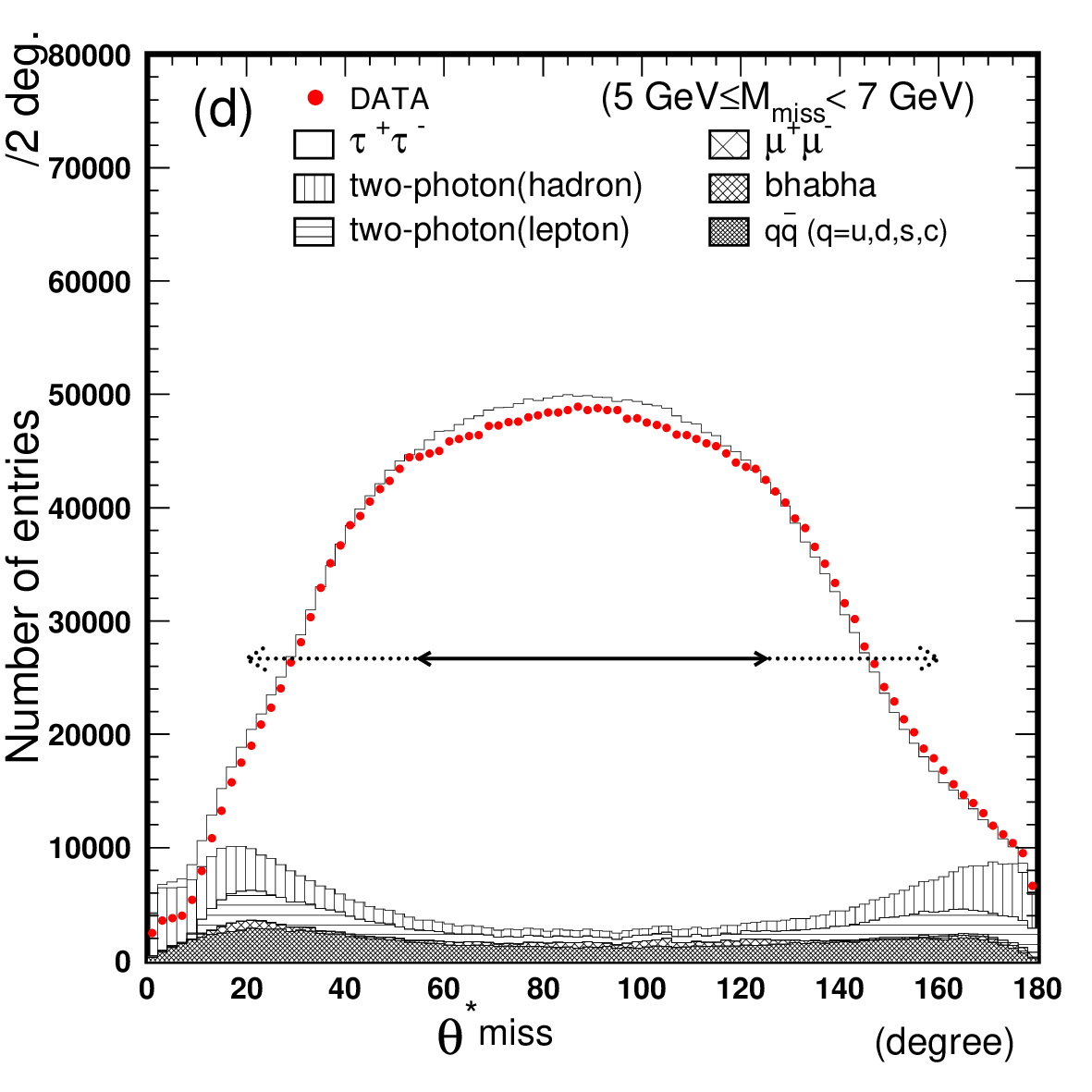}

\includegraphics[width=0.39\textwidth,clip]{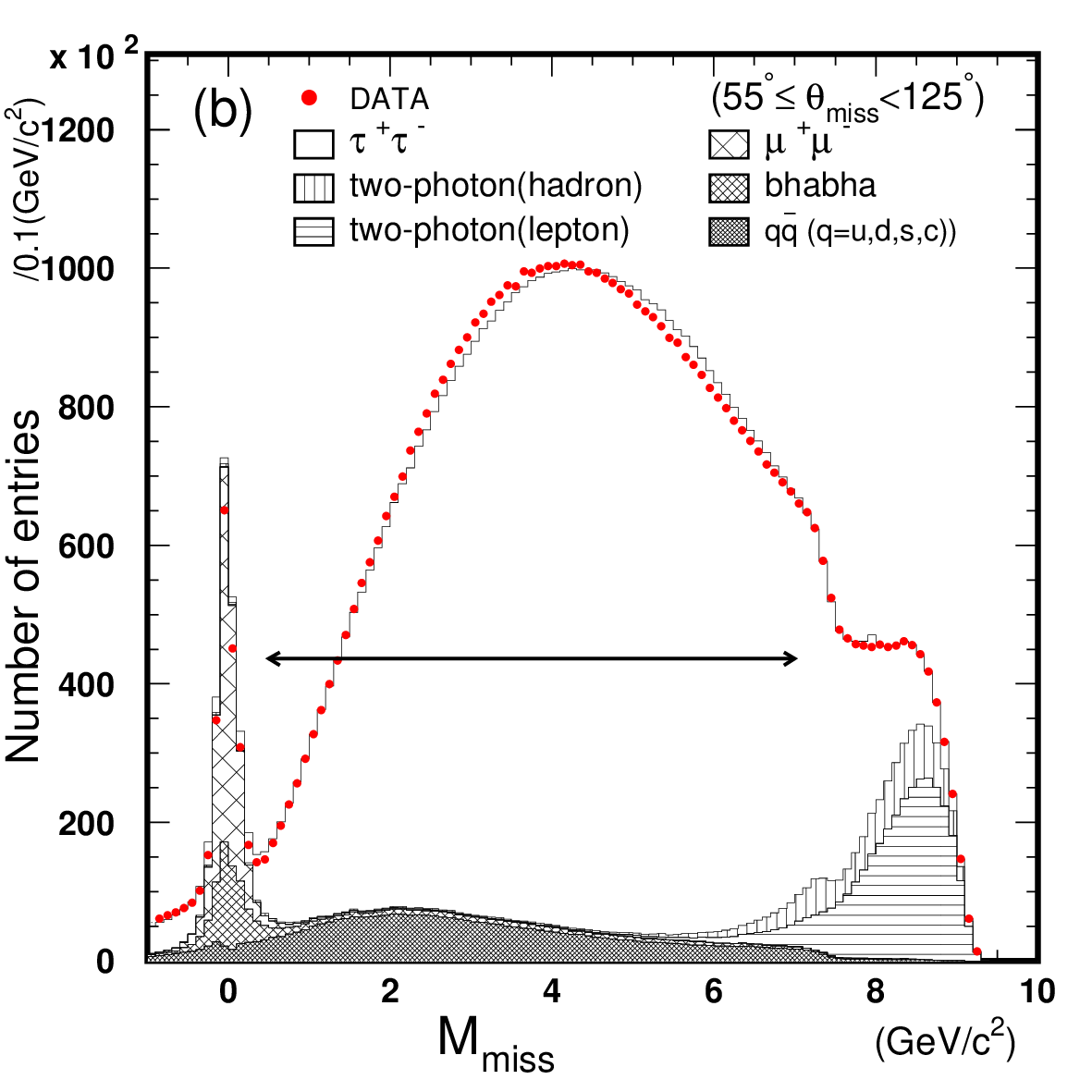}
\includegraphics[width=0.39\textwidth,clip]{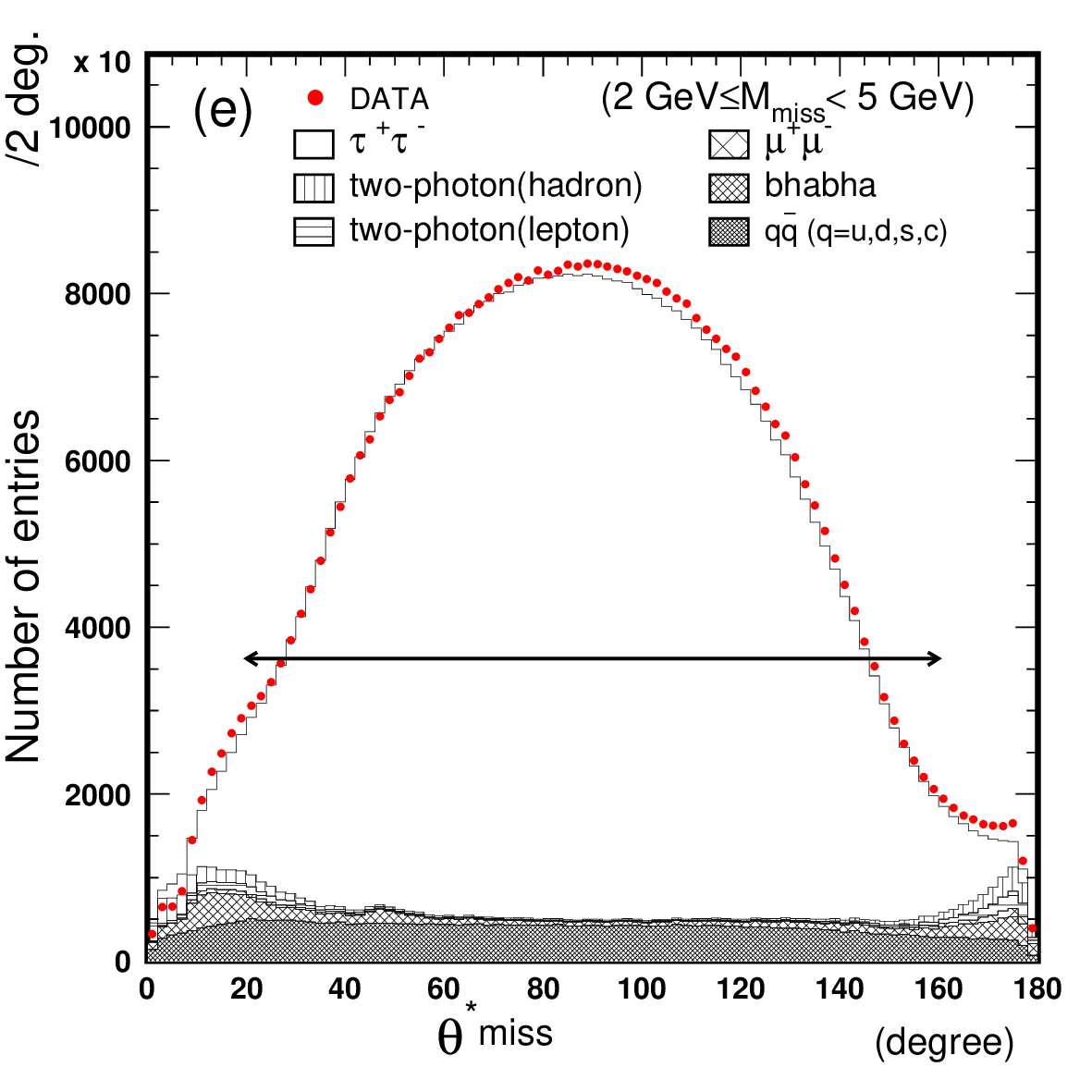}

\includegraphics[width=0.39\textwidth,clip]{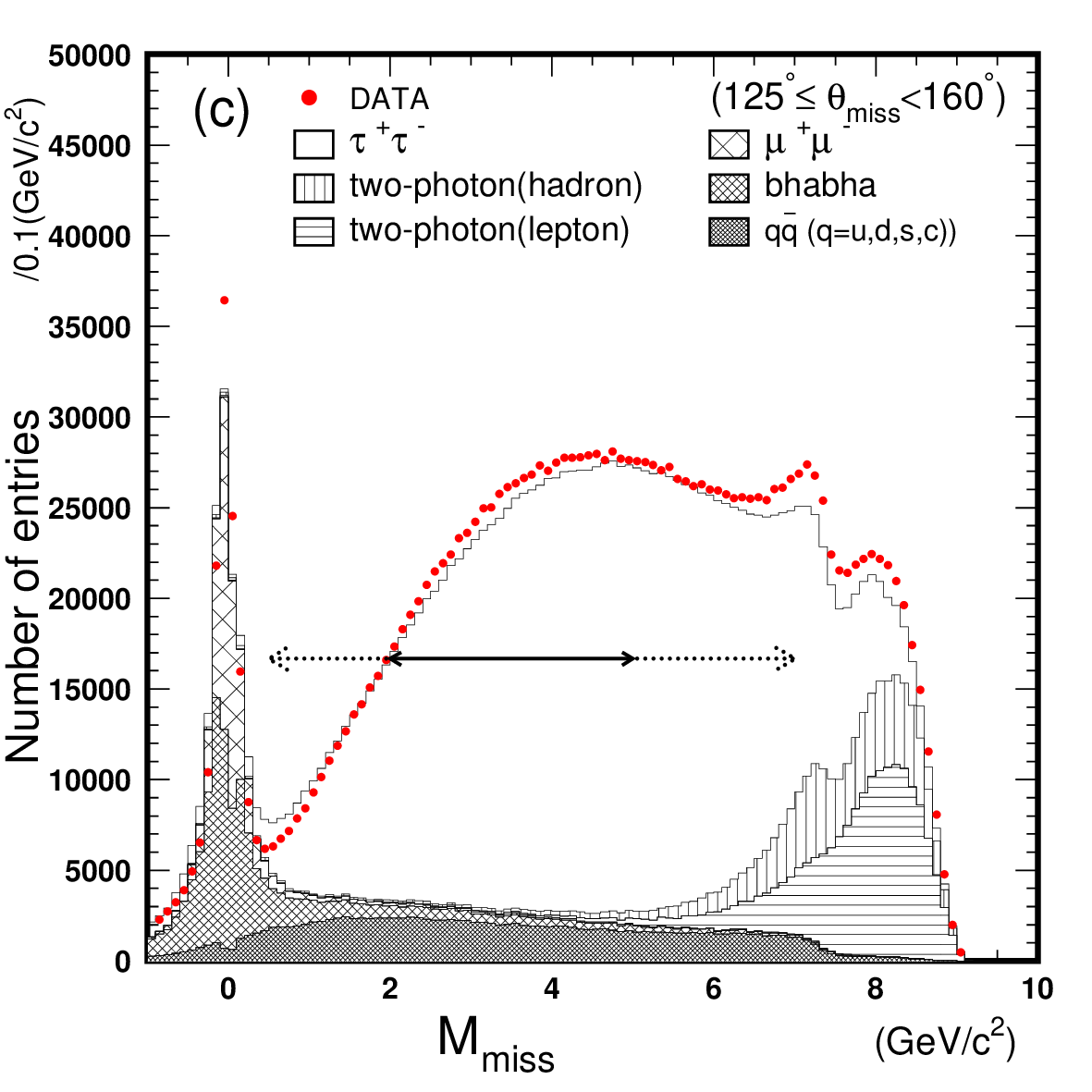}
\includegraphics[width=0.39\textwidth,clip]{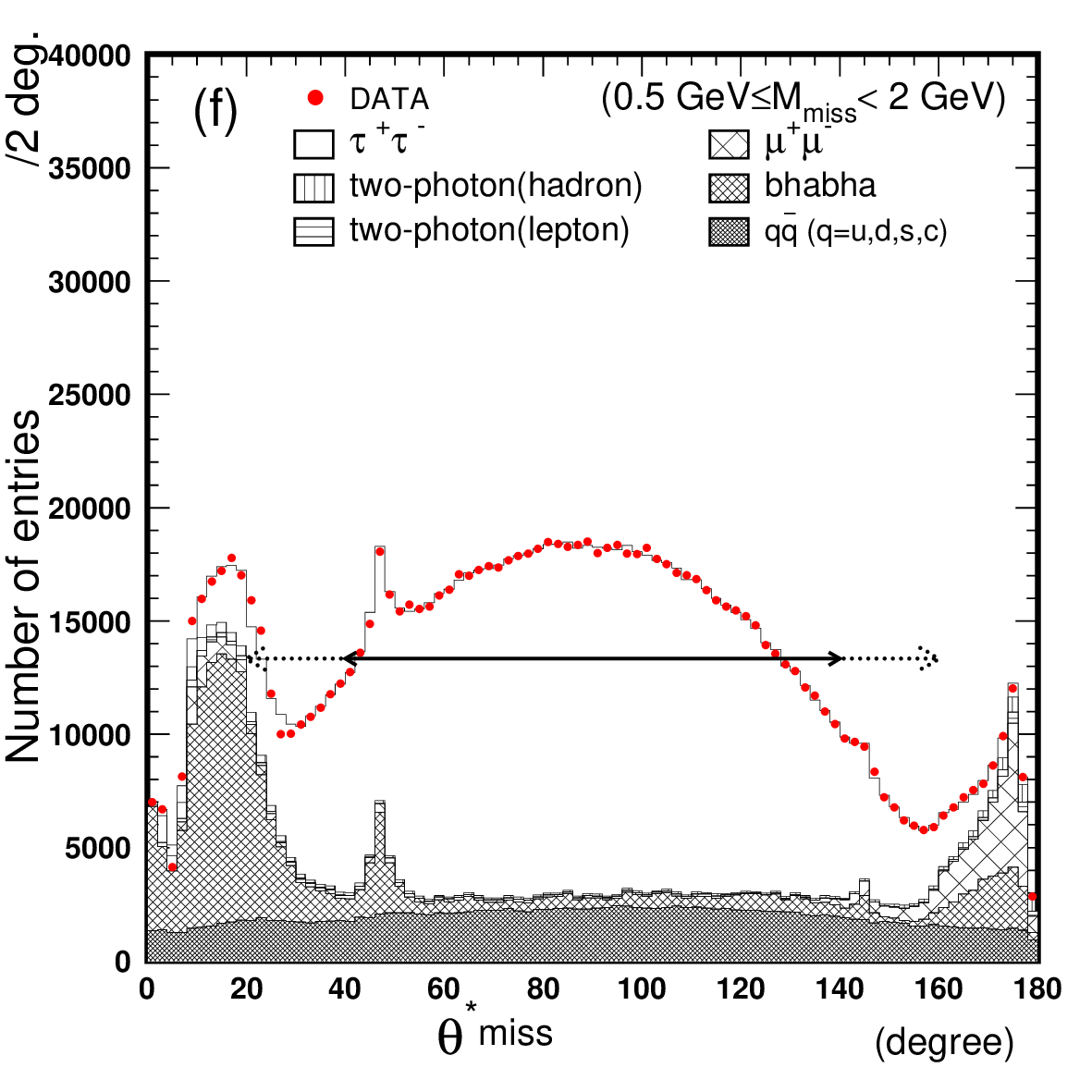}
\caption
{
Projections to the missing mass ($M_{\rm miss}$) and the missing direction ($\theta^{*}_{\rm miss}$): 
 (a)-(c) correspond to the vertical slices from left to right in Fig.\ref{mm-region}.
 (d)-(f) correspond to the horizontal slices from top to bottom. The solid circles represent the data,
 and the histogram represents MC simulation (signal+ background).
 The open histogram shows the contribution from $\tau^+\tau^-$-pair process, 
 the vertical (horizontal) striped area shows that from two-photon leptonic (hadronic)
processes; the wide (narrow) hatched area shows that from Bhabha ($\mu^+\mu^-$) process;
and the shaded area shows that from the $q\bar{q}$ continuum processes. 
 The arrows with solid (dotted) lines indicates the widest (narrowest) region
 corresponding to the octagonal boundary shown in Fig~\ref{mm-region}.
}
\label{mm-sub}
\end{figure}

\begin{figure}[!ht]
\begin{center}
\includegraphics[width=0.40\textwidth,clip]{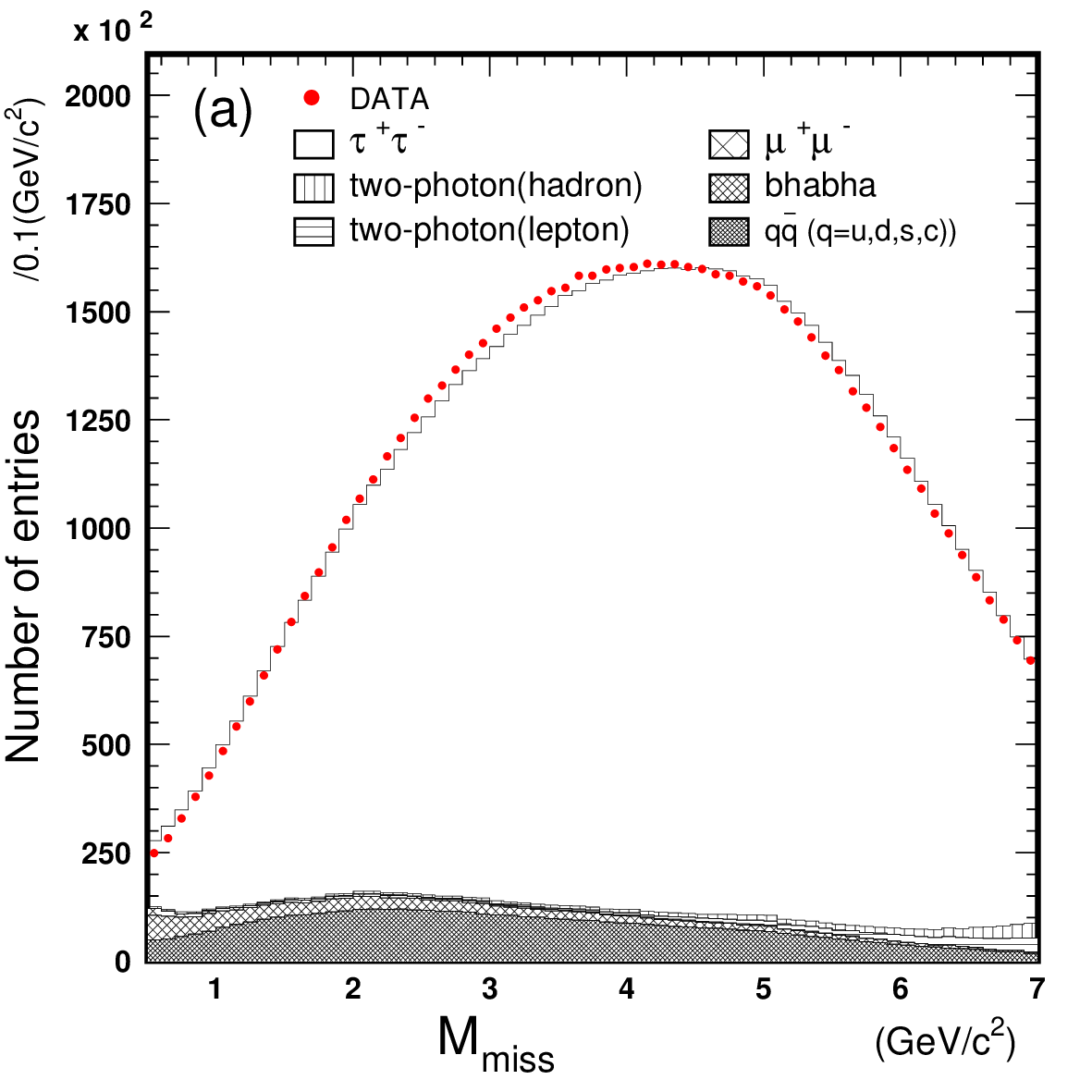}
\includegraphics[width=0.40\textwidth,clip]{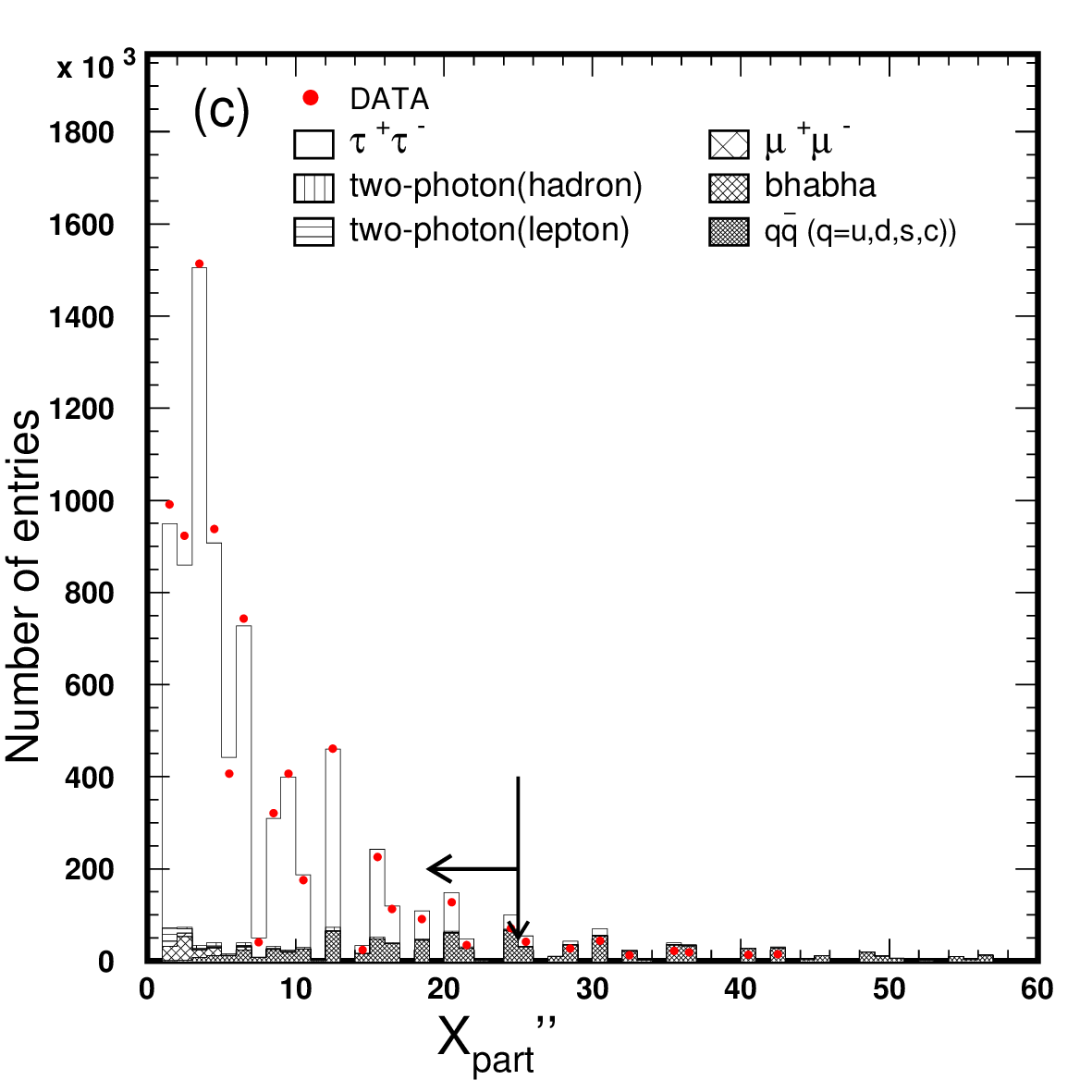}
\includegraphics[width=0.40\textwidth,clip]{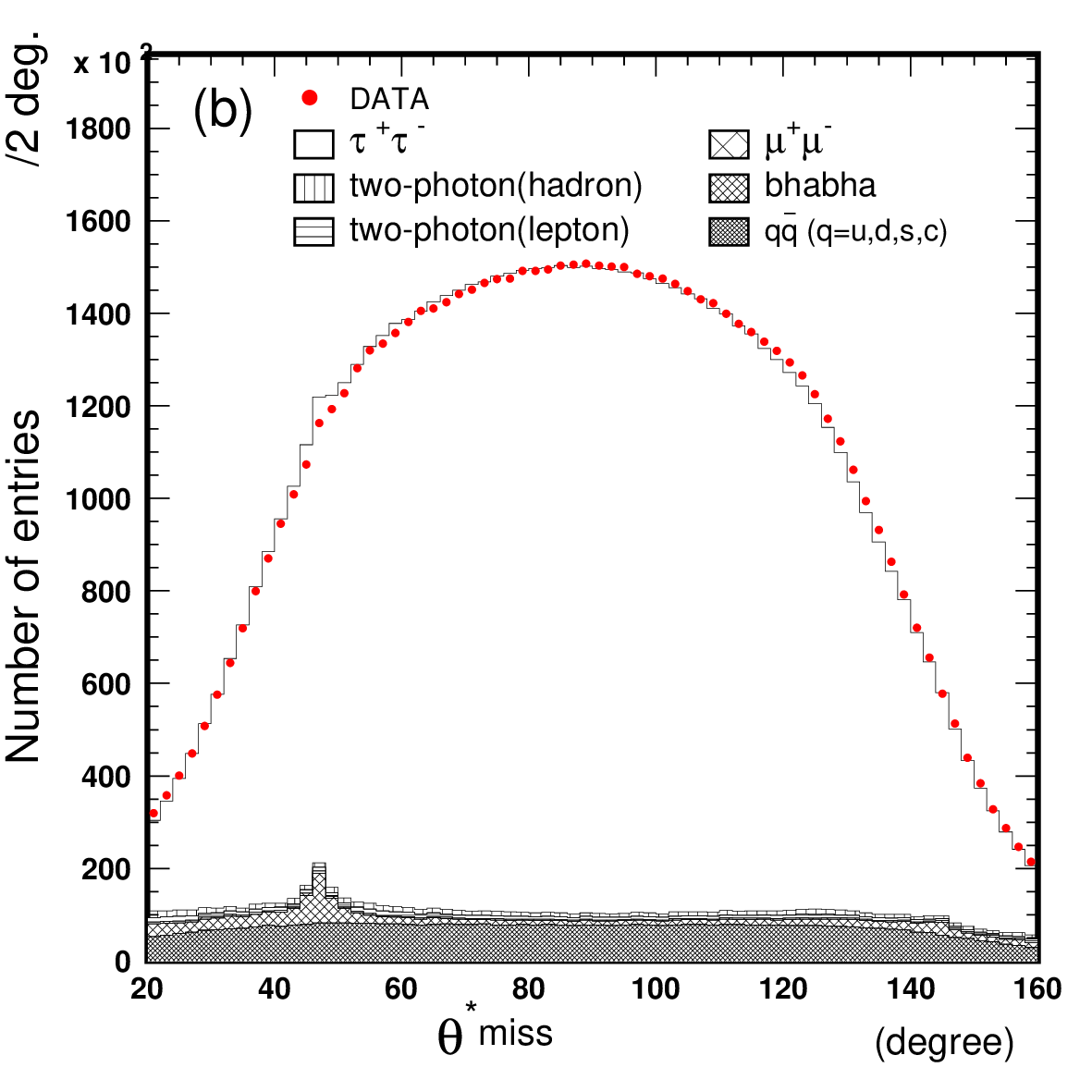}
\includegraphics[width=0.40\textwidth,clip]{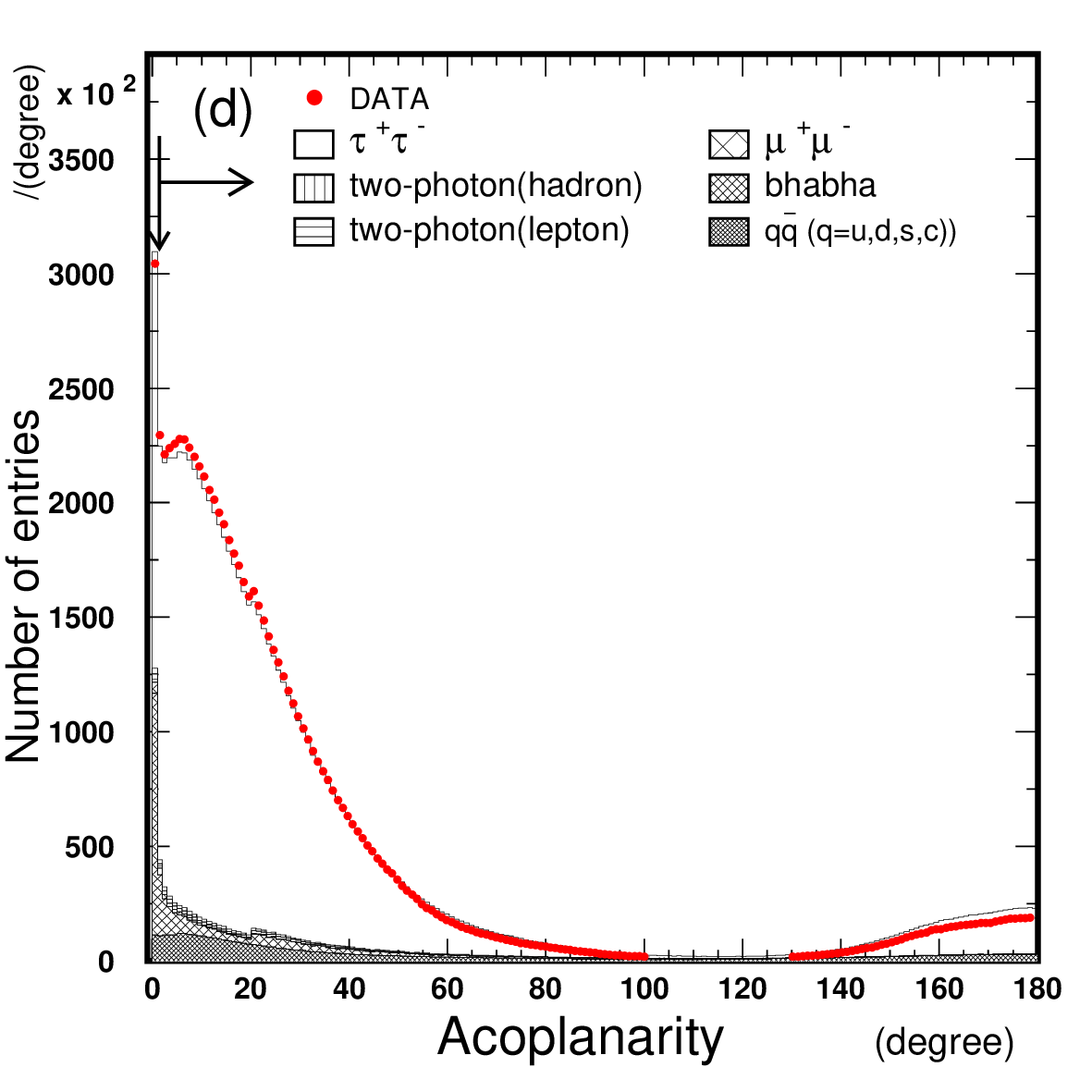}
\caption{
  Characteristic distributions  for surviving $\tau^+\tau^-$ candidates:
 (a) $M_{\rm miss}$, (b) $\theta_{\rm miss}$, (c) particle multiplicity $X_{\rm part} \equiv 
  (n_{\rm tr} + n_{\gamma})_{1} \times 
  (n_{\rm tr} + n_{\gamma})_{2}$,
(d) acoplanarity angle $\xi$.   
The points indicate the data, the open histogram shows 
the $\tau$-pair MC and the hatched histogram shows the background
 from $e^+e^-\to q\bar{q}$ and other sources. 
 All selection criteria are applied for (a) and (b). All criteria except for the quantity 
in  question are applied for (c) and (d).
 The arrows in (c) and (d) indicate the boundary used to select a $\tau$-pair sample.
}
\label{taup}
\end{center}
\end{figure}

To display the $\tau$-pair and background contribution quantitatively, 
 we divide the scatter plots of 
$M_{\rm miss}$ vs \ $\theta_{\rm miss}$ into three vertical and three 
horizontal slices as shown
in Fig.\ref{mm-region}. 
Projections for the six slices are shown in Fig.~\ref{mm-sub}, where 
each process shows a characteristic shape:
the $\tau^+\tau^-$ candidates dominate in the central region in
$M_{\rm miss}$ and $\theta^{*}_{\rm miss}$. Both Bhabha and $\mu^+\mu^-$ show
a prominent peak at $M_{\rm miss}\approx 0$, but the width for the Bhabha is slightly
wider than that of $\mu^+\mu^-$. We use the events in the region $|M_{\rm miss}|< 0.5~\GeVCC$ to
determine the normalization for the Bhabha and $\mu^+\mu^-$.
As two-photon processes dominate in the high-$M_{\rm miss}$ region,
the normalization for the two-photon processes is determined using the events
at $|M_{\rm miss}|>8.0~\GeVCC$.
The arrows with solid (dotted) lines indicate the narrowest (widest) areas used
to select $\tau^+\tau^-$-pairs by the  octagonal selection. 
Although overall features of the data are modeled reasonably well by MC,
some discrepancies are seen, for example, in the regions $M_{\rm miss} \approx 0$ and $> 6$ 
$\GeVCC$ in Fig~\ref{taup}-(c), which are taken into account as the systematic error on
the background estimation.

Candidate events are divided into two hemispheres in the 
CM frame by the plane perpendicular  to 
the highest momentum particle, and the 
remaining background from $e^{+}e^{-}$ annihilation is suppressed by selecting
 events with low multiplicity as characterized by the quantity
$X_{\rm part} \equiv 
  (n_{\rm tr} + n_{\gamma})_{1} \times 
  (n_{\rm tr} + n_{\gamma})_{2}$,
where $n_{{\rm tr},j}$ and  $n_{\gamma,j}$ are the numbers 
of tracks and photons in hemisphere~$j$. 
We require
$X_{\rm part}\leq 25$. 
Finally, in order to eliminate
Bhabha events in which 
one or both electrons  produce a shower in 
material near the interaction region,
the acoplanarity angle $\xi$ between
the first and second highest momentum tracks is required 
to be $\xi>1^{\circ}$, where  
$\xi\equiv||\phi_{1} -\phi_{2}|- \pi|$ 
is defined as the two-track acollinearity in azimuth.
The $X_{\rm part}$ and $\xi$ distributions after applying all selection criteria
except for the quantity in question are shown in  Figs.~\ref{taup}-(c) and (d), respectively.
The selection boundary is shown by the arrows.

After applying all selection criteria,
$22.83\times 10^{6}$  $\tau^+\tau^-$-pairs survive.
The $M_{\rm miss}$ and $\theta_{\rm miss}$ distributions for the surviving events,  
shown in Fig.\ref{taup}-(a) and (b), respectively, demonstrate a low level of the background
and an overall good agreement between 
the data and the MC model.

For surviving events, the dominant background is from the
$e^+e^-\rightarrow q\bar{q}\,\ (q=u,d,s,c)$ continuum and amounts to~$(5.30\pm 0.53)$\% of the
 total number of events. 
The systematic error for the $q\bar{q}$ background is determined from the uncertainty of the
normalization of the events in the region $25<X_{\rm part}<30$, where $q\bar{q}$ processes dominate.
The background from $e^+e^-\rightarrow \Upsilon(4S)\rightarrow B\bar{B}$ is small
(0.1 \%). 
 Backgrounds from Bhabha, $\mu^+\mu^-$, two-photon leptonic and hadronic 
events are  to be 
$0.92\pm 0.09$ \%, $0.28\pm 0.01$\%, $0.62\pm 0.03 $\% and $0.60\pm 0.09 $\%, respectively.
 Here the systematic errors for each background is determined
from the
  uncertainty of the normalization of the events in  the background enhanced
 region mentioned before.

\begin{figure}[t]
\begin{center}
\rotatebox{0}{\includegraphics[width=0.57\textwidth,clip]
{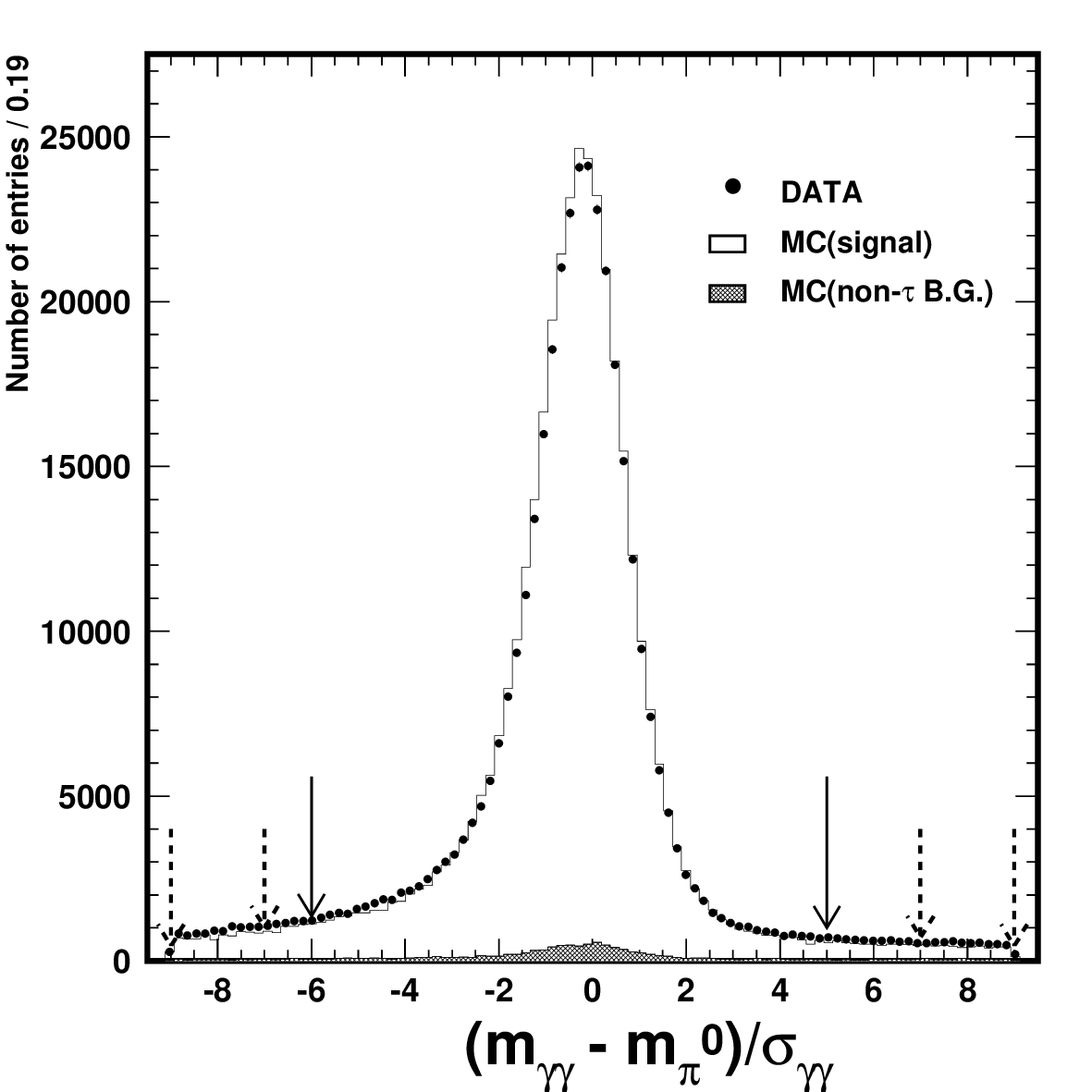}}
%
\caption{
 Normalized $\gamma\gamma$ invariant mass ($S_{\gamma \gamma}$)
spectrum for data (points) and the $\Tauhpi0$ signal MC (open histogram),
for the sample described in the text.
The data plotted here correspond to 6\% of the full data sample used in this analysis.
The arrows indicate the signal region
$-6 < S_{\gamma \gamma} < 5$ and the sideband regions  
$7 < \left|S_{\gamma \gamma}\right| < 9$. The sideband regions 
are used to subtract fake-$\pi^{0}$ background.
The shaded histogram shows the non-$\tau$ background determined from MC simulation.
}
\label{mresol}
\end{center}
\end{figure}

\subsection{$\Taupipi0$ selection}

Within  the $\tau^+\tau^-$-pair sample, $\Taupipi0$ decays are reconstructed 
by requiring that there be both one charged track
and one ${\pi ^{0}}$ in a single hemisphere. 
The $\pi^{0}$ candidate is selected based on the normalized
invariant mass
$S_{\gamma\gamma}\equiv 
(m_{\gamma\gamma} - m^{}_{\pi^0})/\sigma_{\gamma\gamma}$,
where 
$\sigma_{\gamma\gamma}$ is the mass resolution of the $\gamma\gamma$ system.
The value of $\sigma_{\gamma\gamma}$ ranges from 0.005 $\GeVCC$ to 0.008 $\GeVCC$,
depending on the $\pi^{0}$ momentum and polar angle.
Pairs of photons with $|S_{\gamma\gamma}|<9$ are considered as $\pi^{0}$
candidates.
To keep beam-related background at a negligible level, 
we require that the CM momentum of the ${\pi^0}$ be greater than 
0.25~GeV/$c$ and the photon CM energy be greater than 0.08~GeV.

The distribution of  $S_{\gamma\gamma}$
for the selected $\pi^{-}\pi^{0}$ sample, with one charged 
track and one $\pi^{0}$ candidate in a single hemisphere, 
is shown in Fig.~\ref{mresol}.
The lower-side tail of the $S_{\gamma\gamma}$ distribution is 
primarily due to
rear and transverse leakage of electromagnetic showers out of the 
CsI(Tl) crystals and the conversion 
of photons in the material located in front of  the crystals. 
Good agreement between data (points)
 and MC (open histogram) indicates that
these effects are properly modeled by the MC simulation. 
We define the interval ${-6 < S_{\gamma \gamma} < 5}$ as the
$\pi^{0}$ signal region.
Spurious $\pi^{0}$ background 
is small and estimated from the sideband regions
${7 < \left|S_{\gamma \gamma}\right| < 9}$.
To reduce feed-down background from multi-$\pi^{0}$
decays such as 
$\tau^-\rightarrow\pi^- (n\pi^{0})\nu_{\tau}$
( $n\ge 2$),
signal candidates are rejected 
if there are additional $\gamma$'s 
in the same hemisphere with energy greater than 0.2~GeV.

%
\begin{figure}[t]
\begin{center}
\rotatebox{0}{\includegraphics[width=0.57\textwidth,clip]
%
 {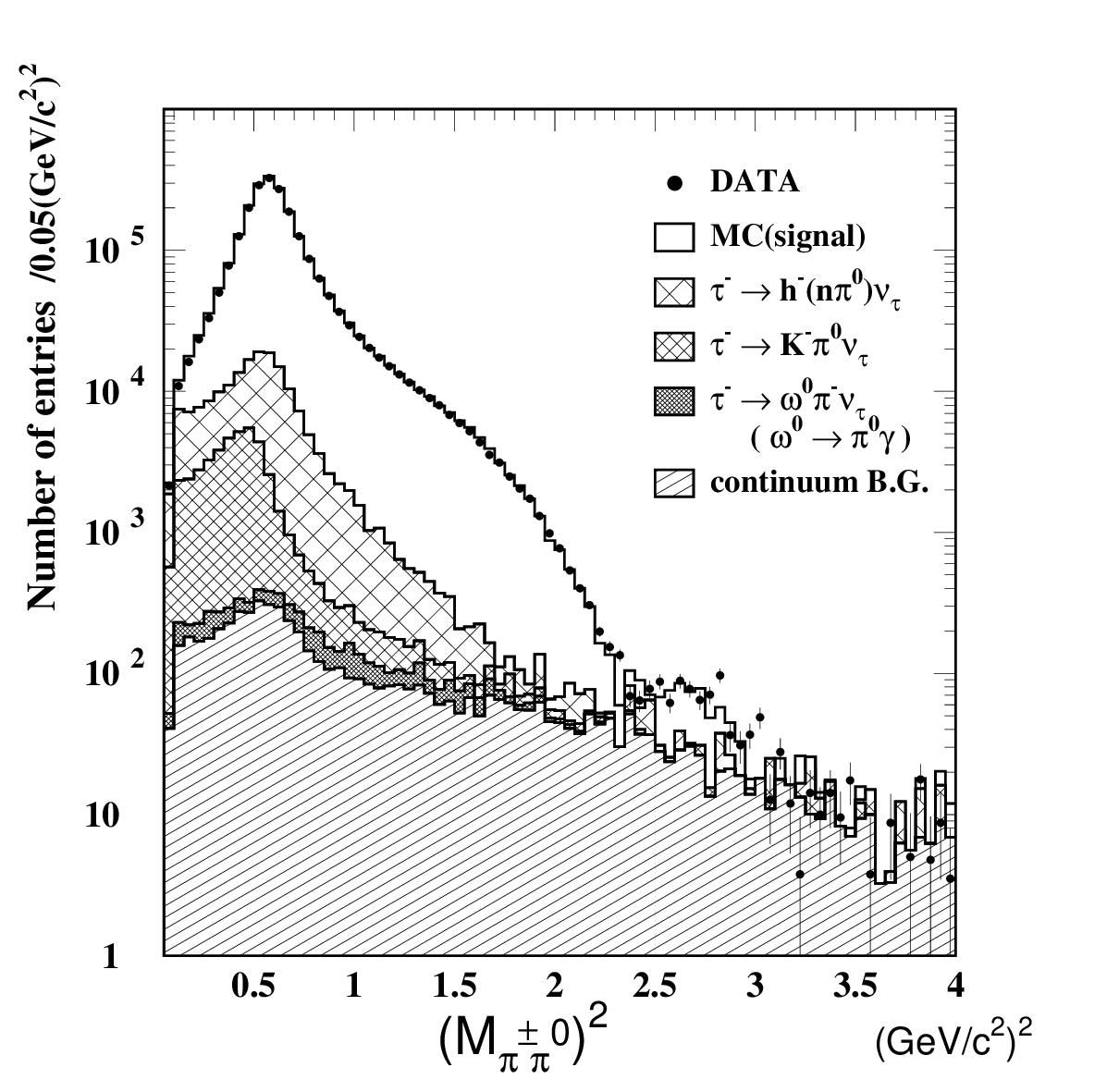}}
\caption
{
Invariant-mass-squared ($\MassSQ$) distribution for $\Taupipi0$
after imposing tight tag-side requirements.
The solid circles with error bars 
represent the data, and
the histogram represents the MC simulation (signal\,+\,background).
The open histogram shows the contribution from $\Taupipi0$; 
the narrow cross-hatched 
area shows that from ${\tau^{-}\ra K^{-} \pi^{0} \nu_{\tau}}$; 
the wide cross-hatched 
area shows that from ${\tau^{-}\ra h^{-}(n\pi^{0})\nu_{\tau}}$;
and the striped area
shows that from the $q\bar{q}$ continuum 
and other non-$\tau$ processes.
}
\label{pipi0_log}
\end{center}
\end{figure}

The $\pi^{-}\pi^{0}$ invariant-mass-squared ($\MassSQ$) spectrum
is obtained assuming the pion mass for the charged track; it is shown
in Fig.~\ref{pipi0_log} along with the MC prediction. To improve the
$\pi^{0}$ energy resolution, a $\pi^{0}$ mass constraint is imposed.
%
%
The spurious $\pi^{0}$ background level 
depends on the $\MassSQ$ region, varying from 4\% to 7\%.
(This is subtracted using $S_{\gamma\gamma}$ sidebands.) 
The final sample contains $5.43\times 10^{6}$ $\Tauhpi0$ candidates
after the ${\pi^{0}}$ background subtraction, where $h^{-}$ denotes
$\pi^{-}$ or $K^{-}$. This sample is 50 times larger than those of previous 
studies.

The spectrum is dominated by the $\rho(770)$ peak and a shoulder due to
the $\rho'(1450)$. A small but clear structure from 
the $\rho''(1700)$ is visible at $\MassSQ \sim 2.7\  \GeVcc2$.

There are two sources of background: feed-down from other
$\tau$ decay modes and the $q\bar{q}$-continuum.
Feed-down background arises mainly from multi-$\pi^{0}$ modes 
such as $\tau^{-}\rightarrow h^{-}(n\pi^{0})\nu_{\tau}$ ($6.02\pm0.08\%$),
$\tau\rightarrow K_{L}h^{-}\pi^{0}\nu_{\tau}$($0.48\pm 0.04\%$) and
$\tau\rightarrow \omega \pi^{-}\nu_{\tau}  (\omega\rightarrow\pi^{0}\gamma)$
($0.10\pm 0.01\%$).
 Here $h^{-}$ denotes either $\pi^-$ or $K^{-}$.
After all modes are included, the total feed-down background level is 
$(7.02\pm 0.08)$\%.
The error given here includes a MC statistical uncertainty 
as well as the uncertainty on relevant branching fractions.
The contribution of these feed-down backgrounds 
dominates at low values of  $\MassSQ$ (Fig.~\ref{pipi0_log}).

The $q\bar{q}$-continuum background level is 
$(2.22\pm 0.05)\%$ in total, 
and is concentrated mostly in the high $\MassSQ$ 
region above 2.0 $\GeVcc2$. 
Since the reduction of this high-mass background is essential 
in the measurement of the mass spectrum,
we impose the stringent requirement that 
the tag side contain only one charged track and no photons.
 This requirement improves the signal-to-noise ratio
in the high-mass region $\MassSQ \ge 2.0$ $\GeVcc2$ by a factor of 3,
although  the total
size of the $\Taupipi0$ sample is reduced by a factor of 2.5. 
The normalization of the continuum MC is validated using data in 
the mass region above the $\tau$ lepton mass: $\MassSQ > m^{2}_{\tau}$.
Background from the other non-$\tau$ processes, 
such as  $B \bar{B}$, Bhabha and $\mu^+\mu^-\gamma$ 
in the final sample
is negligible ($<0.1\% $).

\section{Measurement of the Branching Fraction}
\subsection{Basic Method }

The branching fraction for $\Tauhpi0$ 
($\mathcal{B}_{h\pi^0}$) is determined 
by dividing the signal yield $N_{h\pi^{0}}$ by
 the total number of
selected $\tau$ leptons $2 N_{\tau\tau}$ taking into account various efficiencies and
background corrections:  
\begin{eqnarray}
\mathcal{B}_{h\pi^{0}}   &=& \frac{N_{h\pi^{0}}}{2 N_{\tau\tau}}\times
	 \frac{ (1 - b^{{\rm feed}\mbox{-}{\rm down}}-
	   b^{{\rm non}\mbox{-}\tau})}
	      { (1 - b_{\tau\tau})}  \times
  \left(       \frac{\epsilon_{\tau\tau}}
	     {\epsilon^{\tau}_{h\pi^{0}}}
   \right)
   \times \frac{1}{\epsilon^{ID}_{h\pi^{0}} }\,.
\label{br}
\end{eqnarray}
\noindent
In this formula, $b_{\tau\tau}$ is the background fraction 
in the $\tau^+\tau^-$ sample,
$\epsilon_{\tau\tau}$ is the efficiency of the $\tau^+\tau^-$-pair selection,
$\epsilon^{\tau}_{h\pi^{0}}$ is the efficiency for
$\tau^-\!\rightarrow\!h^-\pi^{0}\nu$ decays to pass 
the $\tau^+\tau^-$-pair selection,
and $\epsilon^{ID}_{h\pi^{0}}$ is the efficiency for
$\tau^-\!\rightarrow\!h^-\pi^{0}\nu$ decays
satisfying the $\tau^+\tau^-$-pair selection 
to pass the $h^-\pi^{0}$ selection.
The product $\epsilon^{\tau}_{h\pi^{0}}\cdot \epsilon^{ID}_{h\pi^{0}}$ 
is the overall detection efficiency for the $h^-\pi^{0}\nu$ final state.
The parameter $b^{{\rm feed}\mbox{-}{\rm down}}$ is the fraction of 
$h^{-}\pi^0\nu$ candidates coming from other $\tau$ decay modes, and 
$b^{{\rm non}\mbox{-}\tau}$ is the fraction coming from non-$\tau$ 
processes. In this formula, several common uncertainties such as 
that on the luminosity, on the cross section for 
$\tau^+\tau^-$-pair production, on the trigger efficiency, and 
on the $\tau^+\tau^-$ selection efficiency cancel in the ratio.
In the measurement of the branching fraction, 
the stringent tag-side condition is not imposed to avoid any possible bias  that it might introduce.
The values for all factors are  listed in Table~\ref{tab:br} along
with the MC statistical error.

\begin{table}[!htb]
\renewcommand{\arraystretch}{1.4} 
\begin{center}   
\caption{ Values of parameters used for the 
branching fraction measurement along with MC statistical errors.}
\label{tab:br}                   
\begin{tabular}{l|c}  \hline
\hline
Parameter &\hspace{1.5cm} Value \hspace{1.5cm}   \\
\hline 
\hline    
${\varepsilon_{\tau\tau}}$  &  
 ${32.59 \pm 0.05~\%}$  \\ 
${\varepsilon^{\tau}_{h\pi^{0}}}$ &  ${36.24\pm 0.07 ~\%}$  \\ 
${\displaystyle f_{b} = 
	     \frac { \varepsilon^{\tau}_{h\pi^{0}} }
			 { \varepsilon_{\tau\tau} }   }$ &
	      ${1.112 \pm 0.003 }$  \\ 
${\varepsilon_{h\pi^{0}}^{ID}}$ &
	      ${41.01 \pm 0.13 ~\%}$ \\ 
${b_{\tau \tau}}$ &
	      ${7.80 \pm 0.03~\%}$  \\ 
${ b^{{\rm feed}\mbox{-}{\rm down}}_{h\pi^{0}} }$ &
	      ${7.02 \pm 0.08 ~\% }$  \\
${ b^{{\rm non}\mbox{-}\tau}_{h\pi^{0}} }$  &
	      ${2.22 \pm 0.05 ~\% }$  \\
\hline
\hline
\end{tabular}
\end{center}
\end{table}

\subsection{Systematic uncertainties}

The sources of systematic uncertainties on $\mathcal{B}_{h\pi^{0}}$ are listed in Table~\ref{tab:br_sys}. 
The uncertainty on the tracking efficiency is estimated using  
$D^{*\,+}\rightarrow D^0\pi^+\rightarrow K^-\pi^+\pi^+$ 
decays to be 1\% per track.
A large part of this uncertainty cancels in the ratio  of
Eq.~(\ref{br}); the resulting relative uncertainty  from this source is
$\Delta \mathcal{B}/\mathcal{B}=0.47$ \%.

\begin{table}[htbp!]
\begin{center}
\caption{ Systematic uncertainties for the $\Tauhpi0$ branching fraction.
}
\label{tab:br_sys}
\begin{tabular}{l c c } 
\hline 
\hline
Source of  uncertainty & $\Delta \mathcal{B}_{h\pi^{0}}$ (\%) &  
$(\Delta \mathcal{B}/\mathcal{B})$ (\%)
 \\
\hline
\hline
Tracking efficiency & 0.12   & 0.47  \\
$\pi^0$ efficiency & 0.32  & 1.27 \\
Background for $\tau^+\tau^-$  &   0.15 & 0.59  \\
Feed-down background for $\Tauhpi0$   & 0.04  & 0.16 \\
Non-$\tau$ background for $\Tauhpi0$  & 0.05  & 0.20  \\
$\gamma$ veto  &  0.05      & 0.20 \\
Trigger            & 0.08  & 0.32  \\
MC statistics &   0.02  &   0.08 \\
\hline
\hline 
Total &  0.39   &  1.52 \\
\hline
\hline
\end{tabular}
\end{center}
\end{table}

The systematic error on the $\pi^0$ detection efficiency has two components: one
 is the uncertainty coming from the $\pi^{0}$ selection criteria and the other is that 
from the absolute efficiency calibration.
For the uncertainty coming from  the $\pi^{0}$  selection, 
we check the uncertainty  by changing the definition of the signal and background region, 
by taking into account the
 uncertainty in the resolution function and by changing the $\pi^{0}$ threshold momentum.
For example,
the relative branching fraction  changes by only 
$\Delta \mathcal{B}/\mathcal{B}= 0.1\% $ if the signal region is changed from the nominal one to $-7< S_{\gamma\gamma} < 7$. 
 Also the uncertainty is  
$\Delta \mathcal{B}/\mathcal{B} = \pm 0.2\% $ for the changes of the $\pi^{0}$ threshold
by $\pm 0.05$ GeV from the nominal value.

 In order to make an absolute efficiency calibration   
independently of the signal process, 
we use the $\eta\rightarrow \gamma\gamma$ and $\eta\rightarrow \pi^{0}\pi^{0}\pi^{0}$
signals, whose branching fractions are known rather precisely.
Combining the PDG world average~\cite{PDG2006} for the $\eta\rightarrow \gamma\gamma$ 
and $\eta\rightarrow \pi^{0}\pi^{0}\pi^{0}$
branching fractions and the recent measurement from the CLEO collaboration~\cite{CLEO2007},
we obtain the ratio of the branching fractions of
$$
       \frac{\mathcal{B}(\eta\rightarrow\gamma\gamma)}
{\mathcal{B}(\eta\rightarrow\pi^{0}\pi^{0}\pi^{0})}
= 0.829 \pm 0.007,
$$
which has 0.84\% relative accuracy.

By comparing the signal ratio 
$R_{i} \equiv N(\eta\rightarrow\pi^{0}\pi^{0}\pi^{0}/ N(\eta\rightarrow \gamma\gamma))$
for the data ($i=1$) and the MC ($i=2$), the correction factor for the detection efficiency 
of one $\pi^{0}$, $\eta_{\rm cor}$, is determined to be $\eta_{\rm cor} = \sqrt{ R_{\rm data}/R_{\rm MC}}
=0.950\pm 0.012$, where the error includes the uncertainties in the $\eta$ signal measurement $(1.2\%)$ and
 the errors on the $\eta$ decay branching fractions $(0.4\%)$.

 This correction factor is also confirmed by a study 
 of electron/positron tracks from photon conversions 
(i.e. $\gamma\ra e^{+}e^{-}$)  
in the SVD region.
It is found that the $E/P$ distribution for those tracks is simulated correctly 
 above 1.0 GeV, but requires some tuning  below 1.0 GeV.
This imperfection of the MC primarily leads to a difference in 
the $\pi^{0}$ signal shape and an efficiency difference between data and MC.

The non-$\tau$ background is dominated by  $q\bar{q}$ 
continuum processes; this is estimated by using the events 
 above the $\tau$ mass: $\MassSQ > m^{2}_{\tau}$.
The statistics of the data and MC sample determine the error.

The uncertainty on the feed-down
background $\Delta b_{h\pi^{0}}^{\rm feed-down}$
comes from the MC statistics  and 
 the uncertainty on the branching fractions for
$\tau^{-}\!\rightarrow\!h^{-}(n\pi^{0})\nu_{\tau}$,
$\tau^-\!\rightarrow\!K^{-}\pi^{0}\nu_{\tau}$ and
$\tau^{-}\!\rightarrow\ \omega \pi^{-}\nu_{\tau}\ (\omega\rightarrow\pi^{0}\gamma)$.

The veto of additional $\gamma$'s is required in the event
selection to reduce background from multi-$\pi^0$ decay channels. 
However, this veto can reject signal itself if
 photons are radiated in the initial or final state and those photons
are detected within the detector fiducial volume.
 In addition, photon candidates
can also appear due to electromagnetic shower fragments and/or 
misreconstructed electrons.
Therefore a precise simulation of the photon radiation as well
as the shower simulation are important.
The uncertainty from these sources is estimated by 
changing the veto threshold  by
$\pm 0.1$~GeV around the nominal value of $0.2$~GeV; the resulting relative change in 
$\mathcal{B}^{}_{h\pi^0}$ 
is only $\pm 0.20$\%.
Signal events are flagged by several trigger conditions
that require two or more CDC tracks with associated TOF hits, 
ECL clusters, or a significant sum of energy in the ECL. 
This redundancy 
allows one to monitor the efficiency of each trigger requirement.
The uncertainty arising from the trigger is estimated by
assuming that there is a $\pm 3$\% uncertainty on the track and 
energy trigger efficiencies, which is the maximum variation 
measured during experimental running. The resulting relative uncertainty 
is small (0.32\%) since the 
$\tau^+\tau^-$ trigger efficiency is high (97\%).

\subsection{Results}

Inserting all values into Eq.~(\ref{br}) we obtain
\begin{eqnarray}
\mathcal{B}_{h\pi^{0}} & = & (25.67\,\pm \,0.01\,\pm\, 0.39 )\%\,,  
\end{eqnarray}
\normalsize
where the first error is statistical and the second is systematic.
This result is in good agreement with previous measurements,
as shown in Table~\ref{tab:br_comp}. Our statistical error is 
significantly lower than those of the other measurements;
 our systematic error is similar to those of CLEO, L3 and OPAL, and
larger than those of ALEPH and DELPHI.

\begin{table}[htbp!]
\begin{center}
\caption{ 
Branching fractions for $\Tauhpi0$ measured 
by different experiments. 
}
\label{tab:br_comp}
\begin{tabular}{l|c |c} 
\hline 
\hline 
Experiment & $\mathcal{B}_{h\pi^{0}}(\%)$ & Reference \\ 
\hline
\hline
CLEO & $25.87 \pm 0.12 \pm 0.42$ & \cite{CLEO94} \\
L3 & $25.05 \pm 0.35 \pm 0.50$   &\cite{L395}       \\
OPAL & $25.89 \pm 0.17 \pm 0.29$  & \cite{OPAL98M}   \\
ALEPH & $25.924 \pm 0.097 \pm 0.085$ & \cite{ALEPH05}   \\
DELPHI & $25.740 \pm 0.201 \pm 0.138$ & \cite{DELPHI06} \\
\hline
This work & $25.67 \pm 0.01 \pm 0.39$       \\
\hline
\hline
\end{tabular}
\end{center}
\end{table}

%
%
We combine the PDG world average for the $\tau^{-}\!\rightarrow\! K^-\pi^0 \nu_{\tau}$ branching
fraction~\cite{PDG2006} with a recent BaBar measurement~\cite{BA2007} to obtain
the result $\mathcal{B}_{ K^-\pi^0}=(0.428\pm 0.015)\%$.
Subtracting this from our $\Tauhpi0$ result gives a
$\Taupipi0$ branching fraction of
\begin{eqnarray}
\mathcal{B}_{\pi\pi^{0}} & = & (25.24 \,\pm\,0.01\,\pm\, 0.39)\%\,,
\end{eqnarray}
which is consistent with the previous 
measurements from
CLEO~\cite{CLEO94} and
ALEPH~\cite{ALEPH05}.


\section{Measurement of the Mass Spectrum}

In order to obtain the true $\pi^{-}\pi^{0}$ mass spectrum,
one must apply corrections for: (1) background, (2) smearing due to finite 
resolution and radiative effects, and (3) mass-dependent acceptance.

\begin{figure}[ht!]
\begin{center}
\rotatebox{0}{\includegraphics[width=0.45\textwidth,clip]
{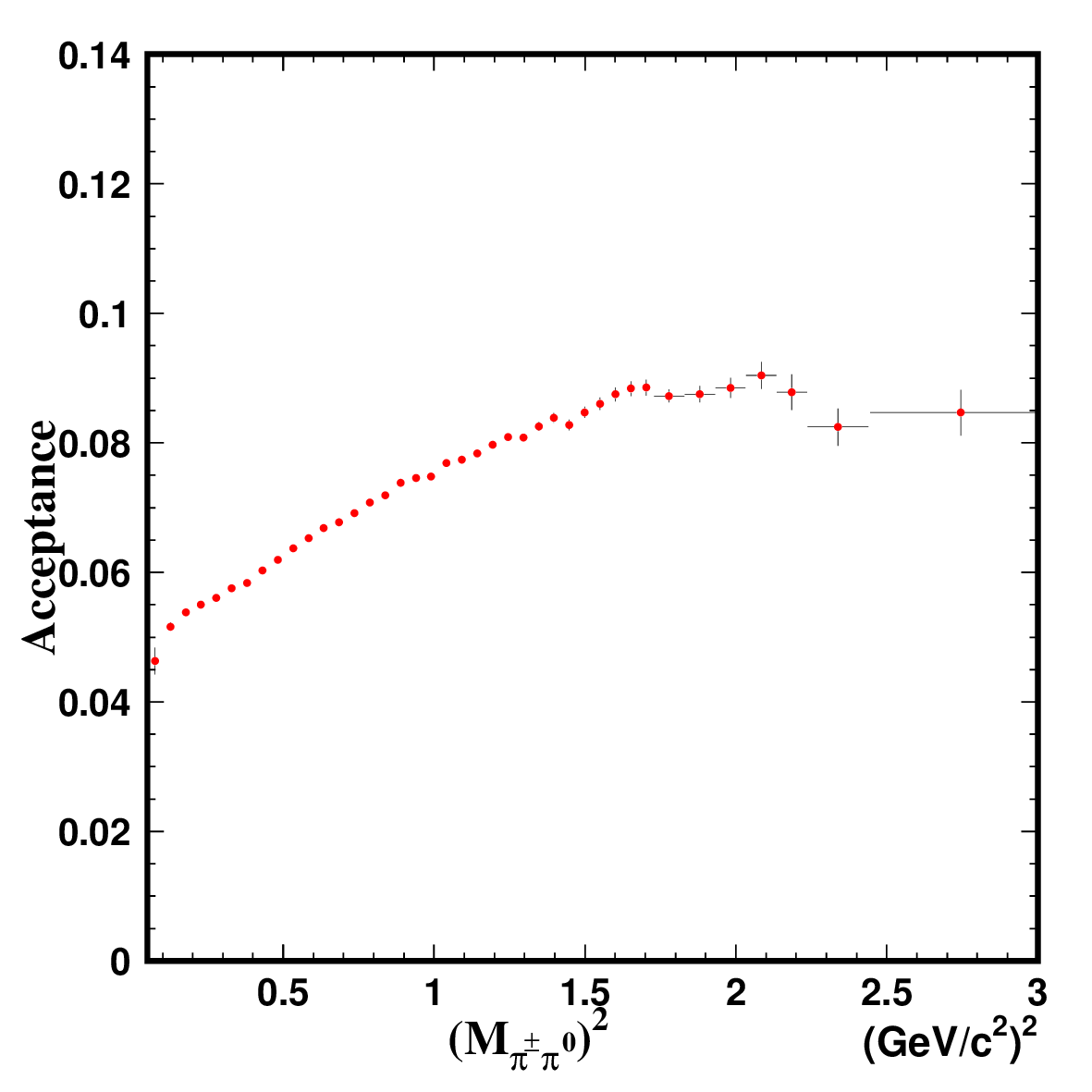}}
%
\caption
{ The acceptance determined from MC simulation 
 as a function of the $\pi^{-}\pi^{0}$ mass squared.
   }
\label{acceptance}
\end{center}
\end{figure}

\subsection{Background Correction}

As noted earlier,
there are three sources of the background that enter the $\Taupipi0$ sample:
(1) fake $\pi^{0}$ background, (2) feed-down background from other
$\tau$ decay channels, and (3) the background from the $q\bar{q}$ continuum.
The total magnitude of these background contributions is 
about 7\% in the $\rho(770))$ peak region,
 but  the fraction of the background varies strongly with $\MassSQ$;
there is
approximately a 4-order-of-magnitude difference between the signal level in
the $\rho(770)$ peak region and that in the high  $\MassSQ$ region 
above 2.5 $\GeVcc2$.
Thus a reliable estimation of the background is important for the measurement of the 
mass spectrum.

 The sidebands of the $M_{\gamma\gamma}$ distribution 
are used to estimate the fake $\pi^{0}$ contribution.
This background dominates at values of $\MassSQ$ less 
than about $0.25~\GeVcc2$.

In the $\pi^{-}\pi^{0}$ system, 
the feed-down background 
dominates  at  similarly low values of $\MassSQ$ 
 while the
$q\bar{q}$-continuum background dominates at high values of $\MassSQ$
 (see Fig.~\ref{pipi0_log}). 
These backgrounds are subtracted bin-by-bin.
%

\begin{figure}[ht!]
\begin{center}
\rotatebox{0}{\includegraphics[width=0.45\textwidth,clip]
 {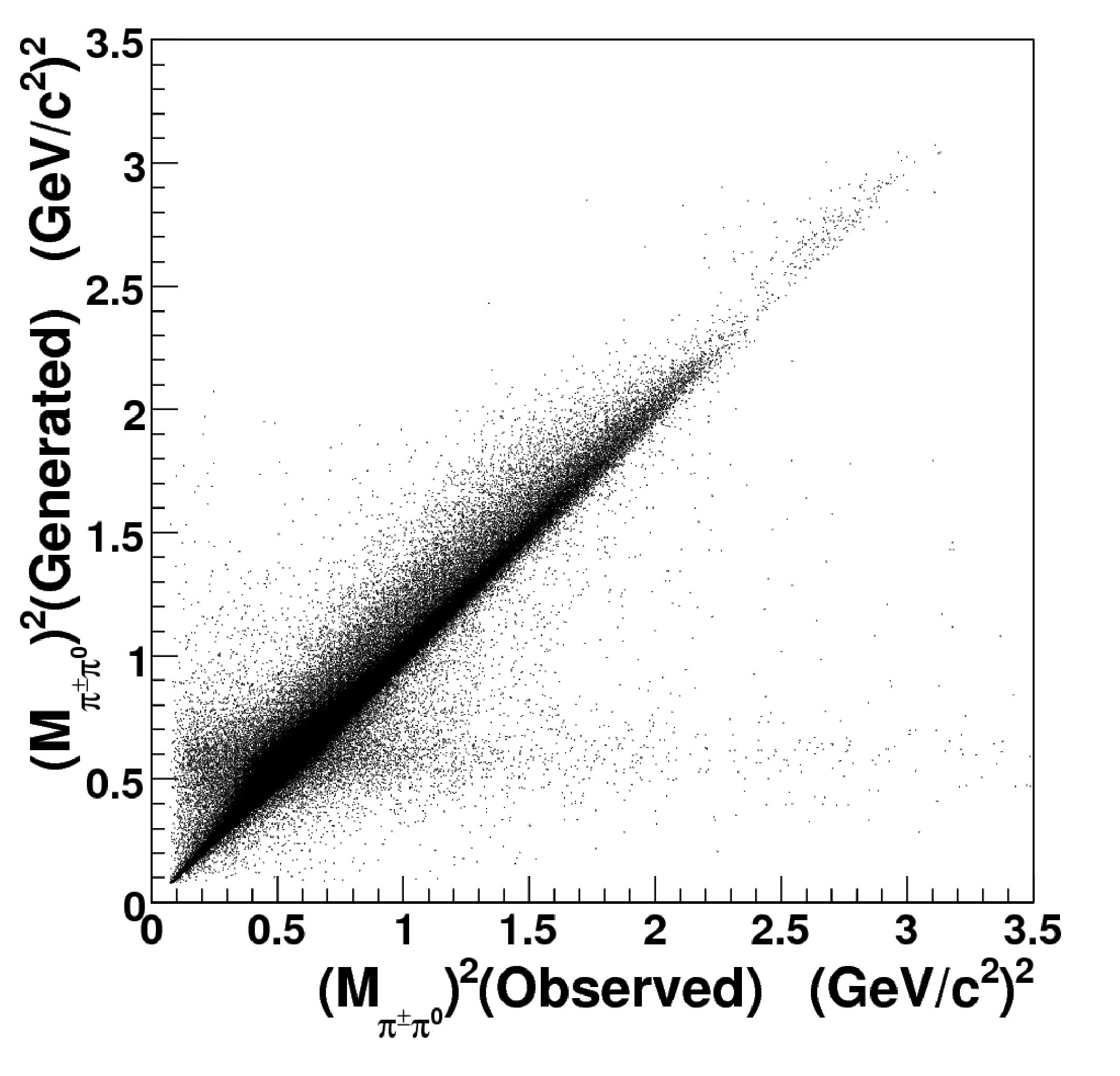}}
\caption
{
Correlation between  
the generated and
observed invariant masses squared of the $\pi^-\pi^{0}$ 
system in the $\Taupipi0$ decay.
}
\label{cor_matrix}
\end{center}
\end{figure}

\subsection{Acceptance  Correction}

The acceptance determined from MC simulation as a function of 
the generated $\pi^{-}\pi^{0}$ mass squared is shown in 
Fig.~\ref{acceptance}. 
The acceptance varies smoothly 
and its average value is ~7\%.
This acceptance includes a factor for the tag-side branching fractions
${\cal B}(\tau^{-} \rightarrow \ell^{-} \bar{\nu}_{\mu} \nu_{\tau})$ $(\ell^{-}=\mu^-,e^-)$,
${\cal B}(\tau^{-} \rightarrow h^{-}  \nu_{\tau})$ $(h^{-}=\pi^{-},K^{-})$,
which does not affect the shape of the mass spectrum.
The acceptance decreases at low values
of $\MassSQgen$ due to the overlap of $\gamma$ clusters
with the $\pi^{-}$ track in the calorimeter.

The detector effects include $\MassSQ$-dependent 
acceptance and bin-by-bin migration caused by the finite mass resolution. 
The radiative decay $\tau^{-}\rightarrow \pi^{-}\pi^{0}\gamma\nu_{\tau}$
also causes some bin migration.
We correct for these effects using an unfolding procedure that makes use of the
MC to characterize the acceptance and the bin migration. These effects can be characterized
by the acceptance matrix $ A $ defined by
$$
  {\bm b}  =  {  A} \, {\bm x}\,,  
$$ 
where ${\bm x}$ is the vector containing the generated 
$\pi^{-}\pi^{0} (\gamma) $ mass-squared spectrum
and ${\bm b}$ is the reconstructed one. 
It is possible to apply the inverse of ${ A}$ to
the spectrum observed in the data to obtain an unfolded spectrum. However, this procedure is not robust with respect to the statistical
fluctuations entering the determination of ${ A}$, 
and can yield
unphysically large point-by-point fluctuations. 
 To cure this problem, we use an unfolding program employed in the ALEPH experiment~\cite{SVD}.
In this program, the unfolding is based on the Singular-Value-Decomposition (SVD) method,
in which 
the acceptance matrix is inverted 
by  constraining the number of singular values to
only those elements that are statistically significant.

The acceptance matrix is determined iteratively using a  
signal MC based on the KKMC/TAUOLA
program. In the second iteration, the $\rho''(1700)$ resonance 
is included in the MC based on our measurement.
Final state radiation in $\tau$ hadronic decays is simulated by the PHOTOS program.
In order to take into account the effects of  $\gamma$ radiation 
in the decay
$\tau^{-}\rightarrow \pi^{-}\pi^{0}\gamma\nu_{\tau}$,
 the 
invariant mass squared of the $\pi^{-}\pi^{0}\gamma$ system is taken
as the generated quantity.
 
The output of the program is the unfolded distribution and 
its covariance matrix.
The correlation between the generated quantity 
 and the measured one 
is shown in Fig.~\ref{cor_matrix}. The figure 
shows a clear correlation between the measured and generated values. 
The resolution in $\MassSQ$ is 0.005~$\GeVcc2$
in the low-mass region and 
0.030~$\GeVcc2$ in the 
high-mass region; thus 
by choosing the bin size to be 
$\Delta M^{2} = 0.050~\GeVcc2$, the off-diagonal 
components of the acceptance matrix are small.

\subsection{ Systematic Uncertainties}

The sources of  systematic errors associated with the unfolded mass spectrum
$(1/N )({\rm d}n/{\rm d}s$) are subdivided into several classes according
to their origin, which   
are  the
unfolding procedure (UNF), the background subtraction (BKG), 
the acceptance correction (ACC), and the energy scale (ENG).
These contributions  
are summarized in Table~\ref{tab:dnsys}  for each $\MassSQ$ region
and are described below.

The systematic error due to the unfolding procedure is determined from MC
by comparing the true and the unfolded results (UNF1). Another estimate of the
uncertainty of the unfolding is made by changing the value of the unfolding 
parameter that determines the optimum number of the 
singular values of the acceptance matrix (UNF2). 

\begin{table*}[htbp!]
\begin{center}
\caption{ 
Relative systematic errors (in $\%$) of the unfolded spectrum for
each $\MassSQ$ region for the different sources of  uncertainty:
 unfolding procedure (UNF1, UNF2), the background subtraction (BKG1, BKG2, BKG3), the acceptance
correction (ACC), and the photon energy scale (PES).
See the text for a more detailed description.}
\label{tab:dnsys}
\begin{tabular}{l| c c c c c c } 
\hline \hline 
$\MassSQ$ region &First bin & Threshold  & $\rho$   &   & $\rho^{\prime}$ & $\rho^{\prime\prime}$ \\ 
               &        & region     &  region  &    &    region  & region  \\ 
($\GeVcc2$)      & (0.08) & (0.2-0.3) & (0.55-0.60) & (1.0-1.2) & (1.9-2.0) & (2.5-2.7) \\ 
\hline\hline
UNF1  & 2.50   & 0.79   &  0.31    & 0.85    & 1.50   &  1.50   \\
UNF2  & 2.60    & 0.53    &   0.09    &  0.27    &0.58   &  9.19    \\
BKG1   & 1.13   & 0.09    &   0.01    &  0.04    & 0.52   & 5.76    \\
BKG2   & 4.90   & 0.65    &   0.10    &  0.10       &...  & 0.50   \\
BKG3 &  25.21    & 4.80    &  ... & ...  & ... & ... \\
ACC  &   5.36    &  1.44    &  0.03    &  0.15   & 0.15     & 0.40    \\
PES   &  1.24    & 1.08    &  0.59    &  0.99    & 0.05   & 0.50    \\
\hline\hline
Total &  26.5    &  5.3    &  0.7   & 1.5   & 1.8   & 11.4   \\
\hline
\end{tabular}
\end{center}
\end{table*}

The uncertainty of the background subtraction is estimated for each 
source. BKG1 is  from  continuum processes. Its 
uncertainty is estimated
using the control  sample in the mass region higher than  the $\tau$ mass. 
The statistics of the data and MC sample determine its error.
BKG2 is the
feed-down background. Its uncertainty is estimated  by varying the 
 branching fraction values~\cite{PDG2006} used in the MC by $\pm 1\sigma$. BKG3 is 
 the non-$\pi^0$ background. The uncertainty of the non-$\pi^{0}$ background is
estimated by changing the $\pi^{0}$ sideband region. 
This uncertainty  dominates 
in the threshold region but is negligible elsewhere.

The acceptance uncertainty is dominated by the 
uncertainty of the $\pi^{0}$ efficiency.
This is estimated by changing the measured values of the photon efficiency
by 1 standard deviation.
In addition, the effect of 
requiring that photons be isolated from charged tracks
is checked by changing
the isolation criteria from 20~cm (default) to 30~cm. 

The uncertainty of the photon energy scale (PES) is estimated from the
$\pi^{0}$ peak position to be $\pm 0.2\%$.    
This uncertainty is important for the peak position of the resonances.
The uncertainty in the charged track momentum scale 
is negligible compared to that of the photon energy scale.

Individual components of the uncertainty are added in quadrature to
obtain total  systematic errors of 
 $5.3\%$  in the threshold region, 
$0.7\%$ near the $\rho(770)$ peak and 
 $1.8\%$ in the vicinity of  the $\rho''(1450)$ (see Table~\ref{tab:dnsys}).

\begin{figure}[ht!]
\begin{center}
\rotatebox{0}{\includegraphics[width=0.60\textwidth,clip]
{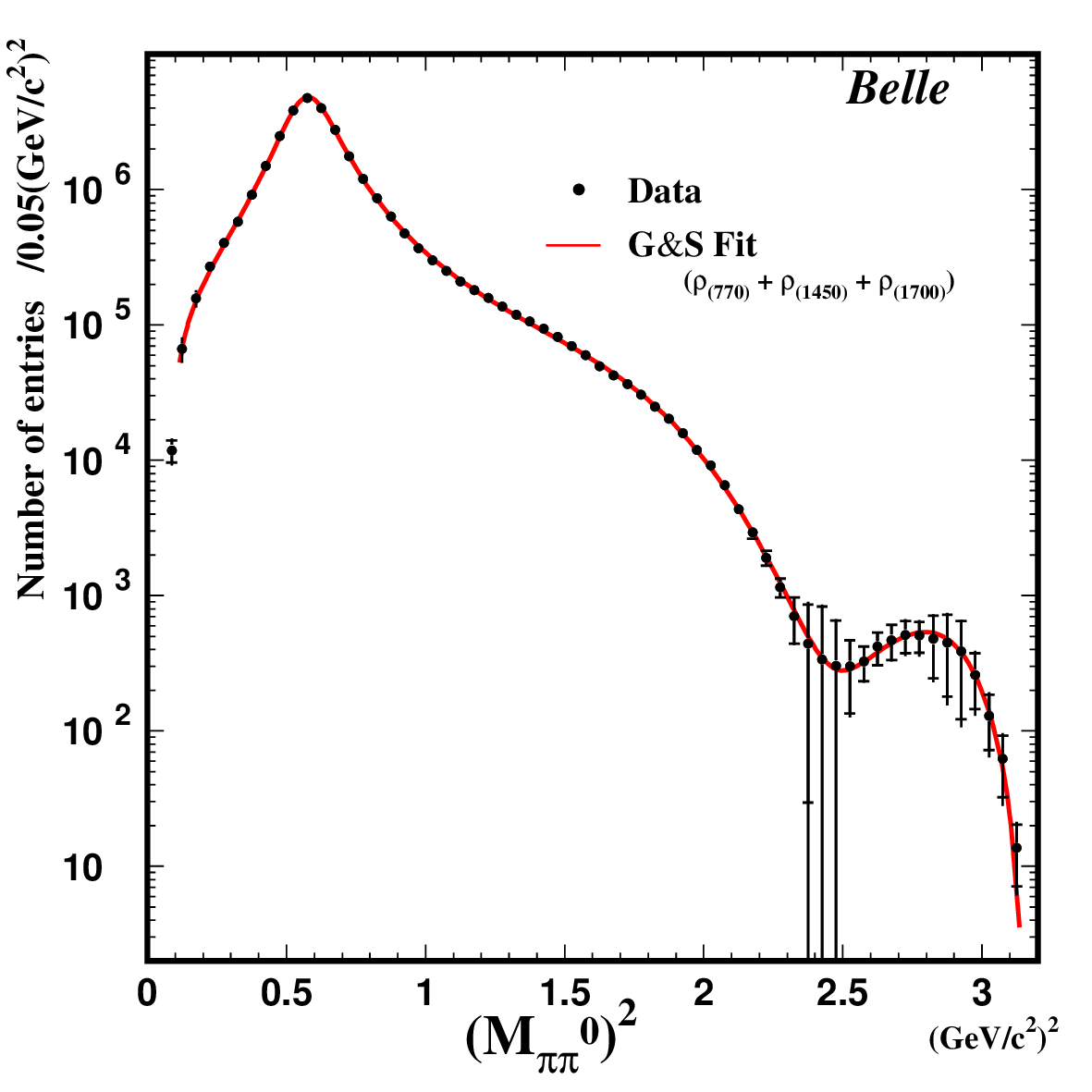}}
\end{center}
\caption
{ The fully corrected $M^2_{\pi\pi^0}$ distribution
for $\Taupipi0$. 
The solid curve is the 
result of a fit to the Gounaris-Sakurai model with the 
$\rho(770)$, $\rho'(1450)$, and $\rho''(1700)$ resonances.
All resonance parameters (mass, width, phase and the normalization factor 
$|F_{\pi}(0)|^{2}$) are allowed to float.
}
\label{unfold_pipi0}
\end{figure}



\begin{table}[ht!]
\begin{center}
\caption{
The unfolded normalized spectrum $(1/N_{\pi\pi}) ({\rm d}N_{\pi\pi}/{\rm d}s)$ as
a function of the invariant mass squared $s=\MassSQ$.
 The square of the diagonal element of the error matrix is
taken for the statistical errors. 
 Note that the scale is different for the left-and right-sides.    
}
\small
\begin{tabular}{lccrrrllccrrrl}
 &  &  &  &  &  &  &  &  &  &  &  &  &  \\ \cline{2-13}
& Bin  & $\MassSQ$ & \multicolumn{1}{l}{~~~~$\frac{1}{N}\frac{{\rm d}N}{{\rm d}s}$} &
 \multicolumn{1}{r}{\phantom{00}Stat.} & \multicolumn{1}{r}{\phantom{00}Syst.} &
 \multicolumn{1}{l|}{} &  
& Bin & $\MassSQ$ & 
\multicolumn{1}{l}{~~~~$\frac{1}{N}\frac{{\rm d}N}{{\rm d}s}$} 
& \multicolumn{1}{r}{\phantom{00}Stat.} & \multicolumn{1}{r}{\phantom{00}Syst.} &  \\
 &  &{ \scriptsize }&
 {\scriptsize  $\times\ 10^{-3}$} & 
 {\scriptsize  $\times\ 10^{-3} $} &
 {\scriptsize  $\times\ 10^{-3} $}  
& \multicolumn{1}{r|}{} &  & No. &
{ \scriptsize }  & 
{\scriptsize  $\times\ 10^{-4} $}  & 
{\scriptsize  $\times\ 10^{-4} $}& 
{\scriptsize  $\times\ 10^{-4} $}& \\ 
 & No. &{ \scriptsize $\GeVcc2$ }&
  {\scriptsize  $ \GeVinvcc2$} & 
 \phantom{0} {\scriptsize  $ \GeVinvcc2$} &
 \phantom{0} {\scriptsize  $ \GeVinvcc2$}  
& \multicolumn{1}{r|}{} &  & No. &
{ \scriptsize $\GeVcc2$ }  & 
  {\scriptsize  $ \GeVinvcc2$}  & 
 \phantom{0} {\scriptsize  $ \GeVinvcc2$}& 
 \phantom{0} {\scriptsize  $ \GeVinvcc2$}& \\ 
\cline{2-13}
 & \phantom{0}1 & 0.088 & \phantom{000}8.1 & 3.1 & \phantom{0}2.1 & \multicolumn{1}{l|}{} &  & 32 & 1.625 & 341.47 & 5.91 & 5.83 &  \\
 & \phantom{0}2 & 0.125 & \phantom{00}45.8 & 2.4 & \phantom{0}9.6 & \multicolumn{1}{l|}{} &  & 33 & 1.675 & 290.94 & 5.56 & 5.39 &  \\
 & \phantom{0}3 & 0.175 & \phantom{0}108.4 & 2.4 & 15.5 & \multicolumn{1}{l|}{} &  & 34 & 1.725 & 250.39 & 4.99 & 4.32 &  \\
 & \phantom{0}4 & 0.225 & \phantom{0}185.1 & 2.6 & 14.3 & \multicolumn{1}{l|}{} &  & 35 & 1.775 & 210.22 & 4.86 & 3.40 &  \\
 & \phantom{0}5 & 0.275 & \phantom{0}278.0 & 2.8 & \phantom{0}8.3 & \multicolumn{1}{l|}{} &  & 36 & 1.825 & 170.63 & 4.36 & 2.75 &  \\
 & \phantom{0}6 & 0.325 & \phantom{0}396.0 & 3.3 & \phantom{0}4.1 & \multicolumn{1}{l|}{} &  & 37 & 1.875 & 139.72 & 3.89 & 2.26 &  \\
 & \phantom{0}7 & 0.375 & \phantom{0}628.6 & 4.2 & \phantom{0}5.0 & \multicolumn{1}{l|}{} &  & 38 & 1.925 & 109.26 & 3.82 & 1.85 &  \\
 & \phantom{0}8 & 0.425 & 1024.2 & 5.5 & 14.7 & \multicolumn{1}{l|}{} &  & 39 & 1.975 & \phantom{0}81.85 & 3.20 & 1.39 &  \\
 & \phantom{0}9 & 0.475 & 1710.8 & 7.2 & 21.5 & \multicolumn{1}{l|}{} &  & 40 & 2.025 & \phantom{0}63.08 & 2.84 & 1.25 &  \\
 & 10 & 0.525 & 2643.0 & 9.0 & 22.5 & \multicolumn{1}{l|}{} &  & 41 & 2.075 & \phantom{0}45.02 & 2.72 & 0.89 &  \\
 & 11 & 0.575 & 3268.0 & 9.7 & 21.4 & \multicolumn{1}{l|}{} &  & 42 & 2.125 & \phantom{0}29.89 & 2.22 & 0.65 &  \\
 & 12 & 0.625 & 2755.5 & 9.0 & 19.4 & \multicolumn{1}{l|}{} &  & 43 & 2.175 & \phantom{0}20.06 & 1.94 & 0.61 &  \\
 & 13 & 0.675 & 1907.2 & 7.4 & 17.0 & \multicolumn{1}{l|}{} &  & 44 & 2.225 & \phantom{0}13.08 & 1.68 & 0.42 &  \\
 & 14 & 0.725 & 1214.0 & 5.8 & \phantom{0}8.4 & \multicolumn{1}{l|}{} &  & 45 & 2.275 & \phantom{00}7.93 & 1.25 & 0.42 &  \\
 & 15 & 0.775 & \phantom{0}826.4 & 4.6 & 10.7 & \multicolumn{1}{l|}{} &  & 46 & 2.325 & \phantom{00}4.85 & 1.82 & 0.51 &  \\
 & 16 & 0.825 & \phantom{0}592.8 & 3.6 & \phantom{0}8.6 & \multicolumn{1}{l|}{} &  & 47 & 2.375 & \phantom{00}3.04 & 2.84 & 1.39 &  \\
 & 17 & 0.875 & \phantom{0}435.6 & 2.9 & \phantom{0}3.5 & \multicolumn{1}{l|}{} &  & 48 & 2.425 & \phantom{00}2.32 & 3.36 & 1.17 &  \\
 & 18 & 0.925 & \phantom{0}327.5 & 2.4 & \phantom{0}4.6 & \multicolumn{1}{l|}{} &  & 49 & 2.475 & \phantom{00}2.09 & 2.39 & 0.76 &  \\
 & 19 & 0.975 & \phantom{0}253.7 & 2.1 & \phantom{0}3.7 & \multicolumn{1}{l|}{} &  & 50 & 2.525 & \phantom{00}2.07 & 1.14 & 0.37 &  \\
 & 20 & 1.025 & \phantom{0}206.3 & 1.8 & \phantom{0}2.7 & \multicolumn{1}{l|}{} &  & 51 & 2.575 & \phantom{00}2.24 & 0.64 & 0.19 &  \\
 & 21 & 1.075 & \phantom{0}172.5 & 1.5 & \phantom{0}3.6 & \multicolumn{1}{l|}{} &  & 52 & 2.625 & \phantom{00}2.88 & 0.78 & 0.17 &  \\
 & 22 & 1.125 & \phantom{0}144.0 & 1.3 & \phantom{0}1.4 & \multicolumn{1}{l|}{} &  & 53 & 2.675 & \phantom{00}3.22 & 0.94 & 0.33 &  \\
 & 23 & 1.175 & \phantom{0}124.0 & 1.2 & \phantom{0}3.7 & \multicolumn{1}{l|}{} &  & 54 & 2.725 & \phantom{00}3.52 & 0.95 & 0.51 &  \\
 & 24 & 1.225 & \phantom{0}108.7 & 1.1 & \phantom{0}3.3 & \multicolumn{1}{l|}{} &  & 55 & 2.775 & \phantom{00}3.49 & 0.89 & 0.57 &  \\
 & 25 & 1.275 & \phantom{00}94.0 & 1.0 & \phantom{0}1.0 & \multicolumn{1}{l|}{} &  & 56 & 2.825 & \phantom{00}3.29 & 1.60 & 0.62 &  \\
 & 26 & 1.325 & \phantom{00}82.0 & 0.9 & \phantom{0}2.4 & \multicolumn{1}{l|}{} &  & 57 & 2.875 & \phantom{00}3.09 & 1.86 & 0.80 &  \\
 & 27 & 1.375 & \phantom{00}72.6 & 0.9 & \phantom{0}1.3 & \multicolumn{1}{l|}{} &  & 58 & 2.925 & \phantom{00}2.65 & 1.81 & 0.65 &  \\
 & 28 & 1.425 & \phantom{00}64.5 & 0.8 & \phantom{0}1.8 & \multicolumn{1}{l|}{} &  & 59 & 2.975 & \phantom{00}1.79 & 0.79 & 0.43 &  \\
 & 29 & 1.475 & \phantom{00}56.1 & 0.8 & \phantom{0}1.8 & \multicolumn{1}{l|}{} &  & 60 & 3.025 & \phantom{00}0.88 & 0.39 & 0.22 &  \\
 & 30 & 1.525 & \phantom{00}48.0 & 0.7 & \phantom{0}1.4 & \multicolumn{1}{l|}{} &  & 61 & 3.075 & \phantom{00}0.43 & 0.20 & 0.12 &  \\
 & 31 & 1.575 & \phantom{00}41.1 & 0.6 & \phantom{0}0.7 & \multicolumn{1}{l|}{} &  & 62 & 3.125 & \phantom{00}0.09 & 0.04 & 0.03 &  \\ \cline{2-13}
 &  &  &  &  &  &  &  &  &  &  &  &  & 
\end{tabular}
\label{tab:dnds}
\end{center}
\end{table}

\begin{table}
\begin{center}
\caption{
The  form factor squared $|F_{\pi}^{-}(s)|^{2}$ as a function of 
the invariant mass squared $s$.
The results are obtained by inserting the measured value 
$(1/N_{\pi\pi}) ({\rm d}N_{\pi\pi}/{\rm d}s)$ into Eq.~(\ref{eq:fpiweak}).
Note that the short-distance radiative correction is already
applied, where  
the value of $S_{\rm EW}$ is taken to be $1.0235\pm 0.0003$
(see the discussion in the Appendix).
}
\small
\begin{tabular}{lcrrrrllcrrrrl}
 &  &  &  &  &  &  &  &  &  &  &  &  &  \\ \cline{2-13}
 & Bin  & $\MassSQ$ & \multicolumn{1}{c}{~~~~$\ |F_\pi|^{2}$} & \multicolumn{1}{c}{\phantom{0000}Stat.} &
  \multicolumn{1}{c}{\phantom{0000}Syst.} & 
 \multicolumn{1}{c|}{} & 
 & Bin  & $\MassSQ$ & \multicolumn{1}{c}{~~~~$\ |F_\pi|^{2}$} & 
 \multicolumn{1}{c}{\phantom{0000}Stat.} &
  \multicolumn{1}{c}{\phantom{0000}Syst.} &  \\
 & No. &\tiny{$\GeVcc2$} &  & &  & \multicolumn{1}{l|}{}   & & No. &\tiny{$\GeVcc2$}&  &  &  &  \\ \cline{2-13}
%
  & \phantom{0}1 & 0.088 & \phantom{0}1.434 & 0.549 & 0.377 & \multicolumn{1}{l|}{} &  & 32 & 1.625 & 0.711 & 0.012 & 0.012 &  \\
 & \phantom{0}2 & 0.125 & \phantom{0}1.707 & 0.091 & 0.358 & \multicolumn{1}{l|}{} &  & 33 & 1.675 & 0.636 & 0.012 & 0.012 &  \\
 & \phantom{0}3 & 0.175 & \phantom{0}2.362 & 0.053 & 0.337 & \multicolumn{1}{l|}{} &  & 34 & 1.725 & 0.576 & 0.011 & 0.010 &  \\
 & \phantom{0}4 & 0.225 & \phantom{0}3.211 & 0.045 & 0.248 & \multicolumn{1}{l|}{} &  & 35 & 1.775 & 0.511 & 0.012 & 0.008 &  \\
 & \phantom{0}5 & 0.275 & \phantom{0}4.260 & 0.042 & 0.127 & \multicolumn{1}{l|}{} &  & 36 & 1.825 & 0.439 & 0.011 & 0.007 &  \\
 & \phantom{0}6 & 0.325 & \phantom{0}5.622 & 0.046 & 0.058 & \multicolumn{1}{l|}{} &  & 37 & 1.875 & 0.382 & 0.011 & 0.006 &  \\
 & \phantom{0}7 & 0.375 & \phantom{0}8.492 & 0.057 & 0.067 & \multicolumn{1}{l|}{} &  & 38 & 1.925 & 0.318 & 0.011 & 0.005 &  \\
 & \phantom{0}8 & 0.425 & 13.392 & 0.072 & 0.193 & \multicolumn{1}{l|}{} &  & 39 & 1.975 & 0.255 & 0.010 & 0.004 &  \\
 & \phantom{0}9 & 0.475 & 21.894 & 0.093 & 0.275 & \multicolumn{1}{l|}{} &  & 40 & 2.025 & 0.211 & 0.009 & 0.004 &  \\
 & 10 & 0.525 & 33.384 & 0.113 & 0.284 & \multicolumn{1}{l|}{} &  & 41 & 2.075 & 0.162 & 0.010 & 0.003 &  \\
 & 11 & 0.575 & 40.996 & 0.122 & 0.269 & \multicolumn{1}{l|}{} &  & 42 & 2.125 & 0.117 & 0.009 & 0.003 &  \\
 & 12 & 0.625 & 34.503 & 0.112 & 0.243 & \multicolumn{1}{l|}{} &  & 43 & 2.175 & 0.085 & 0.008 & 0.003 &  \\
 & 13 & 0.675 & 23.936 & 0.093 & 0.214 & \multicolumn{1}{l|}{} &  & 44 & 2.225 & 0.060 & 0.008 & 0.002 &  \\
 & 14 & 0.725 & 15.324 & 0.074 & 0.106 & \multicolumn{1}{l|}{} &  & 45 & 2.275 & 0.040 & 0.006 & 0.002 &  \\
 & 15 & 0.775 & 10.525 & 0.058 & 0.137 & \multicolumn{1}{l|}{} &  & 46 & 2.325 & 0.027 & 0.010 & 0.003 &  \\
 & 16 & 0.825 & \phantom{0}7.637 & 0.047 & 0.111 & \multicolumn{1}{l|}{} &  & 47 & 2.375 & 0.019 & 0.018 & 0.009 &  \\
 & 17 & 0.875 & \phantom{0}5.693 & 0.038 & 0.046 & \multicolumn{1}{l|}{} &  & 48 & 2.425 & 0.017 & 0.024 & 0.008 &  \\
 & 18 & 0.925 & \phantom{0}4.350 & 0.032 & 0.061 & \multicolumn{1}{l|}{} &  & 49 & 2.475 & 0.017 & 0.019 & 0.006 &  \\
 & 19 & 0.975 & \phantom{0}3.435 & 0.028 & 0.050 & \multicolumn{1}{l|}{} &  & 50 & 2.525 & 0.019 & 0.011 & 0.003 &  \\
 & 20 & 1.025 & \phantom{0}2.851 & 0.024 & 0.037 & \multicolumn{1}{l|}{} &  & 51 & 2.575 & 0.024 & 0.007 & 0.002 &  \\
 & 21 & 1.075 & \phantom{0}2.439 & 0.022 & 0.052 & \multicolumn{1}{l|}{} &  & 52 & 2.625 & 0.036 & 0.010 & 0.002 &  \\
 & 22 & 1.125 & \phantom{0}2.087 & 0.019 & 0.020 & \multicolumn{1}{l|}{} &  & 53 & 2.675 & 0.050 & 0.014 & 0.005 &  \\
 & 23 & 1.175 & \phantom{0}1.847 & 0.018 & 0.056 & \multicolumn{1}{l|}{} &  & 54 & 2.725 & 0.066 & 0.018 & 0.010 &  \\
 & 24 & 1.225 & \phantom{0}1.667 & 0.017 & 0.050 & \multicolumn{1}{l|}{} &  & 55 & 2.775 & 0.083 & 0.021 & 0.013 &  \\
 & 25 & 1.275 & \phantom{0}1.486 & 0.016 & 0.015 & \multicolumn{1}{l|}{} &  & 56 & 2.825 & 0.102 & 0.050 & 0.019 &  \\
 & 26 & 1.325 & \phantom{0}1.339 & 0.015 & 0.039 & \multicolumn{1}{l|}{} &  & 57 & 2.875 & 0.132 & 0.079 & 0.034 &  \\
 & 27 & 1.375 & \phantom{0}1.229 & 0.015 & 0.022 & \multicolumn{1}{l|}{} &  & 58 & 2.925 & 0.165 & 0.113 & 0.040 &  \\
 & 28 & 1.425 & \phantom{0}1.132 & 0.014 & 0.031 & \multicolumn{1}{l|}{} &  & 59 & 2.975 & 0.178 & 0.079 & 0.043 &  \\
 & 29 & 1.475 & \phantom{0}1.025 & 0.014 & 0.032 & \multicolumn{1}{l|}{} &  & 60 & 3.025 & 0.165 & 0.073 & 0.041 &  \\
 & 30 & 1.525 & \phantom{0}0.913 & 0.013 & 0.027 & \multicolumn{1}{l|}{} &  & 61 & 3.075 & 0.203 & 0.098 & 0.056 &  \\
 & 31 & 1.575 & \phantom{0}0.818 & 0.013 & 0.014 & \multicolumn{1}{l|}{} &  & 62 & 3.125 & 0.287 & 0.138 & 0.081 &  \\ \cline{2-13}
 &  &  &  &  &  &  &  &  &  &  &  &  & 
\end{tabular}
\label{tab:fpi2}
\end{center}
\end{table}

\subsection{Results}
The unfolded $s= M^2_{(\pi\pi^0\,{\rm unf.})}$ spectrum
${\rm d}N_{\pi\pi}/{\rm d}s$ is shown in Fig.~\ref{unfold_pipi0}.
The error bars in the figure include both statistical and systematic
errors added in quadrature and in most cases they are smaller than the 
size of the data points shown by closed circles. 
 The results are also presented in terms of the
 normalized unfolded spectrum $(1/N_{\pi\pi})({\rm d}N_{\pi\pi}/{\rm d}s)$ in 
Table~\ref{tab:dnds} and in terms of the pion form factor in Table~\ref{tab:fpi2}.
 In these tables, the statistical and
systematic errors are given separately.
 The statistical errors  in the figure and the table are the square 
roots of the diagonal components of the covariance matrix.
 
In Fig.~\ref{unfold_pipi0},
the $\rho$ peak and a shoulder due to the $\rho'(1450)$
are clearly visible. The dip at $s\approx 2.5~\GeVcc2$ is
caused by destructive interference between the $\rho'(1450)$ 
and $\rho''(1700)$ resonances.

To determine the parameters of  the $\rho$, $\rho'$ and $\rho^{''}$ 
resonances, a  $\chi^{2}$ fit using Breit-Wigner (BW) functions 
is performed. 
The pion form factor   
is parametrized with Breit-Wigner functions corresponding to the 
$\rho$, $\rho^{\prime}(1450)$, and $\rho^{\prime\prime}$(1700) 
resonances:
%
\begin{equation}
F_{\pi}(s) = \frac{1}{1 + \beta +\gamma } 
	  (BW_{\rho} + \beta \cdot  BW_{\rho^{\prime}} 
 +\gamma \cdot BW_{\rho^{\prime\prime}})\,,
\end{equation}
where the parameters $\beta$ and $\gamma$
(denoting the relative magnitude of the two resonances) are 
in general complex.
We use the Gounaris-Sakurai\,(GS) model~\cite{GS} for the
Breit-Wigner shape:
\begin{equation}
BW_{i}^{GS} = \frac {M_{i}^2  + d \cdot M_{i}\Gamma_{i}(s) }
	   {(M_{i}^{2} - s) + f(s) - i \sqrt{s} \Gamma_{i}(s)}\,,
\end{equation}
\noindent
with an energy-dependent width
\begin{equation}
\Gamma_{i}(s) = \Gamma_{i}  \left( \frac{M_{i}^{2}}{s}\right)
\left( \frac{k(s)}{k(M_{i}^2)}\right)^{3}\,.
\label{eq:width}
\end{equation}
Here, $ k(s)  =  \frac{1}{2} \sqrt{s} \beta_{-}(s)$ is the pion
momentum in the $\pi^{-}\pi^{0}$ rest frame.
The functions $f(s)$ and $h(s)$ are defined as

\begin{eqnarray}
f(s) & = & \Gamma^{}_i\,\frac{M^2_i}{k^3(M^2_i)} 
  \left[\, 
     k^2(s) \left( h(s) -h(M_{i}^2) \right)
     + (M_{i}^2 - s) k^2(M_{i}^2) 
	\left.\frac{dh}{ds}\right|_{s=M^2_{i}} 
  \,\right]  
\label{eq:dif} \\
& & \nonumber \\  
h(s) & = & \frac{2}{\pi} \frac{k(s)}{\sqrt{s}} 
\ln \left( \frac{\sqrt{s} + 2k(s)}{2m_{\pi}}\right) \,,
\end{eqnarray}
\noindent
with  
$ \left. dh/ds\right|_{M_{i}^{2}} = 
h(M_{i}^{2})  \left [  
		 \left( 8k^2(M_{i}^{2}) \right)^{-1}
		  - (2 M_{i}^{2})^{-1}    \right ]
+ (2 \pi M_{i}^{2})^{-1}
$
and  

\begin{equation}
    d = \frac{3}{\pi} \frac{m_{\pi}^2} {k^2(M_{i}^2)} 
   \ln\left( \frac{M_{i} + 2 k(M_{i}^2)} {2 m_{\pi}} \right)
		 +
  \frac{M^{}_i} {2 \pi k(M^2_{i})} -
  \frac{m_{\pi}^2 M^{}_i} {\pi k^3(M^2_{i})}\,.          
\end{equation}
\noindent
Note that the function $d$ is chosen so that the $BW^{GS}$ 
function is unity at $s=0$~\cite{GS}.

Since the unfolded mass spectrum has bin-by-bin 
correlations, the off-diagonal components of the covariance matrix
$X$ are included in the $\chi^{2}$ evaluation:
\begin{equation}
\chi^{2}= \sum_{i,j} \left( y _{i} - f(s_{i}; \alpha) \right)
	(X^{-1})_{ij}
	\left( y _{j} - f(s_{j}; \alpha) \right)\,,
\label{eq:chi2def} 
\end{equation}
where $y_{i}$ is the measured value at the $i$-th bin,
$f(s; \alpha)$ is the value of the function for parameters $\alpha$,
and $(X^{-1})_{ij}$ is the inverse of the covariance matrix. 

 There are 10 parameters in this formula: 
the masses~($M_{i}$) and widths~($\Gamma_{i}$) for the 
$\rho,\,\rho^{\prime}$,\, and $\rho^{\prime\prime}$ resonances,
their relative amplitudes $|\beta|$,\,$|\gamma|$,\, and phases
$\phi_{\beta}$ and $\phi_{\gamma}$.
In addition,
 as an overall normalization factor, we introduce 
$|F_{\pi}(0)|^{2}$ as an additional parameter.
In the BW form, this value should be unity. 
However, in order to take into account
a possible deviation from the form of the fitting function, 
two kinds of fits, in which this parameter is either fixed or floated, are carried out.
The other 10 parameters are floated in the fit.

\begin{figure}[ht!]
\begin{center}
\rotatebox{0}{\includegraphics*[width=0.85\textwidth,clip]
{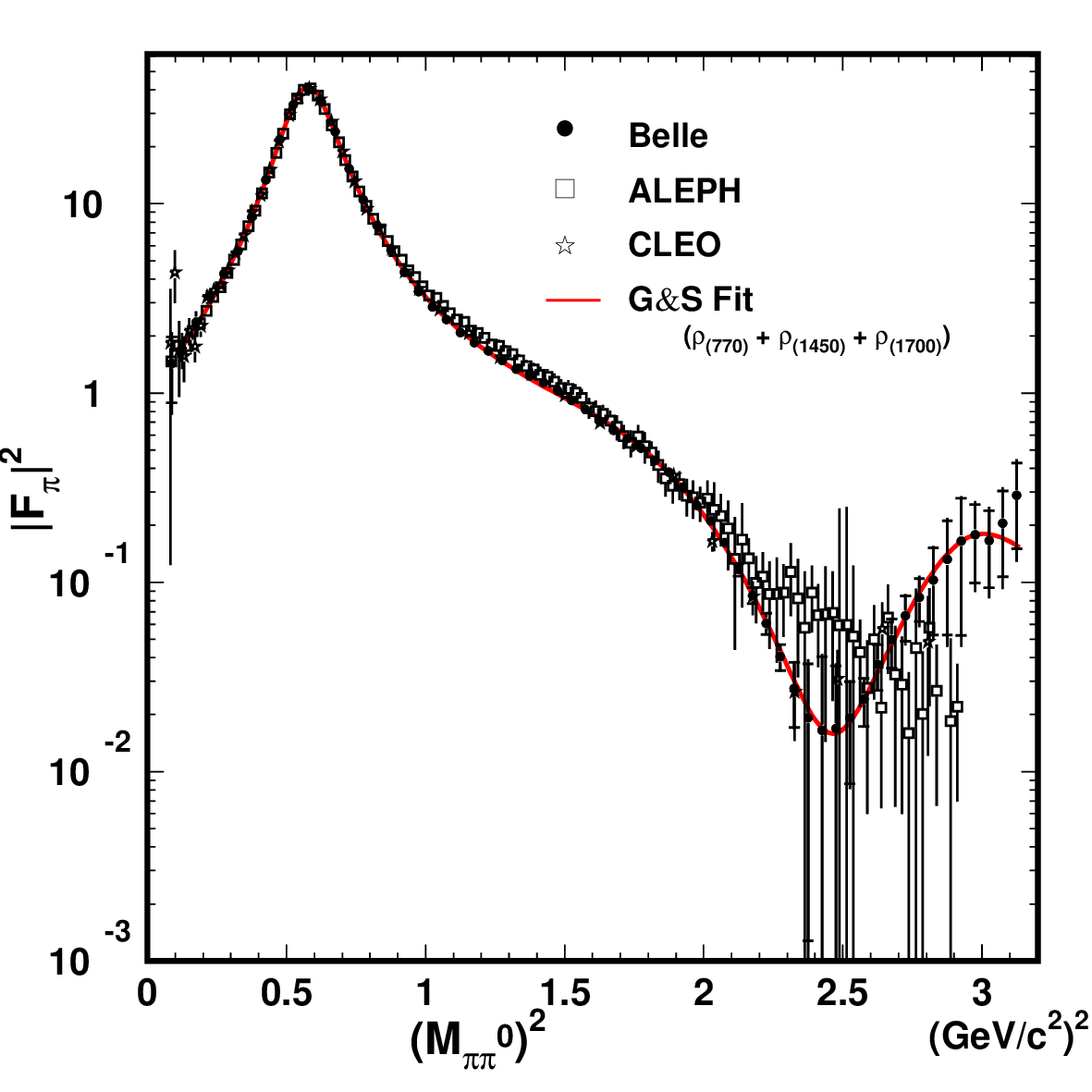}}

%
%
%

\caption
{ Pion form factor for $\Taupipi0$.
The solid circles are the Belle result while the
 squares and stars show the result of ALEPH ~\protect{\cite{ALEPH05}} and 
 CLEO~\protect{\cite{CLEO2000}}, respectively. 
The error bars for the Belle data include both
statistical and systematic errors added in quadrature.
The solid curve is the 
result of a fit to the Gounaris-Sakurai model with 
the
$\rho(770)$, $\rho'(1450)$, and $\rho''(1700)$ resonances,
where all parameters are floated.
}
\label{Fpi}
\end{center}
\end{figure}

\begin{figure}[!]
\begin{center}
\rotatebox{0}{\includegraphics[width=0.63\textwidth,clip]
{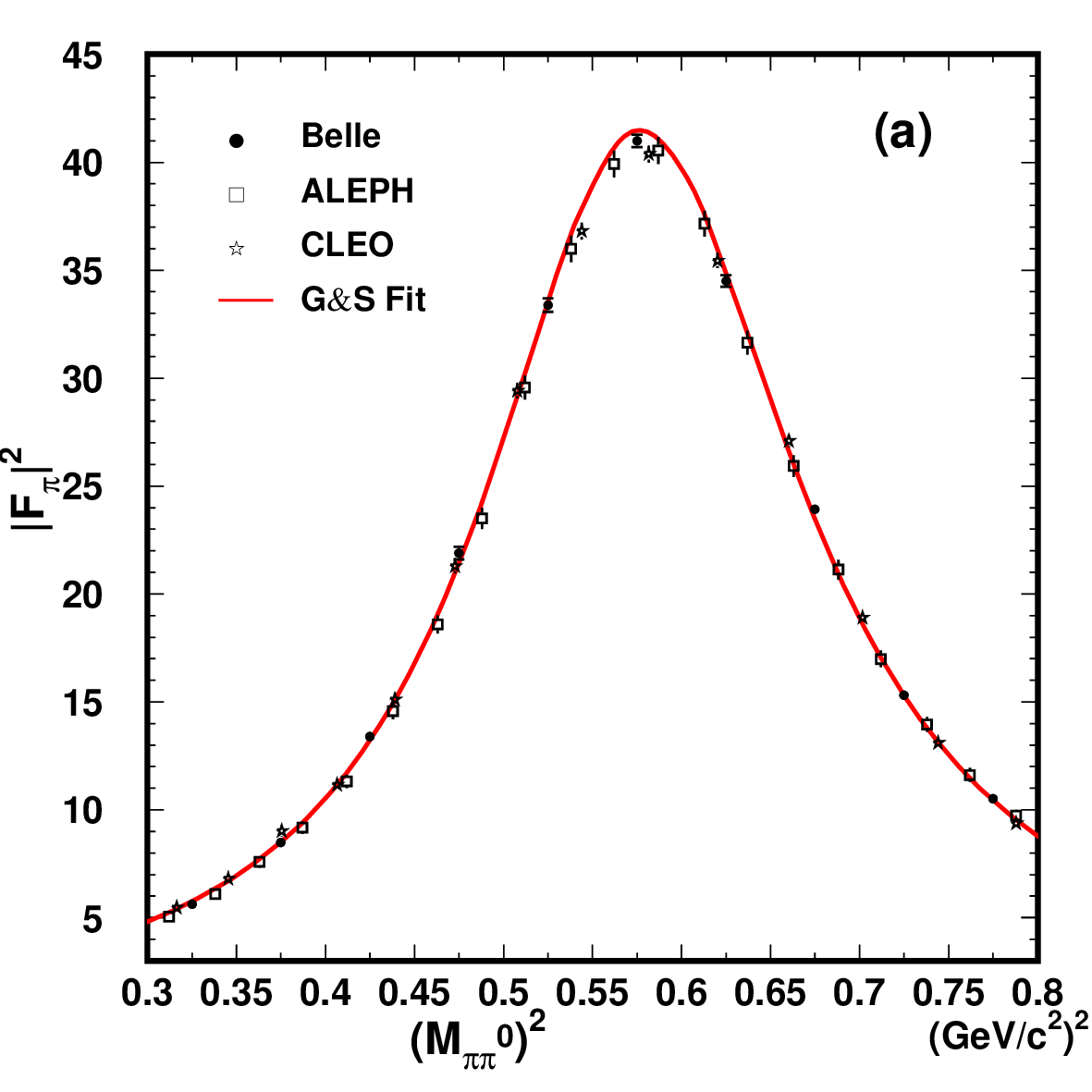}}
%
%

\rotatebox{0}{\includegraphics[width=0.63\textwidth,clip]
{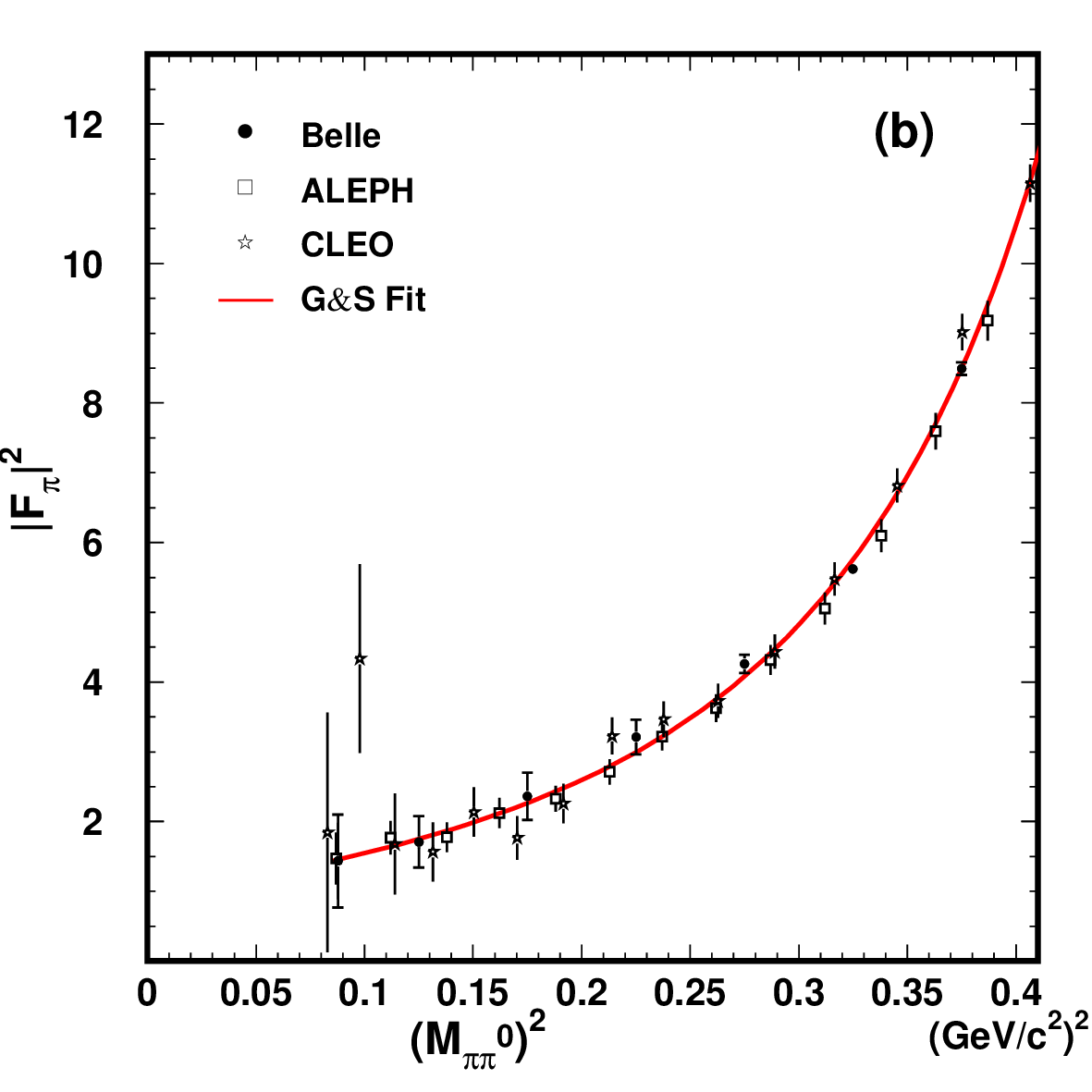}}

\caption
{ (a) Pion form factor for $\Taupipi0$  in the $\rho(770)$ mass region 
and (b) in the threshold region.
The convention of the plots is the same as in Fig.~\ref{Fpi}.
The error bars for the Belle data include both
statistical and systematic errors added in quadrature.
The solid curve is the 
result of a fit to the Gounaris-Sakurai model, where
all parameters are floated.
See the text for details.
}
\label{Fpirho}
%
%
%
\end{center}
\end{figure}

The results
 of the fit are shown as the solid line in Fig.~\ref{unfold_pipi0} 
for the $\MassSQ$ distribution as well
as in Fig.~\ref{Fpi} and  Fig.~\ref{Fpirho}(a) and (b), where 
the results are compared directly to  the weak form factor squared   
 $|F_{\pi}^{-}(s)|^{2}$,  
derived  bin-by-bin  
from Eq.~(\ref{eq:fpiweak}).   
In Eq.~(\ref{eq:fpiweak}),
we use the world average value (including our measurement)  for the branching fraction
$\mathcal{B}_{\pi\pi}=(25.24\pm 0.10)\%$ 
and  for the CKM matrix element 
$V_{ud}=0.97377\pm 0.00027$~\cite{PDG2006}.
 For the short-distance radiative correction 
$S_{\rm EW}= S_{\rm EW}^{\pi\pi}/S_{\rm EW}^{e}$,  
we take the value $1.0235\pm 0.0003$, to be  consistent with  
the isospin breaking
correction discussed in Ref.~\cite{ISB2001,DEHZ} 
(see the appendix for more details).

 The fitted results  
are summarized in Table~\ref{tab:fit_param}
for the cases when $|F_\pi(0)|^2$ is fixed to unity (the second column) and 
is allowed to float (the third column).
In the table, the first error is statistical and the second is systematic. 
The value of the $\chi^{2}$ per degree of freedom (NDF) is 80/52 for the fixed and
65/51 for the floated cases.
It is  found that $|F_\pi(0)|^2$ is close to unity
($|F_\pi(0)|^2=1.02\pm0.01\pm 0.04$) even when it is allowed to float.
It should be noted that the data can be fitted using BW resonances only, 
 without any additional background terms.
The fit quality for the fixed case is 
slightly 
worse than that for the floated one.
The curves shown in Figs.~\ref{unfold_pipi0}, \ref{Fpi} and 
\ref{Fpirho}(a) and \ref{Fpirho}(b),  correspond to the case when
the parameter $|F_{\pi}(0)|^{2}$ is floated, but the differences  
between the floated and the fixed cases are small.

The significance of the $\rho''(1700)$ signal is given in
the last row of Table~\ref{tab:fit_param}. 
The significance is determined from
the change in the $\chi^{2}$ when the signal and its associated degree
of freedom are removed from the fit.
If the $\rho''(1700)$ signal is excluded from the fit,
the $\chi^{2}$ for the fit increases by 55 (60) units,
in the case that $|F(0)|^{2}$ is fixed to unity (allowed to float).  
This increase in the $\chi^2$ for the fit, with the joint estimation of
four removed parameters (mass, width, $|\gamma|$, $\phi_{\gamma}$),
corresponds to a 6.5$\sigma$ (7.0$\sigma$) significance for
the $\rho''(1700)$ signal~\cite{PDGstat}. 
\begin{table}[ht!]
\begin{center}
\caption{
Results of fitting the $M^2_{\pi\pi^0}$ distribution
for $\Taupipi0$ to the Gounaris-Sakurai model with
the $\rho(770)$, $\rho'(1450)$, and $\rho''(1700)$ resonances.
The results are shown for two cases, fixed $|F_\pi(0)|^2=1$ (the second column) 
and all parameters are allowed to float (the third column).
The first error is statistical and the second one is systematic. 
The systematic errors include the uncertainty in the backgrounds,
unfolding as well as the uncertainty of the photon energy scale.
The last row gives the significance of the $\rho''(1700)$ signal.
}
\label{tab:fit_param}
\begin{tabular}{l|c|c}
\hline
\hline
\phantom{0000} Parameter \phantom{0000} &\phantom{000000}  Fit result \phantom{000000} & \phantom{000000} Fit result \phantom{000000} \\
  & (fixed $|F(0)|^2$)  &  (all free) \\ 
\hline\hline
{\small $M_{\rho}$, \,${\rm  MeV/}c^{2}$} & {\small $774.6\pm 0.2\pm 0.5$} & {\small $774.9\pm 0.3\pm 0.5$}\\
{\small $\Gamma_{\rho}$, \,${\rm MeV}$} & {\small $148.1\pm 0.4\pm 1.7$} & {\small $148.6\pm 0.5\pm 1.7 $}\\
{\small $M_{\rho^{\prime}}$, \,${\rm  MeV}/c^{2}$} & {\small $1446\pm 7\pm 28$} & {\small $1428\pm 15\pm 26$}\\
{\small $\Gamma_{\rho^{\prime}}$, \,${\rm  MeV}$} & {\small $434\pm 16\pm 60$} & {\small $413 \pm 12\pm 57$}\\
{\small $|\beta|$} & {\small $0.15 \pm 0.05 ^{+0.15}_{-0.04}$} & {\small $0.13 \pm 0.01 ^{+0.16}_{-0.04}$} \\
{\small $\phi_{\beta}$, \,degree} & {\small $202 \pm 4 ^{+41}_{-\ 8}$} & {\small $197 \pm 9 ^{+50}_{-\ 5}$} \\
{\small $M_{\rho^{\prime\prime}}$, \,${\rm MeV}/c^{2}$} & {\small $1728\pm 17\pm 89$} & {\small $1694 \pm 41\pm 89$} \\
{\small $\Gamma_{\rho^{\prime\prime}}$, \,${\rm  MeV}$} & {\small $164\pm 21 ^{+89}_{-26} $} & 
{\small $135\pm 36 ^{+50}_{-26}$} \\
{\small $|\gamma|$} & {\small $0.037 \pm 0.006 ^{+0.065}_{-0.009}$} &
 {\small $0.028 \pm 0.020 ^{+0.059}_{-0.009}$} \\
{\small $\phi_{\gamma}$, \,degree} & {\small $24 \pm 9 ^{+118}_{-\ 28}$} & {\small $-3\pm 13 ^{+136}_{-\ 29}$} \\
{\small $|F(0)|^2$} & {\small $ [$1.0$] $} & {\small $1.02\pm 0.01\pm 0.04$}\\
\hline
{\small $\chi^{2}/{\rm NDF}$} & {\small 80/52} & {\small 65/51} \\
\hline
{\small $\rho''(1700)$ signif.},\, $\sigma$ &  6.5    &           7.0     \\
\hline\hline
\end{tabular}
\end{center}
\end{table}

The fit parameters are correlated, with the correlation matrix:
\begin{equation}
\left(
\begin{array}{@{\,} ccccccccccc@{\,} }
 1.00 &            &          &           &      &     &     &     &    &    &       \\
 0.58 & 1.00     &          &           &       &     &     &    &    &    &       \\
 0.45 &  0.39    & 1.00   &           &       &    &     &    &    &    &       \\
 -0.30 & -0.13  &  0.24  &  1.00  &        &         &         &         &       &          &            \\
  -0.14 & -0.28  & -0.15  &  0.36  &  1.00  &         &         &         &       &          &            \\
   -0.31 & -0.24  & -0.06  &  0.47  &  0.13 &  1.00   &         &         &       &          &            \\
    -0.42 & -0.08  &  0.25  &  0.34  & -0.61  & 0.38    & 1.00    &         &       &          &            \\
     0.31 &  0.22  &  0.08  & -0.10  & -0.41  & 0.37    & 0.29    & 1.00    &       &          &            \\
     0.32  &  0.27  &  0.13  & -0.13  & -0.44  & 0.48    & 0.28   & 0.64    & 1.00  &          &            \\
    0.54.& 0.29   & -0.01  &  -0.07 & 0.05   & 0.52    & -0.27   & 0.54    & 0.67  & 1.00     &            \\
        -0.19 & -0.06  &  0.13  &  0.61  & 0.23   & 0.49    & 0.57    & 0.53    & 0.17  &  0.07    & 1.00       \\
 \end{array}
\right),
\end{equation}
where the parameters are $|F_{\pi}(0)|^{2}$, $M_{\rho}$,
{ $\Gamma_{\rho}$,
{ $M_{\rho^{\prime}}$,
{ $\Gamma_{\rho^{\prime}}$,
{ $|\beta|$},
{ $\phi_{\beta}$,
{ $M_{\rho^{\prime\prime}}$,
{ $\Gamma_{\rho^{\prime\prime}}$,
{ $|\gamma|$} and
{ $\phi_{\gamma}$.

%
%
\begin{table}[htbp!]
\begin{center}
\caption{ 
 Systematic errors for resonance parameters from the
  unfolding procedure (UNF), the background subtraction (BKG), the acceptance
correction (ACC) and the photon energy scale (PES). 
}
\label{tab:sys-resparam}

\begin{tabular}{lccccccccccc} 
\hline 
 & $M_{\rho}$ & $\Gamma_{\rho}$ & $M_{\rho^{\prime}}$ & $\Gamma_{\rho^{\prime}}$ & $|\beta|$ & $\phi_{\beta}$ & $M_{\rho^{\prime\prime}}$ & $\Gamma_{\rho^{\prime\prime}}$ & $|\gamma|$ & $\phi_{\gamma}$ \\
 & $({\rm MeV}/c^{2})$ & $({\rm MeV})$ & $({\rm MeV}/c^{2})$ & $({\rm MeV})$  & &(deg.) &$({\rm MeV}/c^{2})$  & $({\rm MeV})$  & & (deg.)  \\
\hline
Fit Bias & 0.3 & 1.6 & 25 &  49 & 0.028 &   4 &  75 & 10 & 0.006 & 13 \\
UNF      & 0.3 & 0.3 &  4 &  24 & 0.020 &   4 &  11 & 14 & 0.002 & 12 \\
BKG1     & 0.3 & ... & 11 &  25 & $^{+0.143}_{-0.031}$ 
            &  $^{+41}_{-\ 5}$ &  13 & $^{+86}_{-10}$ & $^{+0.053}_{-0.020}$ & $^{+117}_{-\ 22}$ \\
BKG2     & ... & ... &  1 & ... &  ...  & 2  &   2 &  2 & 0.001 & ... \\
ACC      & ... & 0.1 &  1 &   4 &  ...  & ... & ... &  7 &  ...  & 1 \\
PES      & 0.3 & 0.6 &  2 &   1 &  ...  &   2 &  45 & 15 &  ...  & 1 \\
\hline
Total    & 0.5 & 1.7 & 28 &  60 & $^{+0.147}_{-0.047}$ & 
         $^{+41}_{-\ 8}$ &  89 & $^{+89}_{-26}$ & $^{+0.053}_{-0.021}$ & $^{+118}_{-\ 28}$ \\
\hline
\end{tabular}
\end{center}

\end{table}

The systematic uncertainties for the parameters are summarized 
in Table~\ref{tab:sys-resparam},
where the sources of the systematic errors from the unfolding procedure (UNF), 
the $q\bar{q}$ background subtraction (BKG1),
the feed-down background subtraction (BKG2), the acceptance correction (ACC) and 
the photon energy scale (PES) are shown separately.
The uncertainty in the $\rho$ mass ($0.7~{\rm MeV}/c^{2}$) is 
mainly due to the uncertainty in the photon energy scale. 
The uncertainty in the $q\bar{q}$ background (BKG1) dominates for the 
$\rho'(1450)$ and $\rho''(1700)$ resonance parameters.

The values of the $\rho(770)$ mass and width are 
consistent with the results of the previous measurements in $\tau$ decay.
For the  $\rho^{\prime}(1450)$, a slightly higher mass value than in the 
previous measurement is obtained. It is found that this value is 
sensitive to the value of the resonance parameters for the 
$\rho''(1700)$ since they both interfere. This is the first time that all 
the parameters for the $\rho(770)$, $\rho'(1450)$ and $\rho''(1700)$ 
are determined in a single fit. 
 Production of the $\rho''(1700)$ 
in $\tau^-$ decays has been unambiguously demonstrated 
and its parameters determined.



\subsection{ Comparison of Belle and previous $\tau$ data}

Comparisons of the pion form factor squared $|F_{\pi}^{-}(s)|^{2}$ 
measured in Belle to those measured by CLEO~\protect{\cite{CLEO2000}}
 and ALEPH~\protect{\cite{ALEPH05}} 
experiments are given in Fig.~\ref{Fpirho}(a) for the $\rho(770)$ region
and in Fig.~\ref{Fpirho}(b) for the low-mass region $\MassSQ < 0.4 \GeVcc2$.
Figure \ref{Fpi_comp} shows a more detailed comparison,
the difference in  $|F_{\pi}(s)|^{2}$ of Belle
and CLEO, ALEPH data for the fit of the Belle data divided by the fit value,
 for the mass-squared range 0.11--1.20~$\GeVcc2$.
It can be seen that over the entire mass range shown, the $|F_{\pi}(s)|^{2}$
values from Belle are consistent within errors with those of CLEO.
Agreement is worse when Belle data are compared to ALEPH data.
 Below $0.5~({\rm GeV}/c^2)^2$ our points are mostly higher
than those of ALEPH, while above $0.7~({\rm GeV}/c^2)^2$  
they are systematically and
 significantly lower.

\begin{figure}[ht!]
\begin{center}
\rotatebox{0}{\includegraphics*[width=0.60\textwidth,clip]
{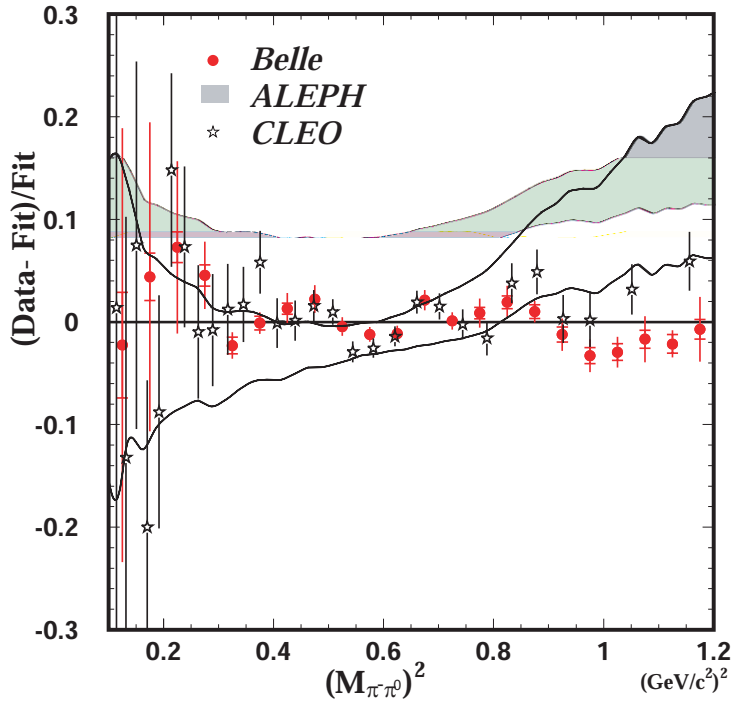}}
\caption
{ 
Comparison of the pion form factor squared $|F_{\pi}(s)|^{2}$ measured
 with the Belle detector to results from the CLEO~\protect{\cite{CLEO2000}}
 and ALEPH~\protect{\cite{ALEPH05}
experiments
 in the $\rho(770)$ 
and $\rho^{\prime}(1450)$ mass region.}  The difference from the fit of  
 the Belle $\Taupipi0$ data divided by the fit value is plotted.
 The solid circles show the Belle result, the open stars show the 
CLEO result~\protect{\cite{CLEO2000}} 
and the hatched band shows the ALEPH
 result~\protect{\cite{ALEPH05}}.
 Both Belle and ALEPH results include the systematic errors. 
The inner error bars in the Belle data indicate the statistical errors.
}
\label{Fpi_comp}
\end{center}
\end{figure}

\section{Implications for the muon anomalous magnetic moment}

\subsection{Basic Formula}
As  described in the introduction,  the hadronic vacuum polarization 
term $a_{\mu}^{\rm had,LO}$
plays an important role in the standard model prediction for  the muon anomalous magnetic moment $a_{\mu}$;
 the error on this contribution
is the most significant source of the uncertainty in $a_{\mu}$.
In this section, we discuss the implication of our measurement for the determination
of $a_{\mu}^{\rm had. LO}$.
 
The value of $a_{\mu}^{\rm had,LO}$ is related to the
$e^+e^-$ annihilation cross section via the dispersion integral
\begin{eqnarray}
a^{\rm had,LO}_{\mu}
= \frac{1}{4\pi^{3}}\int_{4m_{\pi}^{2}}^{\infty} 
\sigma_{\rm had}^{0}(s)
\left( \frac{m_{\mu}^{2}}{3s} \right) \hat{K}(s)\,{\rm d}s\,,
\label{eq:amu}
\end{eqnarray}  
where $\sigma_{\rm had}^{0}$ is the total cross section
for $e^+e^-\!\rightarrow\! {\rm hadrons}$ at the center-of-mass energy $\sqrt{s}$.
The superscript in $\sigma^{0}_{\rm had}$ denotes the "bare" hadronic cross section,
which is defined as the measured cross section, corrected for
 QED radiative corrections
such as
initial state radiation, electron-vertex correction and the vacuum 
polarization in the photon
propagator.
However, the final state radiation (FSR) photons coming from the process 
$\sigma^{0}(e^+e^-\rightarrow V \gamma \rightarrow \pi^+\pi^-\gamma )$
should be included in the $\sigma_{\rm had}^{0}$ term 
(See the detailed discussion on page 12 of  Ref.~\cite{DEHZ}).
The kernel function $\hat{K}(s)$  is given by
\begin{eqnarray}
 \hat{K}(s)&=& \frac{3s}{m_{\mu}^{2}} \left\{  x^2\left(1-\frac{x^2}{2}\right)   
  + (1+x)^2\left( 1+\frac{1}{x^2} \right)
 \biggl( \ln ( 1 + x ) 
 \biggr. \right. 
 \nonumber \\
  &-&  \left. \left. x   
+ \frac{x^2}{2}\right) 
 +   \left( \frac{1+x}{1-x}\right)x^2\ln x \right\}.
\nonumber
\end{eqnarray}
with 
$x=(1-\beta_\mu)/(1+\beta_\mu)$ and
$\beta_\mu\equiv \sqrt{1-4m_\mu^2/s}$\,. 
$\hat{K}(s)$ is 
 a smooth function increasing from 0.63 at
the threshold $s=4m_{\pi}^{2}$ to unity at $s\!=\!\infty$.
Because of  the $1/s$ dependence of $\sigma_{\rm had}^{0}(s)$ and the additional $1/s$ factor
in the integral in Eq.~(\ref{eq:amu}), low-mass hadronic final states 
dominate the contribution to $a_{\mu}^{\rm had,LO}$; 
in fact about 70\% of $a_{\mu}^{\rm had,LO}$ is due to the two-pion 
final state with $4m_{\pi}^{2}\le s\le 0.8$  $\GeVcc2$.
Consequently, the $2\pi$ mass spectrum from
$\tau$ data is useful 
 for obtaining predictions for $a_\mu^{\rm had,LO}$ using CVC.

%
%
%
\begin{table*}[!ht]
\begin{center}
\caption {   Summary of the $a_{\mu}^{\rm \pi\pi}$ contribution from $\tau$ data 
from Belle, ALEPH~\cite{ALEPH05} and CLEO~\cite{CLEO2000} experiments. 
The errors are only shown for the
Belle data. 
In this table, the isospin breaking correction is not applied except for the short-distance radiative correction given by the term $S_{\rm EW}$.  \\
The errors in the  Belle data are only statistical. 
The errors for other experiments are similar or 
slightly worse than the Belle value in the same mass range.
}
\label{tab:amucomp}
\begin{tabular}{c|rrr|rrr}
\hline \hline
\phantom{00}$\MassSQ$ range \phantom{00} & \multicolumn{3}{c|}{$a_{\mu}^{\rm \pi\pi}\, 
(10^{-10})$}& \multicolumn{3}{c}{Integrated $a_{\mu}^{\rm \pi\pi}\,(10^{-10})$}   \\
($\GeVcc2$) &\phantom{00000000} Belle\phantom{0} &\phantom{0} ALEPH & 
\phantom{0}CLEO$^{(*)}$\phantom{0} &
\phantom{0000000} Belle\phantom{0} &\phantom{0} ALEPH & \phantom{0}CLEO$^{(*)}$ \\
\hline
\hline

0.075-0.200	 & 39.55{ $\pm $ 0.97}  &	38.20  &	43.81\phantom{0}	&	 39.55{ $\pm $ 0.97}	& 38.20        & 	    43.81\\
0.200-0.350	  &70.62{ $\pm $ 0.46}   &	66.84	& 75.71\phantom{0}	    &	110.3{ $\pm $ 1.07}	&     105.0	&119.5 \\
0.350-0.500   &123.25{ $\pm $ 0.28}  &	119.10 &	106.10\phantom{0}		&	233.5{ $\pm $ 1.11}&	 224.1     &	225.6 \\
0.500-0.650	  & 196.78{ $\pm $ 0.23} &	194.00 &	197.80\phantom{0}		&	430.3{ $\pm $ 1.13}	& 418.2       &	423.3\\
0.650-0.800    & 62.35{ $\pm $ 0.10}   & 62.35   &	73.03\phantom{0}	   &		492.6{ $\pm $ 1.14}&	 480.5 &	496.3\\
0.800-0.950	  & 15.64{ $\pm $ 0.04}   &	16.40 &	14.31\phantom{0}	     &		508.3{ $\pm $ 1.14}&	 496.9 &	510.6\\
0.950-1.100    &	5.74{ $\pm $ 0.02}  &	6.50  &	6.01\phantom{0}	        &		514.0{ $\pm $ 1.14}&	503.4 &	516.6\\
1.100-1.250    & 2.86{ $\pm $ 0.01}     &	3.27  &	2.32\phantom{0}	     &		516.9{ $\pm $ 1.14}&	506.7  &	519.0\\
1.250-1.400    &	1.65{ $\pm $ 0.01}  &	1.89  &	2.65\phantom{0}	     &		518.5{ $\pm $ 1.14}&	  508.6&	521.6\\
1.400-1.550	  &1.03{ $\pm $ 0.01}      &	1.13  &	0.77\phantom{0}	     &		519.5{ $\pm $ 1.14}&	509.7&	522.4\\
1.550-1.700    &	0.60{ $\pm $ 0.00}  &	0.66   &	0.49\phantom{0}	    &		520.2{ $\pm $ 1.14}&	510.3&	522.9\\
1.700-1.850  &	0.36{ $\pm $ 0.00}     &	0.37  &	0.33\phantom{0}	    &		520.5{ $\pm $ 1.14}&	510.7&	523.2\\
1.850-2.000  &	0.19{ $\pm $ 0.00}     &	0.19  &	0.21\phantom{0}	    &		520.7{ $\pm $ 1.14}&	  510.9&	523.4\\
2.000-2.600     &  0.13{ $\pm $ 0.00}     &   0.21         &      0.15\phantom{0}	            &      521,8{ $\pm $ 1.14}    & 511.1  & 523.6           \\
2.600-3.200     &  0.13{ $\pm $ 0.00}     &   0.21        &       0.04\phantom{0}	           &     522.0{ $\pm $ 1.14}     & 511.1  & 523.6            \\                        
\hline
\hline
\end{tabular}
\begin{minipage}{43em}
\begin{center}
$^{(*)}$\, 
For the CLEO data, the boundary of the mass range is slightly different from the one shown in the first column,
since $\sqrt{s}$ bins are used.
\end{center}
\end{minipage}
\end{center}
\end{table*}

\subsection{Results} 

Details of our determination of $a_{\mu}^{\pi\pi}$, the $2\pi$ 
contribution to $a_{\mu}^{\rm had,LO}$, are given
in the appendix, where the basic formulas, the corrections applied
for the isospin-violating effects, and discussions on the error estimation are presented.

Our result on $a_{\mu}^{\pi\pi}$  over the mass range
$\sqrt{s}=2m_{\pi}-1.8~$$\GeVCC$  is
\begin{eqnarray}
a_{\mu}^{\pi\pi}[2m_{\pi}, 1.8~\GeVCC] = (523.5 \pm 1.5\,{\rm (exp.)} 
 \pm 2.6\, {\rm (Br.)} 
 \pm 2.5 \, 
{\rm (isospin)}  )
\times 10^{-10},
\nonumber 
\end{eqnarray}
where the first error is due to the experimental uncertainties, {\it i.e.} the
statistical error ($1.1\times 10^{-10}$) and experimental systematic  error
($1.0\times 10^{-10}$) added in quadrature. 
The second error comes from  the uncertainties in the branching fractions. 
 The  third one is the error on the isospin-violating 
corrections. These sources of error are discussed in the appendix.

This result can be compared to those from previous ALEPH, CLEO, and OPAL $\tau$ data.
The combined result given in Ref.~\cite{DEHZ} is 
\begin{eqnarray*}
 a_{\mu}^{\pi\pi}[2m_{\pi},1.80~\GeVCC]  = 
( 520.1 \pm 2.4\,{\rm (exp.)} \pm 2.7 ({\rm  Br.}) \pm 2.5\,{\rm (isospin)})
 \times 10^{-10} \quad (\tau ) \,.
%
%
%
\end{eqnarray*}
In terms of  the  experimental error 
(i.e. the first uncertainty),  
our result improves the previous combined result  by 40\%.
A detailed comparison from our results and those of ALEPH~\cite{ALEPH05} 
and CLEO~\cite{CLEO2000}   is
given in Table~\ref{tab:amucomp} for the $a_{\mu}^{\pi\pi}$ contribution.
As seen in Table~\ref{tab:amucomp}, the contribution from the mass-squared region  
$0.8~\GeVcc2<s < 1.25~\GeVcc2$, 
 where one observes a deviation between
the Belle and ALEPH data in Fig.~9, is only 4.6\% of the total $2\pi$ contribution. 
In this region,
 the contribution
from  the ALEPH measurements is higher than that from Belle but
 it is compensated by the opposite tendency  in the region  $s < 0.50~\GeVcc2$.
Moreover, since the contribution from CLEO is between Belle and ALEPH,
the difference between our result and the combined previous-$\tau$ result
becomes smaller.
Consequently,
our result agrees well with that of the combined result 
given from the previous $\tau$ data within $< 1\sigma$ of the experimental error.


On the other hand,
the value of $a_{\mu}^{\pi\pi}$ in the same $\sqrt{s}$ region evaluated from 
the $e^+e^-$ cross section measurements is~\cite{DAV2007} 
\begin{eqnarray*}
 a_{\mu}^{\pi\pi}[2m_{\pi},1.80~\GeVCC] = 
 ( 504.6 \pm 3.1\,{\rm (exp.)} \pm 0.9\,{\rm (rad.)})
 \times 10^{-10} \quad\quad (e^+e^-:{\rm CMD2,SND})
\,,
\end{eqnarray*}
where the first error includes both statistical and experimental systematic 
errors added in  quadrature.
The second error  is due
to radiative corrections.

Our $\tau$ result is
noticeably higher than the $e^+e^-$ result. 
This confirms the longstanding difference between the 
spectral functions of the $2\pi$ systems produced in $\tau$-decay and 
$e^+e^-$ annihilation~\cite{maruce}.

In summary, we have studied the decay $\Taupipi0$ using a
 high-statistics data sample
taken with the Belle detector at the KEKB $e^+e^-$ collider. 
The branching fraction is measured with 1.5\% accuracy.
From Table~\ref{tab:br_comp} we can calculate the accuracy of the
previous experiments:  CLEO 1.7\%, L3 2.4\%, OPAL 1.3\%,
ALEPH 0.5\%, DELPHI 0.9\%.  These comparison shows that the accuracy of the Belle result is better than 
CLEO and L3, similar to OPAL and worse than ALEPH and DELPHI.
The result is in good agreement with  previous measurements. 
In the unfolded $\pi^-\pi^0$ mass spectrum, in addition to the 
$\rho(770)$ and $\rho^{\prime}(1450)$ mesons, 
the production of the $\rho^{\prime\prime}(1700)$ in $\tau^{-}$ 
decays has been unambiguously demonstrated and its parameters determined. 
The unfolded spectrum is used to evaluate  
the 2$\pi$ contribution to the muon anomalous magnetic
moment $a_{\mu}^{\pi\pi}$  in the region 
$\sqrt{s}=2m_{\pi}-1.80~{\rm GeV}/c^{2}$.
Our results agree well 
with the previous $\tau$ based results but 
are higher than those from $e^+e^-$ annihilation.

\section*{Acknowledgments}

We thank M.~Davier and J.~H.~K\"{u}hn for
their advice and encouragement during this analysis.
We thank  S.~Dubni$\breve{\rm c}$ka for a useful discussion about the
BW resonance form.  
We are grateful to G.~L\'opez Castro  for providing  the table of the long-distance
 radiative corrections for  $\Taupipi0$. 
 We thank the KEKB group for the excellent operation of the
accelerator, the KEK cryogenics group for the efficient
operation of the solenoid, and the KEK computer group and
the National Institute of Informatics for valuable computing
and SINET3 network support. We acknowledge support from
the Ministry of Education, Culture, Sports, Science, and
Technology of Japan and the Japan Society for the Promotion
of Science; the Australian Research Council and the
Australian Department of Education, Science and Training;
the National Natural Science Foundation of China under
Contracts No.~10575109 and No. 10775142; the Department of
Science and Technology of India; 
the BK21 program of the Ministry of Education of Korea, 
the CHEP SRC program and Basic Research program 
(Grant No.~R01-2005-000-10089-0) of the Korea Science and
Engineering Foundation, and the Pure Basic Research Group 
program of the Korea Research Foundation; 
the Polish State Committee for Scientific Research; 
the Ministry of Education and Science of the Russian
Federation and the Russian Federal Agency for Atomic Energy;
the Slovenian Research Agency;  the Swiss
National Science Foundation; the National Science Council
and the Ministry of Education of Taiwan; and the U.S.\
Department of Energy.

\newpage
\section*{Appendix}

\subsection{Determination of  $a_{\mu}^{\pi\pi}$ from Belle data}

In this appendix, we list the details of our determination of $a_{\mu}^{\pi\pi}$
over the range $\sqrt{s}=2m_{\pi} - 1.8~ \GeVCC$.
We discuss the basic formula, the integration procedure, the corrections for
 the isospin-violating effects, and the
evaluation of the experimental errors.

\subsubsection{Basic formula}

In the isospin symmetry limit,  CVC relates the quantities in 
$\Taupipi0$ decay to the cross section  $e^+ e^- \rightarrow
\pi^+\pi^-$ ($\sigma^{0}_{\pi\pi}$) through  the relation~\cite{ISB2001}
\begin{eqnarray}
\sigma^{0}_{\pi\pi}|_{\rm CVC}
&=&
\frac{1}{\mathcal{N}(s)\Gamma_{e}^{0}}
\frac{{\rm d}\Gamma(\Taupipi0)}{{\rm d}s}   \nonumber \\
&=&\frac{1}{\mathcal{N}(s)} \times 
\left( \frac{\mathcal{B}_{\pi\pi}}{B_{e}} \right)
\times \left( \frac{1}{N_{\pi\pi}}\frac{{\rm d}N_{\pi\pi}}{{\rm d}s} \right) \,,
\label{eq:cvcsigma}
\end{eqnarray}
where $\mathcal{N}(s)$ is given by
\begin{equation}
\mathcal{N}(s)= \frac{3 |V_{ud}|^{2}}
        {2\pi \alpha^{2}_{0} m_{\tau}^2} 
     s \left( 1 - \frac{s}{m_{\tau}^{2}} \right)^{2}
   \left( 1 + \frac{2s}{m_{\tau}^{2}}  \right) \,.
\label{eq:norm}
\end{equation}
 
A more precise link between hadronic spectral functions from $\tau$
decays and the $e^+e^-$ hadronic cross section requires a calculation of radiative
corrections as well as the inclusion of the isospin breaking effects (both of kinematic
and dynamic origin). These effects have recently been discussed 
 by Cirigliano {\it et al.}~\cite{ISB2001} and by 
Flores-Tlalpa {\it et al.}~\cite{LOP2006}. 
According to them,
the formula in Eq.~(\ref{eq:cvcsigma}) is modified to
\begin{eqnarray}
\sigma^{0}_{\pi^{+}\pi^{-}} |_{\rm CVC}
&=&\frac{1}{\mathcal{N}(s)} \times
\left( \frac{\mathcal{B}_{\pi\pi}}{B_{e}} \right)
\times \left( \frac{1}{N_{\pi\pi}}\frac{{\rm d}N_{\pi\pi}}{{\rm d}s} \right) 
\left(
\frac{R_{\rm IB}(s)}{S_{\rm EW}} \right)\,,
\label{eq:cvcsigma1}
\end{eqnarray}
with
\begin{equation}
R_{\rm IB}(s) = 
 \frac{1}{G_{\rm EM}(s)} \frac{\beta^{3}_{0}(s)}
 {\beta_{-}(s)} \left| \frac{F_{0}(s)}{F_{-}(s)}\right|^{2} \,,
 \label{eq:isocor}
\end{equation}
where  $S_{\rm EW}$ is the dominant short-distance electroweak correction and
 $R_{\rm IB}(s)$ takes care of the isospin-violating corrections.  $R_{\rm IB}(s)$  
 includes the long-distance QED correction $G_{\rm EM}(s)$, the phase space correction
factor $\beta^{3}_{0}(s)/\beta^{3}_{-}(s)$, and the ratio of the pion form factors $|F_{0}(s)/F_{-}(s)|^{2}$.

\subsubsection{ Integration procedure}
Using the measured distribution $(1/N_{\pi\pi})({\rm d}N_{\pi\pi}/{\rm d}s)$,
the moment $a_{\mu}^{\pi\pi}$ can be obtained by inserting the bare cross section
 Eq.~({\ref{eq:cvcsigma1}) to Eq.~(\ref{eq:amu}) and integrating  over 
 $s$.
The integration in  Eq.~({\ref{eq:amu})
is carried out numerically 
by taking the sum of the integrand evaluated at the center of each bin.
The statistical error on $a_{\mu}^{\pi\pi}$ is calculated
including the off-diagonal elements of the covariance error 
matrix $X_{ij}$:
\begin{eqnarray}
\Delta a^{\pi\pi}_\mu & = & \sum_{i,j} 
\left( \frac{\partial a_{\mu}}{\partial\alpha_{i}} \right)
X_{ij}                             
\left( \frac{\partial a_{\mu}}{\partial\alpha_{j}} \right).
\end{eqnarray}

There are several external parameters in these equations; 
the values used for them are listed in Table~\ref{tab:amuext}.
For $m_{\tau}$, $V_{ud}$, and $\mathcal{B}_{e}$, 
PDG~\cite{PDG2006} values are used.
For the $\pi^{-}\pi^{0}$ branching fraction, 
our measurement is consistent with the world average given
in Ref.~\cite{PDG2006}. 
Including our result and the recent ALEPH $\mathcal{B}_{\pi\pi^{0}}$
measurement~\cite{ALEPH05}, the new world average is 
\begin{equation}
\mathcal{B}_{\pi\pi^{0}} = (25.42\pm 0.10)\%.
\end{equation}
We use this new world average for the evaluation of
$a_{\mu}^{\pi\pi}$.

The errors on $a_{\mu}^{\pi\pi}$
arising from external parameters are
summarized in Table \ref{tab:amuext}; the total systematic
error from these sources is $\pm\,  2.7\times 10^{-10}$
 (dominated by $\Delta \mathcal{B}_{\pi\pi^{0}}$).

\begin{table*}[!ht]
\begin{center}
\caption {  Values of the external parameters and  
systematic errors on $a_\mu^{\pi\pi}$ arising from 
these sources.}
\label{tab:amuext}
\begin{tabular}{l|c|c|c|c}
\hline \hline
Source & Value & Relative error   &  $\Delta a_{\mu}^{\pi\pi}$ & Reference \\
&       &       (\%)    &   $(10^{-10})$             &       \\
\hline
\hline
$V_{ud}$    &  $0.97377 \pm 0.00027$      & 0.027  &   $\pm$ 0.26   & 
\cite{PDG2006}  \\
$\mathcal{B}_{e}$    & ($17.84\pm\,0.05$)\%  &  0.28 & $\pm$ 1.45   & 
\cite{PDG2006}  \\
$\mathcal{B}_{\pi\pi^{0}}$ & ($25.42\pm\,0.10$)\% & 0.41 & $\pm$ 2.13   &
\\
\hline
Total external\ \ &                       &           & $\pm$ 2.6 &    \\
\hline
\hline
\end{tabular}

\end{center}
\end{table*}

\begin{table}[!ht]
\begin{center}
\caption { Systematic errors on $a_\mu^{\pi\pi}$ arising from
internal sources (specific to this measurement).}
\label{tab:amuerror2}

\begin{tabular}{l|c}
\hline \hline
Source    &     $\Delta a_{\mu}^{\pi\pi} \times 10^{10}$   \\
\hline\hline
Background:                      &        \\
\ \ \  non-$\tau$\,($e^+e^-\ra\bar{q}q$)   &    $\pm\  0.11$    \\
\ \ \ feed-down $h(n\pi^{0})\nu$     &  $\pm\  0.09$   \\
\ \ \ feed-down $K^{-}\pi^{0}\nu$       & $\pm\   0.15$  \\
Energy scale                     &  $\pm\  0.10 $     \\
$\pi^{0}/\gamma$ selection       &   $\pm\   0.24$    \\
$\gamma$ veto                         & $\pm\  0.93 $   \\
Efficiency:                      &       \\
\ \ \  $\pi^{0}/\gamma$                 &  $\pm\   0.35 $     \\
\ \ \ \ charged track                   &    $ < 0.10$      \\ 
Integration procedure\ \            &    $<  0.10$     \\
\hline
Total internal\ \                                 & 
$\pm$ 1.04    \\
\hline\hline
\end{tabular}
\end{center}

\end{table}

\subsubsection{Experimental systematic uncertainty }

The systematic errors  on $a_\mu^{\pi\pi}$ arising from internal sources (specific 
to this measurement) are listed in Table~\ref{tab:amuerror2} 
 and discussed below.
There are two sources of background in the $\pi^{-}\pi^{0}$ sample:
(i) feed-down from  $\tau^{-}\rightarrow h^{-} (n\pi^{0})\nu_{\tau}$, 
$\tau^{-}\rightarrow K^{-}\pi^{0}\nu_{\tau}$, 
$\tau^{-}\rightarrow \omega \pi^{-}\nu_{\tau} (\omega\rightarrow \pi^{0}\gamma)$
and
(ii) non-$\tau$ background.
In the first case, MC statistics and the uncertainty on the branching fraction
are used to estimate the error.
In the second case,
the uncertainty on the background as estimated from the control
samples is assigned as an error.
As mentioned earlier, the fake-$\pi^{0}$ background 
is subtracted using sideband events and the uncertainty 
is determined by varying the signal and sideband regions.

It is found that the shape of the mass spectrum is insensitive 
to uncertainties in the $\pi^{0}$ efficiency, as it is only
at a few \% level. 
The uncertainty of the integration procedure comes from the binning
effects.
Adding all individual errors in quadrature we obtain a total
error on $a^{\pi\pi}_{\mu}$ arising from 
internal sources of 
$\pm 1.0\times 10^{-10}$.

To check the stability of $a_{\mu}^{\pi\pi}$, we perform 
the following tests:
\begin{enumerate}
\item 
The sample is divided into subsamples
based on the tag-side topology, i.e., 
one electron, one-prong, or three-prong.
The values of $a^{\pi\pi}_{\mu}$ obtained from these 
subsamples are consistent within the statistical errors.
\item 
The sample is divided into subsamples based on the
running period.
Again, the values of $a^{\pi\pi}_\mu$ obtained are 
consistent within the statistical errors.
\item The sample might be  sensitive to the 
requirement on the overlap region between
 the charged track and $\gamma$ clusters. To 
estimate this sensitivity, we select events with a tighter isolation 
requirement on $\gamma$'s and on the track extrapolation: 
30~cm instead of 20~cm.

The resulting variation in $a^{\pi\pi}_{\mu}$ is small 
and is included as an additional systematic error.
\end{enumerate}

\subsubsection{ Isospin-violating corrections for $a_{\mu}^{\rm \pi\pi}$}

Three identifiable sources of isospin breaking corrections are the mass difference
of the charged and neutral pions, $\rho-\omega$ interference effects and the
radiative corrections, which are included in the factor $R_{\rm IB}(s)/S_{\rm EW}$ 
in  Eq.~(\ref{eq:cvcsigma}).
The size of the isospin-violating corrections from these sources and the possible uncertainties from other
sources are summarized 
 in Table~\ref{tab:isospin}.

\begin{table}[!t]
\begin{center}
\caption {   Sources of the isospin violation  between the $e^+e^-$ and $\tau$ spectral function
in the $2\pi$ channel and the corrections to $a_{\mu}^{\pi\pi}$.
The correction is based on the procedure given in
 Refs.~\cite{ISB2001, LOP2006}.
 }
\label{tab:isospin}
\begin{tabular}{lccc}
\hline \hline
Source of  &  Correction to  &\phantom{00} Uncertainty on   &    References \\
 isospin violation & $a_{\mu}^{\pi\pi}\ (10^{-10})$     & \phantom{00} $a_{\mu}^{\pi\pi}\ (10^{-10})$          &               \\
\hline
\hline
Short-distance rad. cor. ($S_{\rm EW}$) &  $-$ 12.0         &   $\pm$ 0.2            & ~\cite{Marciano:1988vm},~\cite{Braaten:1990ef},~\cite{Erler:2002mv},~\cite{DEHZ}    \\
Long-distance rad. cor. ($G_{EM}(s)$)&  $-$ 1.0       &    $--$              & ~\cite{ISB2001},~\cite{LOP2006} \\
$m_{\pi^{-}} \neq m_{\pi^{0}}$ ($\beta^{3}$ in phase space) & $-$ 7.0   &        $--$          &                 \\ 
$\rho-\omega$ interference  &     $+$ 3.5               &     $\pm$ 0.6        &    \\
 $m_{\pi^{-}} \neq m_{\pi^{0}}$  ($\beta^{3}$ in the decay width) &  $+$ 4.2  & $--$ &       \\
  Electromagnetic decay modes
                                                                      &     $-$ 1.4      & $\pm $ 1.4     &                          \\
$ m_{\rho^{0}} \neq  m_{\rho^{-}}$   &      $--$             &  $\pm $ 2.0    &     \\
\hline
 Total                                       &      $-$  13.7            &  $ \pm $ 2.5     &    \\
\hline
\hline
\end{tabular}
\end{center}
\end{table}

\begin{enumerate}
\item[(i)] The dominant contribution from
electroweak radiative corrections comes from the short-distance corrections.
 In the leading logarithmic order, the short-distance radiative corrections to the decays 
 $\tau^-\rightarrow (d\bar{u})\nu_{\tau}$
are enhanced by the factor~\cite{Marciano:1988vm},~\cite{Braaten:1990ef},
\begin{equation}
S(m_{\tau}, m_{Z})
 = 1+ \frac{3\alpha(m_{\tau})}{4\pi} ( 1 + 2 \bar{Q}) \ln \frac{m_{Z}^{2}}{m_{\tau}^{2}} = 1.01878,
\label{eq:SEW1}
\end{equation}
 where $m_{Z}$ is the $Z$ boson mass, and $\alpha(m_{\tau}) = 1/133.50(2)$ is the QED coupling at the $\tau$ lepton mass.
 $\bar{Q}$ is the average quark-doublet charge. Therefore, $\bar{Q}=\frac{1}{2}  (\frac{2}{3} - \frac{1}{3})  = \frac{1}{6}$ for the semileptonic 
decays, $\tau\rightarrow \nu_{\tau} \bar{u}d$.
Since $\bar{Q}= -\frac{1}{2}$ for leptons, there are no leading logarithmic 
corrections for leptonic decays.

  We can go further and sum up all short-distance logarithms of the $\alpha^{n}\ln^{n}m_{Z}$ via  the renormalization group.
This procedure replaces Eq.~(\ref{eq:SEW1})  by
\begin{eqnarray}
   S(m_{\tau},m_{Z}) &=&
    \Bigl[  \frac{ \alpha(m_{b}) }{\alpha( m_{\tau})}    \Bigr]^{\frac{9}{19}}
     \Bigl[  \frac{ \alpha(m_{W}) }{\alpha( m_{m_{b}})}    \Bigr]^{\frac{9}{20}}
      \Bigl[  \frac{ \alpha(m_{Z}) }{\alpha( m_{W})}    \Bigr]^{\frac{36}{17}}  \nonumber \\
      &   \quad  &
       \Bigl[  \frac{ \alpha_{s}(m_{b}) }{\alpha_{s}( m_{\tau})}    \Bigr]^{\frac{3}{25}\frac{\alpha(m_{\tau})}{\pi}}
    \Bigl[  \frac{ \alpha_{s}(m_{Z}) }{\alpha_{s}( m_{b})}    \Bigr]^{\frac{3}{23} \frac{\alpha(m_{b}) } {\pi}  }
    = 1.01907\,, 
 \end{eqnarray}
 where the last two terms are the short-distance QCD corrections~\cite{Erler:2002mv}. 

  Taking into account the subleading correction for the leptonic decay,
 the short-distance Electroweak correction $S_{\rm EW}$  is
 given by
 \begin{equation}
 S_{\rm EW} =  S(m_{\tau},m_{Z}) \frac{1}{S_{\rm EW}^{e,sub}} 
=1.0235\pm 0.0003\,,
 \label{eq:SEW2}
 \end{equation}
where  $    S_{\rm EW}^{e, sub} =\left( 
          1+   \frac{\alpha(m_{\tau}) } {2\pi} \left(  \frac{25}{4} -\pi^{2} 
         \right)
       \right)   = 0.9957 $.
   The difference between the resummed value (1.01907)  and the lowest-order estimate (1.01878)
is taken as the error on $S_{\rm EW}$ and the resummed $S_{EW}$   value is used
throughout this paper.
The shift  of $a_{\mu}^{\pi\pi}$ from this correction is
 $\bigtriangleup a_{\mu}^{\rm \pi\pi} = (12.0\pm 0.2)\times 10^{-10}$.
 Note that this correction is already applied in Eq.~(\ref{eq:fpiweak}) and
the pion form factor given in Table~\ref{tab:fpi2}.
%
    
\item[(ii)]  
The long-distance QED radiative correction $G_{\rm EM}(s)$ 
 was computed in the framework of chiral perturbation theory 
by Cirigliano {\it et al.}  in Ref.~\cite{ISB2001}\,.
 Recently, it was reevaluated based on a meson dominance model
 by  Flores-Tlalpa {\it et al.} in Ref.~\cite{LOP2006}.
 It is found that 
the predictions of both models coincide   if the contribution
of the $\omega(782)$ intermediate state 
$\tau\rightarrow \omega\pi^{-}\nu_{\tau} (\omega\rightarrow \pi^0\gamma)$,
 is excluded in the
latter model calculation. 
Since in our data 
this intermediate $\omega(782)$ contribution is already subtracted at  the analysis level,
we use the $s$-dependent correction factor provided by Flores-Tlalpa, 
which does not include the $\omega$ contribution.
This correction produces a shift of $\bigtriangleup a_{\mu}^{\rm \pi\pi}=-1.0\times 10^{-10}$,

\item[(iii)]    
The $\pi^-$ and $\pi^{0}$ mass difference 
in the ratio of the phase space 
$\beta^{3}_{0}/\beta^{3}_{-}$
results in a shift of $\bigtriangleup a_{\mu}^{\rm \pi\pi }=-7.0 \times 10^{-10}$\,.

\item[(iv)]  
  The $\pi^-$ and $\pi^{0}$ mass difference also affects 
   the $\beta^{3}$
  factor in the energy-dependent decay width (Eq.~\ref{eq:width}), which provides a positive shift
 $\bigtriangleup a_{\mu}^{\rm \pi\pi}=+4.2\times 10^{-10}$\,.

  \item[(v)]    
  The effect of the
 $\rho-\omega$ interference is estimated using the 
 interference amplitude parameterized in the following form~\cite{ISB2001}:
 \begin{equation}
    F_{\pi}^{0}(s)_{\rho-\omega} = - \frac{\theta_{\rho\omega}}
{3m_{\rho}^{2}}
\frac{ s}{ m_{\omega}^{2} -s - i m_{\omega}\Gamma_{\omega}} \,.
\label{eq:rho-omega-cir}
 \end{equation}
  For the numerical evaluation, we take 
$\theta_{\rho\omega} = (-3.3\pm 0.7)\times 10^{-3}~{\rm GeV^{2}}$,
$m_{\omega}=0.783$ GeV and $\Gamma_{\omega}=0.00844$~GeV.
 The net effect
is $\bigtriangleup a_{\mu}^{\rm \pi\pi} = (+3.5\pm 0.6)\times 10^{-10}$\,.
 
  \item[(vi)]  
   The largest source of the uncertainty for the isospin-violating effects 
  is from the $\rho^-$
   and $\rho^{0}$ mass difference.
   The mass difference is consistent with zero within  about 1 MeV,
   which gives an uncertainty on $a_{\mu}^{\rm \pi\pi}$ of
   $\pm 2.0\times 10^{-10}$.
  
  \item[(vii)]   
  Finally, electromagnetic decay of the $\rho$ meson is the source of the isospin violation.
 The decay $\pi\pi\gamma$ deserves particular attention. The shift and its 
uncertainty are estimated
to be   $\bigtriangleup a_{\mu}^{\rm had} = (-1.4\pm 1.4)\times 10^{-10}$
 from the width difference 
$\Gamma(\rho^{0}\rightarrow \pi^+\pi^-\gamma)-\Gamma(\rho^+\rightarrow \pi^+\pi^0\gamma)=
(0.45\pm 0.45)$~MeV~\cite{ISB2001}.           
\end{enumerate}

Summing all  these corrections,
the overall isospin-violating correction and its uncertainty are estimated to be 
$(-13.7   \pm 2.5) \times 10^{-10}$.
These corrections relate the $\tau$ spectral function to the 
"pure" $e^+ e^-\rightarrow \pi^+\pi^-$ with all QED corrections switched off.
Since the $\sigma_{\rm had}^{0}$ term in Eq.(\ref{eq:amu}) must
include processes with the final state radiation (FSR) photons 
coming from the process
$e^+e^-\rightarrow V \gamma \rightarrow \pi^+\pi^-\gamma$,
we must reintroduce the FSR contributions~\cite{DEHZ}. 
After including those contributions ($+ 4.2\times 10^{-10}$)~\cite{DEHZ}, 
the total correction
becomes $(-9.5 \pm 2.5)\times 10^{-10}$.

With  the external parameters in Table~\ref{tab:amuext} and  corrections 
discussed above,
we obtain  $a_{\mu}^{\pi\pi}$ over the range
$\sqrt{s}=2m_{\pi}-1.80~$$\GeVCC$
\begin{eqnarray}
a_{\mu}^{\pi\pi}[2m_{\pi}, 1.80~\GeVCC] = (523.5 \pm 1.1\,{\rm (stat.)} \pm 1.0\,{\rm (sys.)} 
 \pm 2.6\, {\rm (Br.)} 
 \pm 2.5 \, 
{\rm (isospin)}  )
\times 10^{-10},
\nonumber 
\end{eqnarray}
where the first error is statistical,  the second is the experimental systematic (Table~\ref{tab:amuerror2}),
the third comes from the uncertainties on the branching fractions (Table~\ref{tab:amuext}), 
and the fourth is  from isospin-violating corrections (Table~\ref{tab:isospin}).

%

\begin{thebibliography}{99}
\bibitem{CCG} Throughout this paper, 
the inclusion of the charge-conjugate decay mode is implied
unless stated otherwise.
\bibitem{KINE}
The formula for $\beta_{-}(s)$ is an approximate one where 
the term proportional to $m_{\pi^{-}}^{2}- m_{\pi^{0}}^{2}$ is ignored. 
In this approximation, the $\pi^{-}$ and $\pi^{0}$ energies
in the ($\pi^{-}\pi^{0}$) rest-frame are given
by $E_{\pi^{-}}^{*} = E_{\pi^{0}}^{*} = \sqrt{s}/2$.
\bibitem{PDGstat}
See Eq.(32.11), p 302 in Ref.~\cite{PDG2006}. 
For the cases when the difference in the number of degrees of freedom is greater than one, 
the significance can be evaluated by the routine GAUSIN in the CERN library.
\end{thebibliography}
\end{document}